\documentclass[peerreviewcs,romanappendices]{IEEEtran} 
\usepackage{cite}

\ifCLASSINFOpdf
\else
\fi
\usepackage{tikz}
\usetikzlibrary{arrows,decorations,shapes,backgrounds}
\usepackage{pgfplots}
\pgfplotsset{compat=newest}

\usepackage[cmex10]{amsmath}
\usepackage{amsfonts}
\usepackage{amssymb}
\usepackage{bm}
\usepackage{array}

\usepackage{mdwmath}
\usepackage{mdwtab}

\usepackage{booktabs}
\def\0{\mathbf{0}}
\def\A{\mathbf{A}}
\def\B{\mathbf{B}}
\def\C{\mathbf{C}}
\def\c{\mathbf{c}}
\def\Cbb{\mathbb{C}}
\def\eqdef{\triangleq}
\def\D{\mathbf{D}}
\def\diag{\mathrm{diag}}

\def\e{\mathbf{e}}
\def\EE{\mathop{E\/}}
\def\eqq{\!=\!}

\def\f{\mathbf{f}}
\def\G{\mathbf{G}}
\def\Ga{{\bm{\Gamma}}}

\def\g{\mathbf{g}}
\def\h{\mathbf{h}}
\def\herm{\mathsf{H}}
\def\hht{\hat{\mathbf{h}}}
\def\H{\mathbf{H}}
\def\Ht{\hat{\mathbf{H}}}
\def\I{\mathbf{I}}
\def\i{\mathbf{i}}
\def\inn{\!\in\!}

\def\J{\mathbf{J}}

\def\n{\mathbf{n}}

\def\Psib{\underline{\Psi}}
\def\P{\mathbf{P}}
\def\Pb{{\bm{\Psi}}}

\def\q{\mathbf{q}}
\def\Q{\mathbf{Q}}
\def\r{\mathbf{r}}
\def\R{\mathbf{R}}
\def\Rbb{\mathbb{R}}

\def\Rsum{R_{\rm sum}}
\def\Rsumh{\hat{R}_{\rm sum}}

\def\SNR{\rho}
\def\SINR{\gamma}
\def\SINRkrzf{\gamma_{k,\mathrm{rzf}}}
\def\dSINRkrzf{\gamma^\circ_{k,\mathrm{rzf}}}
\def\SINRkzf{\gamma_{k,\mathrm{zf}}}
\def\dSINRkzf{\gamma^\circ_{k,\mathrm{zf}}}
\def\SINRrzf{\gamma_{\mathrm{rzf}}}
\def\dSINRrzf{\gamma^\circ_{\mathrm{rzf}}}

\def\S{\mathbf{S}}

\def\toa{\overset{\alpha\to 0}{\longrightarrow}}
\def\tom{\overset{M\to\infty}{\longrightarrow}}
\def\ton{\overset{N\to\infty}{\longrightarrow}}

\def\tr{\mathrm{tr}}
\def\trans{\mathsf{T}}
\def\T{\mathbf{T}}
\def\Tt{{\bm{\Theta}}}

\def\U{\mathbf{U}}

\def\V{\mathbf{V}}
\def\v{\mathbf{v}}
\def\W{\mathbf{W}}
\def\w{\mathbf{w}}
\def\Wt{\hat{\mathbf{W}}}
\def\WW{\mathcal{W}}
\def\x{\mathbf{x}}

\def\X{\mathbf{X}}

\def\y{\mathbf{y}}

\def\z{\mathbf{z}}
\def\Z{\mathbf{Z}}
\def\zh{\hat{\mathbf{z}}}

\newcounter{ccorollary}
\newcounter{cassumption}

\newcounter{clemma}
\newcounter{cproposition}
\newcounter{cremark}

\newtheorem{corollary}[ccorollary]{Corollary}
\newtheorem{assumption}[cassumption]{Assumption}

\newtheorem{lemma}[clemma]{Lemma}
\newtheorem{proposition}[cproposition]{Proposition}
\newtheorem{remark}[cremark]{Remark}
\newtheorem{theorem}{Theorem}

\definecolor{green}{RGB}{32,127,43}
\pgfcreateplotcyclelist{\defListORZF}{%
  {red,solid},
  {blue,dashed},
  {red,solid},
  {blue,dashed},
  {red,only marks,mark=*},
  {blue,only marks,mark=*},
  {red,only marks,mark=*},
  {blue,only marks,mark=*},
}

\begin{document}


\title{Large System Analysis of Linear Precoding in Correlated MISO
  Broadcast Channels under Limited Feedback}

\author{
Sebastian~Wagner,~\IEEEmembership{Student Member,~IEEE,}
Romain~Couillet,~\IEEEmembership{Member,~IEEE,}
M{\'e}rouane~Debbah,~\IEEEmembership{Senior Member,~IEEE,}
and~Dirk~T.~M.~Slock,~\IEEEmembership{Fellow,~IEEE}
\thanks{S. Wagner is with ST-ERICSSON, Sophia-Antipolis}
\thanks{R. Couillet is with Sup{\'e}lec, Gif sur Yvette, France, e-mail:romain.couillet@supelec.fr.}
\thanks{S. Wagner and D.T.M. Slock are with EURECOM, Sophia-Antipolis, 06904,
Route des Cr\^etes, B.P. 193, France, e-mail:\{sebastian.wagner,~dirk.slock\}@eurecom.fr.}
\thanks{M. Debbah is with the Alcatel-Lucent Chair on Flexible Radio,
SUP{\'E}LEC, Gif sur Yvette, France e-mail: merouane.debbah@supelec.fr.}}%
\maketitle


\begin{abstract}
  In this paper, we study the sum rate performance of zero-forcing (ZF) and regularized ZF (RZF) precoding in large MISO broadcast systems under the assumptions of imperfect channel state information at the transmitter and per-user channel transmit correlation. Our analysis assumes that the number of transmit antennas $M$ and the number of single-antenna users $K$ are large while their ratio remains bounded. We derive deterministic approximations of the empirical signal-to-interference plus noise ratio (SINR) at the receivers, which are tight as $M,K\to\infty$. In the course of this derivation, the per-user channel correlation model requires the development of a novel deterministic equivalent of the empirical Stieltjes transform of large dimensional random matrices with generalized variance profile. The deterministic SINR approximations enable us to solve various practical optimization problems. Under sum rate maximization, we derive (i) for RZF the optimal regularization parameter, (ii) for ZF the optimal number of users, (iii) for ZF and RZF the optimal power allocation scheme and (iv) the optimal amount of feedback in large FDD/TDD multi-user systems. Numerical simulations suggest that the deterministic approximations are accurate even for small $M,K$.
\end{abstract}



%
\IEEEpeerreviewmaketitle

\section{Introduction}
\label{sec:introduction}


\IEEEPARstart{T}{he} pioneering work in \cite{foschini1998lwc} and \cite{telatar1999cmg} revealed that the capacity of a point-to-point (single-user (SU)) multiple-input multiple-output (MIMO) channel can potentially increase linearly with the number of antennas. However, practical implementations quickly demonstrated that in most propagation environments the promised capacity gain of SU-MIMO is unachievable due to antenna correlation and line-of-sight components \cite{gesbert2007single}. In a multi-user (MU) scenario, the inherent problems of SU-MIMO transmission can largely be overcome by exploiting multi-user diversity, i.e., sharing the spatial dimension not only between the antennas of a single receiver, but among multiple (non-cooperative) users. The underlying channel for MU-MIMO transmission is referred to as the MIMO broadcast channel (BC) or MU downlink channel. Although much more robust to channel correlation, the MIMO-BC suffers from inter-user interference at the receivers which can only be efficiently mitigated by appropriate (i.e., channel-aware) pre-processing at the transmitter.


It has been proved that dirty-paper coding (DPC) is a capacity achieving precoding strategy for the Gaussian MIMO-BC \cite{caire2003achievable,viswanath2003sum,yu2004sum,vishwanath2003duality,weingarten2006crg}. However, the DPC precoder is non-linear and to this day too complex to be implemented efficiently in practical systems. It has been shown in \cite{caire2003achievable,peel2005vpt,yoo2006optimality,wiesel2008zero}, that suboptimal \textit{linear} precoders can achieve a large portion of the BC rate region while featuring low computational complexity. Thus, a lot of research has recently focused on linear precoding strategies.

In general, the rate maximizing linear precoder has no explicit form. Several iterative algorithms have been proposed in \cite{christensen2008weighted,shi2008rate}, but no global convergence has been proved. Still, these iterative algorithms have a high computational complexity which motivates the use of further suboptimal
linear transmit filters (i.e., precoders), by imposing more structure into the filter design. A straightforward technique is to precode by the inverse of the channel. This scheme is referred to as channel inversion or zero-forcing (ZF) \cite{caire2003achievable}.


Although \cite{christensen2008weighted,shi2008rate,peel2005vpt} assume perfect channel state information at the transmitter (CSIT) to determine theoretically optimal performance, this assumption is untenable in practice. It is indeed a particularly strong assumption, since the performance of all precoding strategies is crucially depending on the CSIT quality. In practical systems, the transmitter has to acquire the channel state information (CSI) of the downlink channel by feedback signaling from the uplink. Since in practice the channel coherence time is finite, the information of the instantaneous channel state is inherently incomplete. For this reason, a lot of research has been carried out to understand the impact of imperfect CSIT on the system behavior, see \cite{love2008olfs} for a recent survey. 



In this contribution, we focus on the multiple-input single-output (MISO) BC, where a central transmitter equipped with $M$ antennas communicates with $K$ \textit{single}-antenna non-cooperative receivers. We assume $M\!\geq\!K$, i.e., we do not account for user scheduling, and consider ZF and regularized ZF (RZF) precoding under imperfect CSIT (modeled as a weighted sum of the true channel plus noise) as well as per-user channel correlation, i.e., the vector channel $\h_k\inn\Cbb^M$ of user $k$ ($k=1,\dots,K$) satisfies $\EE[\h_k]=0$ and $\EE[\h_k\h_k^\herm]=\Tt_k$. To obtain insights into the system behavior, we approximate the signal-to-interference plus noise ratio (SINR) by a deterministic quantity, where the novelty of this study lies in the \textit{large} system approach. More precisely, we approximate the SINR $\SINR_k$ of user $k$ by a \textit{deterministic equivalent} $\SINR^\circ_k$ such that $\SINR_k - \SINR^\circ_k \to 0$ almost surely, as the system dimensions $M$ and $K$ go jointly to infinity with bounded ratio $1\leq\lim_{M,K\to\infty}\frac{M}{K}=\beta<\infty$. Hence, $\SINR^\circ_k$ becomes more accurate for increasing $M,K$. To derive $\SINR^\circ_k$, we apply tools from the well-established field of large dimensional random matrix theory (RMT) \cite{couillet2011rmm,tulino2004random}. Previous work considered SINR approximations based on \textit{bounds} on the average (with respect to the random channels $\h_k$) SINR. The deterministic equivalent $\SINR^\circ_k$ is not a bound but is a \textit{tight} approximation, for asymptotically large $M,K$. Furthermore, the RMT tools allow us to consider advanced channel models like the per-user correlation model, which are usually extremely difficult to study exactly for finite dimensions. Interestingly, simulations suggest that $\SINR^\circ_k$ is very accurate even for small system dimension, e.g., $M=K=16$. Currently, the 3GPP LTE-Advanced standard \cite{3gpp2008tsg} already defines up to $M=8$ transmit antennas further motivating the application of large system approximations to characterize the performance of wireless communication systems. Subsequently, we apply these SINR approximations to various practical optimization problems.

\subsection{Related Literature}
\label{sec:relevant-literature}

To the best of the authors' knowledge, Hochwald et al. \cite{hochwald2002space} were the first to carry out a large system analysis with $M,K\to\infty$ and finite ratio for linear precoding under the notion of ``channel hardening". In particular, they considered ZF precoding, called channel inversion (CI), for $M>K$ under perfect CSIT, and showed that the SINR for independent and identical distributed (i.i.d.) Gaussian channels converges to $\SNR(\beta -1)$, where $\SNR$ is the signal-to-noise ratio (SNR), independent of the applied power normalization strategy. They go on to derive the sum rate maximizing system loading $\beta^{\star\circ}$ for a fixed $M$. Their results are a special case of our analysis in Section \ref{sec:zero-forc-prec} and Section \ref{sec:sum-rate-maximizing-2}. The authors in \cite{hochwald2002space} conclude by showing that for $\beta>1$, ZF achieves a large fraction of the linear (with respect to $K$) sum rate growth. The work in \cite{peel2005vpt} extends the analysis in \cite{hochwald2002space} to the case $M=K$ and shows that the sum rate of ZF is constant in $M$ as $M,K\to\infty$, i.e., the linear sum rate growth is lost. The authors in \cite{peel2005vpt} counter this problem by introducing a regularization parameter $\alpha$ in the inverse of the channel matrix. Under the assumption of large $M,K$, perfect CSIT and for any rotationally-invariant channel distribution, \cite{peel2005vpt} derives the regularization parameter $\alpha=\alpha^{\star\circ}=\frac{1}{\beta\SNR}$ that maximizes the SINR. Note here that \cite{peel2005vpt} does not apply the classic tools from large dimensional RMT to derive their results but rather find the solution by applying various expectations and approximations. In the present contribution, the RZF precoder of \cite{peel2005vpt} is referred to as \textit{channel distortion-unaware} RZF (RZF-CDU) precoder, since its design assumes perfect CSIT, although in practice, the available CSIT is erroneous or distorted. It has been observed in \cite{peel2005vpt} that the RZF-CDU precoder is very similar to the transmit filter derived under the minimum mean square error (MMSE) criterion \cite{joham2002transmit} and both become identical in the large $M,K$ limit. Likewise, we will observe some similarities between RZF and MMSE filters when considering imperfect CSIT. The RZF precoder in \cite{peel2005vpt} has been extended in \cite{hwang2009rci} to account for channel quantization feedback under random vector quantization (RVQ). The authors in \cite{hwang2009rci} do not apply tools from large RMT but use the same techniques as in \cite{peel2005vpt} and obtain different results for the optimal regularization parameter and SINR compared to our results in Section \ref{sec:optim-feedb-large}.

The first work applying tools from large RMT to derive the asymptotic SINR under ZF and RZF precoding for correlated channels was \cite{couillet2008sfp}. However, in \cite{couillet2008sfp} the regularization parameter of the considered RZF precoder was set to fulfill the total average power constraint. Similar work \cite{nguyen2008mtb} was published later, where the authors considered the RZF precoder in \cite{peel2005vpt} and derived the asymptotic SINR for \textit{uncorrelated} Gaussian channels. Moreover, they derived the asymptotically optimal regularization parameter $\alpha^{\star\circ}=\frac{1}{\beta\SNR}$, already derived in \cite{peel2005vpt}, which is a special case of the result derived in Section \ref{sec:sum-rate-maximizing}. Another work \cite{muharar2011dbt}, reproducing our results, noticed that the optimal regularization parameter in \cite{peel2005vpt,nguyen2008mtb} is independent of transmit correlation when the channel correlation is identical for all users.

In the large system limit and for channels with i.i.d.\ entries, the cross correlations between the user channels, and therefore the users' SINRs, are identical. It has been shown in \cite{wiesel2006lpc} that for this symmetric case and equal noise variances, the SINR maximizing precoder is of closed form and coincides with the RZF precoder. Recently, the authors in \cite{zakhour2010bsc} claimed that indeed the RZF precoder structure emerges as the optimal precoding solution for $M,K\to\infty$. This asymptotic optimality further motivates a detailed analysis of the RZF precoder for large system dimensions.

\subsection{Contributions of the Present Work}
\label{sec:contr-pres-work}




In this paper, we provide a concise framework that directly extends and generalizes the results in \cite{hochwald2002space,peel2005vpt,nguyen2008mtb,muharar2011dbt,muharar2009dbt} by accounting for per-user correlation and imperfect CSIT. Furthermore, we apply our SINR approximations to several limited-feedback scenarios that have been previously analyzed by applying bounds on the ergodic rate of finite dimensional systems. Our main contributions are summarized as follows:
\begin{itemize}
\item Motivated by the channel model, we derive a deterministic equivalent of the empirical Stieltjes transform of matrices with \textit{generalized variance profile}, thereby extending the results in \cite{girko2001tsc,hachem2007deterministic}.
\item We propose deterministic equivalents for the SINR of ZF ($\beta>1$) and RZF ($\beta\geq 1$) precoding under imperfect CSIT and channel with per-user correlation, i.e., deterministic approximations of the SINR, which are independent of the individual channel realizations, and (almost surely) exact as $M,K\to\infty$. 
\item Under imperfect CSIT and \textit{common} correlation ($\Tt_k=\Tt~\forall k$), we derive the sum rate maximizing RZF precoder called \textit{channel-distortion aware} RZF (RZF-CDA) precoder.
\item For ZF and RZF, under common correlation and \textit{different} CSIT qualities, we derive the optimal power allocation scheme which is the solution of a water-filling algorithm.
\end{itemize}

For \textit{uncorrelated} channels, we obtain the following results:

\begin{itemize}
\item Under ZF precoding and imperfect CSIT, a closed-form approximate solution of the number of users $K$ maximizing the sum rate per transmit antenna for a fixed $M$.
\item In large frequency-division duplex (FDD) systems, under RVQ, for $\beta\eqq 1$ and high SNR $\SNR$, to exactly maintain an instantaneous per-user rate gap of $\log_2b$ bits/s/Hz, almost surely as $M,K\to\infty$, the number of feedback bits $B$ per user has to scale with
\begin{itemize}
{\setlength\itemindent{-5pt}
\item RZF-CDA: $B \eqq (M-1)\log_2\SNR - (M-1)\log_2(b^2-1)$
\item RZF-CDU/ZF: $B \eqq (M\!-\!1)\log_2\SNR - (M\!-\!1)\log_22(b\!-\!1)$}
\end{itemize}
That is, the RZF-CDA precoder requires $(M-1)\log_2\frac{b+1}{2}$ bits \textit{less} than RZF-CDU and ZF.
\item In large time-division duplex (TDD) systems with channel coherence interval $T$, at high uplink SNR and downlink SNR $\SNR_{dl}$, the sum rate maximizing amount of channel training scales as $\sqrt{T}$ and $1/\sqrt{\log(\SNR_{dl})}$ for a fixed $\SNR_{dl}$ and $T$, respectively under both RZF-CDA and ZF precoding.
\end{itemize}

The remainder of the paper is organized as follows. Section \ref{sec:system-model} presents the transmission model and channel model. In Section \ref{sec:determ-equiv-sinr}, we propose deterministic equivalents for the SINR of RZF and ZF precoding. In Section \ref{sec:sum-rate-maximizing}, we derive the sum rate maximizing regularization under RZF precoding. Section \ref{sec:sum-rate-maximizing-1} studies the sum rate maximizing number of users for ZF precoding and the optimal power allocation when the CSIT quality of the users is unequal. Section \ref{sec:optim-feedb-large} analyses the optimal amount of feedback in a large FDD system. In Section \ref{sec:optim-train-large}, we study a large TDD system and derive the optimal amount of uplink channel training. Finally, in Section \ref{sec:conclusion}, we summarize our results and conclude the paper.

Most technical poofs are presented in the appendix. In these proofs, we apply several lemmas collected in Appendix \ref{sec:coll-import-lemm}.

\textit{Notation}: In the following, boldface lower-case and upper-case characters denote vectors and matrices, respectively. The operators $(\cdot)^{\herm}$, $\tr(\cdot)$ and $\EE[\cdot]$ denote conjugate transpose, trace and expectation, respectively. The $N\!\times\!N$ identity matrix is denoted $\I_N$, $\log(\cdot)$ is the natural logarithm and $\Im(z)$ is the imaginary part of $z\inn\Cbb$. $\|\X\|$ and $\lambda_{\min}(\X)$ are the spectral radius and the minimum eigenvalue of the Hermitian matrix $\X$, respectively. The imaginary unit is denoted $\i$. The sets $\Rbb^+$ and $\Cbb^+$ are defined as $\{x: x>0\}$ and $\{x\eqq r+\i v: r\inn\Rbb, v>0\}$. A random vector $\x\sim\mathcal{CN}(\mathbf{m},\Tt)$ is complex Gaussian distributed with mean vector $\mathbf{m}$ and covariance matrix $\Tt$.


\section{System Model}
\label{sec:system-model}

This section describes the transmission model as well as the underlying channel model.

\subsection{Transmission Model}
\label{sec:transmission-model}

Consider a MISO broadcast channel composed of a central transmitter equipped with $M$ antennas and of $K$ single-antenna non-cooperative receivers. We assume $M\geq K$, thus user scheduling is not taken into account. Furthermore, we suppose narrow-band transmission. The signal $y_k$ received by user $k$ at any time instant reads
\begin{equation*}
  \label{eq:system-model}
  y_k = \h_k^\herm\x + n_k, \quad k=1,2,\dots,K,
\end{equation*}
where $\h_k\inn\Cbb^M$ is the random channel from the transmitter to user $k$, $\x\inn\Cbb^{M}$ is the transmit vector and the noise terms $n_k\sim \mathcal{CN}(0,\sigma^2)$ are independent. We assume that the channel $\h_k$ evolves according to a block-fading model, i.e., the channel is constant at every time instant but varies \textit{independently} from one time instant to another.

The transmit vector $\x$ is a linear combination of the independent user symbols $s_k$ and can be written as
\begin{equation*}
  \label{eq:linear-transformation}
  \x = \sum_{k=1}^K \sqrt{p_k}\g_k s_k,
\end{equation*}
where $\g_k\inn\Cbb^{M}$ and $p_k\!\geq\!0$ are the precoding vector and the signal power of user $k$, respectively. Subsequently, we assume that user $k$ has perfect knowledge of $\h_k$ and the effective channel $\h_k^\herm\g_k$. In particular, an estimate of $\h_k^\herm\g_k$ can be obtained through dedicated downlink training by precoding the pilots of user $k$ by $\g_k$. The precoding vectors are normalized to satisfy the \textit{average} total power constraint
\begin{equation}
  \label{eq:transmit-power-constraint}
  \EE[\|\x\|^2] = \tr(\P\G^\herm\G) \leq P,
\end{equation}
where $\G\!\eqdef\![\g_1,\g_2,\dots,\g_K]\inn\Cbb^{M\times K}$, $\P\eqq \diag(p_1,\dots,p_K)$ and $P>0$ is the total available transmit power.

Denote $\SNR\!\eqdef\!P/\sigma^2$ the SNR. Under the assumption of Gaussian signaling, i.e., $s_k\sim\mathcal{CN}(0,1)$ and single-user decoding with perfect channel state information at the receivers, the SINR $\SINR_k$ of user $k$ is defined as \cite{cover2006elements}
\begin{equation}
  \label{eq:sinr-linear-precoding}
  \SINR_k = \frac{p_k|\h_k^{\herm}\g_k|^2}{\displaystyle\sum_{j=1,j\neq k}^Kp_j|\h_k^{\herm}\g_j|^2 + \sigma^2}.
\end{equation}
The rate $R_k$ of user $k$ is given by
\begin{equation}
  \label{eq:1}
  R_k = \log\left(1 + \SINR_k \right)
\end{equation}
and the ergodic sum rate is defined as
\begin{equation}
  \label{eq:sum-rate}
  \Rsum = \sum_{k=1}^K{\EE}\left[R_k\right],
\end{equation}
where the expectation is taken over the random channels $\h_k$.

\subsection{Channel Model}
\label{sec:channel-model}

Each user channel $\h_k$ is modeled as
\begin{equation}
  \label{eq:2}
  \h_k = \sqrt{M}\Tt_k^{1/2}\z_k, 
\end{equation}
where $\Tt_k$ is the channel correlation matrix of user $k$ and $\z_k$ has i.i.d.\ complex entries of zero mean and variance $1/M$. The channel transmit correlation matrices $\Tt_k$ are assumed to be slowly varying compared to the channel coherence time and thus are supposed to be perfectly known to the transmitter, whereas receiver $k$ has only knowledge about $\Tt_k$. Moreover, only an imperfect estimate $\hht_k$ of the true channel $\h_k$ is available at the transmitter which is modeled as \cite{makouei2011mia,ding2009mmt,wang2006adm,yoo2004mcc}
\begin{equation}
  \label{eq:3}
  \hht_k = \sqrt{M}\Tt_k^{1/2}\left(\sqrt{1 - \tau_k^2}\z_k + \tau_k
    \q_k\right) = \sqrt{M}\Tt_k^{1/2}\zh_k,
\end{equation}
where $\zh_k = \sqrt{1 - \tau_k^2}\z_k + \tau_k\q_k$, $\q_k$ has i.i.d.\ entries of zero mean and variance $1/M$ independent of $\z_k$ and $n_k$. The parameter $\tau_k\inn[0,1]$ reflects the accuracy or quality of the channel estimate $\hht_k$, i.e., $\tau_k=0$ corresponds to perfect CSIT, whereas for $\tau_k=1$ the CSIT is completely uncorrelated to the true channel. The variation in the accuracy of the available CSIT $\hht_k$ between the different user channels $\h_k$ arises naturally. Firstly, there might be low mobility users and high mobility users with large or small channel coherence intervals, respectively. Therefore, the CSIT of the high mobility users will be outdated quickly and hence be very inaccurate. On the other hand, the CSIT of the low mobility users remains accurate since their channel does not change significantly from the time of the channel estimation until the time of precoding and coherent data transmission. Secondly, different CSIT qualities arise when the feedback rate varies among the users. For instance, if the CSIT is obtained from uplink training, the training length of each user could be different, leading to different channel estimation errors at the transmitter. Similarly, if the users feed back a quantized channel, they could use channel quantization codebooks of different sizes depending on their channel quality and the available uplink resources. However, for simplicity, we assume identical CSIT qualities $\tau_k = \tau~\forall k$ for the optimization problems considered in Section \ref{sec:optim-feedb-large} and Section \ref{sec:optim-train-large}.

\begin{remark}
  The model for imperfect CSIT in \eqref{eq:3}, is adequate for instance in a FDD system, where the channel $\h_k$ is \textit{finely} quantized using a random codebook of i.i.d.\ vectors. Since the correlation matrices $\Tt_k$ are known at both ends, user $k$ solely quantizes the fast fading channel component $\z_k$ to the closest codebook vector $\zh_k$, which can be accurately approximated as $\zh_k=\sqrt{1 - \tau_k^2}\z_k + \tau_k\q_k$. Subsequently, the user sends the codebook index back to the transmitter, where the estimated downlink channel is reconstructed by multiplying with $\sqrt{M}\Tt_k^{1/2}$. For uncorrelated channels, this specific FDD system is studied in Section \ref{sec:optim-feedb-large}. 
\end{remark}

Define the compound estimated channel matrix $\Ht\eqdef [\hht_1,\hht_2,\dots,\hht_K]^\herm\inn\Cbb^{K\times M}$. Therefore, the matrix $\frac1M\Ht^\herm\Ht$ can be written as
\begin{equation}
  \label{eq:136}
  \frac1M\Ht^\herm\Ht = \sum_{k=1}^K\Tt_k^{1/2}\zh_k\zh_k^\herm\Tt_k^{1/2}.
\end{equation}

The per-user channel correlation model (also called generalized variance profile) is very general and encompasses various propagation environments. For instance, all channel coefficients $h_{k,i}$ of the vector channel $\h_k$ may have different variances $\sigma^2_{k,i}$ resulting from different attenuation of the signal while traveling to the receivers. This so called variance profile of the vector channel is obtained by setting $\Tt_k = \diag(\sigma_{k,1}^2,\sigma^2_{k,2},\ldots,\sigma^2_{k,M})$, see \cite{girko2001tsc,hachem2007deterministic,hachem2008clt}. Another possible scenario consists of an environment where all user channels have identical transmit correlation $\Tt$, but where the users are heterogeneously scattered around the transmitter and hence experience different channel gains $d_k$. Such a setup can be modeled with $\Tt_k=d_k\Tt$. From a mathematical point of view, a homogeneous system with common user channel correlation $\Tt_k=\Tt~\forall k$ is very attractive. In this case, the user channels are statistically equivalent and the deterministic SINR approximations can be computed by solving a single implicit equation instead of multiple systems of coupled implicit equations. A further simplification occurs when the channels are uncorrelated $\Tt_k=\I_M~\forall k$, in which case the approximated SINRs are given explicitly.

The model in \eqref{eq:136} has never been considered in large dimensional RMT and therefore no results are available. The most general model studied, assumes a variance profile, first treated in \cite{girko2001tsc} and extended in \cite{hachem2007deterministic}, which is a special case of the model in \eqref{eq:136}. Therefore, to be able to derive deterministic equivalents of the SINR, we need to extend the results in \cite{girko2001tsc,hachem2007deterministic} to account for the per-user correlation model in \eqref{eq:136}, which is done in the next section.

\section{A Deterministic Equivalent of the SINR}
\label{sec:determ-equiv-sinr}

This section introduces deterministic approximations of the SINR under RZF and ZF precoding for various assumptions on the transmit correlation matrices $\Tt_k$. These results will be used in Sections \ref{sec:sum-rate-maximizing}-\ref{sec:optim-train-large} to solve practical optimization problems.

The following theorem extends the results in \cite{silverstein1995empirical,girko2001tsc,hachem2007deterministic} by assuming a generalized variance profile. This theorem is required to cope with the channel model in \eqref{eq:2} and forms the mathematical basis of the subsequent large system analysis of the MISO BC under RZF and ZF precoding.

\begin{theorem}
  \label{th:1}
  Let $\B_N\eqq \X_N^\herm\X_N + \S_N$ with $\S_N\inn\Cbb^{N\times N}$ Hermitian nonnegative definite and $\X_N\inn \Cbb^{n\times N}$ random. The $i$th column $\x_i$ of $\X_N^\herm$ is $\x_i\eqq\Pb_i\y_i$, where the entries of $\y_i\inn\Cbb^{r_i}$ are i.i.d.\ of zero mean, variance $1/N$ and have eighth order moment of order $O\left(\frac{1}{N^4}\right)$. The matrices $\Pb_i\inn\Cbb^{N\times r_i}$ are deterministic. Furthermore, let $\Tt_i\eqq\Pb_i\Pb_i^\herm\inn\Cbb^{N\times N}$ and define $\Q_N\inn\Cbb^{N\times N}$ deterministic. Assume $\limsup_{N\to\infty}\sup_{1\leq i\leq n}\|\Tt_i\| < \infty$ and let $\Q_N$ have uniformly bounded spectral norm (with respect to $N$). Define
  \begin{equation}
     \label{eq:det-eq}
     m_{\B_N,\Q_N}(z)\eqdef \frac1N\tr\Q_N\left(\B_N - z\I_N\right)^{-1}.
  \end{equation}
  Then, for $z\inn\Cbb\setminus\Rbb^+$, as $n,N$ grow large with ratios $\beta_{N,i}\!\eqdef\!N/r_i$ and $\beta_N\!\eqdef\! N/n$ such that $0\!<\!\lim\inf_N\beta_N\!\leq\!\lim\sup_N\beta_N\!<\!\infty$ and $0\!<\!\lim\inf_N\beta_{N,i}\!\leq\!\lim\sup_N\beta_{N,i}\!<\!\infty$, we have that
  \begin{equation}
    \label{eq:mN-mNo}
    m_{\B_N,\Q_N}(z) - m^\circ_{\B_N,\Q_N}(z) \ton 0,
  \end{equation}
  almost surely, with $m^\circ_{\B_N,\Q_N}(z)$ given by
  \begin{align}
    \label{eq:mBk}
    m^\circ_{\B_N,\Q_N}(z)\! =\!
    \frac1N\tr\Q_N\!\left(\frac1N\!\sum_{j=1}^n\frac{\Tt_j}{1\! +\! e_{N,j}(z)}\! +\! \S_N\! -\! z\I_N\right)^{-1}
  \end{align}
  where the functions $e_{N,1}(z),\dots,e_{N,n}(z)$ form the unique solution of
  \begin{align}
    \label{eq:eKz}
    e_{N,i}(z) = \frac1N\tr\Tt_i\left(\frac1N\sum_{j=1}^n\frac{\Tt_j}{1\! +\! e_{N,j}(z)}\! +\! \S_N\! -\!z\I_N\right)^{-1}
  \end{align}
  which is the Stieltjes transform of a nonnegative finite measure on $\Rbb^+$. Moreover, for $z\!<\!0$, the scalars $e_{N,1}(z),\dots,e_{N,n}(z)$ are the unique nonnegative solutions to \eqref{eq:eKz}.
\end{theorem}
Note that \eqref{eq:eKz} forms a system of $n$ coupled equations, from which \eqref{eq:mBk} is given explicitly. 
\begin{IEEEproof}
  The proof of Theorem \ref{th:1} is given in Appendix
  \ref{sec:proof-theorem:1}.
\end{IEEEproof}

\begin{proposition}[Convergence of the Fixed Point Algorithm]
  \label{sec:main-result-1}
  Let $z\inn\Cbb\!\setminus\!\Rbb^+$ and $\{e_{N,i}^{(k)}(z)\}$
  ($k\!\geq\!0$) be the sequence defined by $e_{N,i}^{(0)}(z)\eqq
  -\frac1z$ and
  \begin{equation}
    \label{eq:33}
    e_{N,i}^{(k)}(z) = \frac1N\tr\Tt_i\left(\frac1N\sum_{j=1}^n\frac{\Tt_j}{1 + e_{N,j}^{(k-1)}(z)} + \S_N -z\I_N\right)^{-1}
  \end{equation}
  for $k\!>\!0$. Then, $\lim_{k\to\infty}e_{N,i}^{(k)}(z)\eqq
  e_{N,i}(z)$ defined in \eqref{eq:eKz} for $i\inn\{1,2,\dots,n\}$.
\end{proposition}
\begin{IEEEproof}
  The proof of Proposition \ref{sec:main-result-1} is given in
  Appendix \ref{sec:proof-exist-uniq} and \ref{sec:proof-conv-determ}.
\end{IEEEproof}

To derive a deterministic equivalent of the SINR under RZF and ZF precoding, we require the following assumptions on the correlation matrices $\Tt_k$ and the power allocation matrix~$\P$.

\begin{assumption}
  \label{as:0}
  All correlation matrices $\Tt_k$ have uniformly bounded spectral norm on $M$, i.e.,
  \begin{equation}
    \label{eq:134}
    \limsup_{M,K\to\infty}\sup_{1\leq k\leq K}\|\Tt_k\| < \infty.
  \end{equation}
\end{assumption}


\begin{assumption}
  \label{as:0ap}
  The power $p_{\max} = \max(p_1,\dots,p_K)$ is of order $O(1/K)$, i.e.,
  \begin{equation}
    \label{eq:135ap}
    \|\P\| = O(1/K).
  \end{equation}
\end{assumption}

\subsection{Regularized Zero-forcing Precoding}
\label{sec:regul-zero-forc}

Consider the RZF precoding matrix 
\begin{equation}
  \label{eq:precoding-matrix-rci}
  \G_{\rm rzf} = \xi\left(\Ht^\herm\Ht + M\alpha\I_M \right)^{-1}\Ht^{\herm},
\end{equation}
where $\Ht\!\eqdef\![\hht_1,\hht_2,\dots,\hht_K]^\herm\inn\Cbb^{K\times M}$ is the channel estimate available at the transmitter, $\xi$ is a normalization scalar to fulfill the power constraint \eqref{eq:transmit-power-constraint} and $\alpha\!>\!0$ is the regularization parameter. Here, $\alpha$ is scaled by $M$ to ensure that $\alpha$ itself converges to a constant, as $M,K\to\infty$.


From the total power constraint \eqref{eq:transmit-power-constraint}, we obtain $\xi^2$ as
\begin{equation*}
  \label{eq:1a}
  \xi^2 = \frac{P}{\tr\P\Ht(\Ht^\herm\Ht + M\alpha\I_M)^{-2}\Ht^{\herm}} = \frac{P}{\Psi},
\end{equation*}
where we defined $\Psi\eqdef \tr\P\Ht(\Ht^\herm\Ht + M\alpha\I_M)^{-2}\Ht^{\herm}$. Denoting $\Wt\!\eqdef\!(\Ht^\herm\Ht + M\alpha\I_M)^{-1}$, the SINR $\SINRkrzf$ of user $k$ in \eqref{eq:sinr-linear-precoding} under RZF precoding takes the form
\begin{equation}
  \label{eq:sinr}
  \SINRkrzf = \frac{p_k|\h_k^{\herm}\Wt\hht_k|^2}{\h_k^{\herm}\Wt\Ht_{[k]}^{\herm}\P_{[k]}\Ht_{[k]}\Wt\h_k
    + \frac{\Psi}{\SNR}},
\end{equation}
where $\Ht_{[k]}\!\eqdef\![\hht_1,\dots,\hht_{k-1},\hht_{k+1},\dots,\hht_K]^{\herm}\inn\Cbb^{K-1\times
  M}$ and $\P_{[k]}\!\eqdef\!\diag(p_1,\dots,p_{k-1},p_{k+1},\dots,p_K)$.

To derive a deterministic equivalent $\SINRkrzf^\circ$ of the SINR $\SINRkrzf$ defined in \eqref{eq:sinr} such that $\SINRkrzf - \SINRkrzf^\circ\tom 0$, almost surely, we require the following assumption.

\begin{assumption}
  \label{as:1a}
  The random matrix $\frac1M\Ht^\herm\Ht$ has uniformly bounded spectral norm on $M$ with probability one, i.e.,
  \begin{equation}
    \label{eq:5as}
    \limsup_{M,K\to\infty}\left\|\frac1M\Ht^\herm\Ht\right\| < \infty,
  \end{equation}
  with probability one.
\end{assumption}
\begin{remark}
  \label{re:1a}
  Assumption \ref{as:1a} holds true if $\sup_K|\{\Tt_k:k=1,2,\dots,K\}|<\infty$, where $|\mathcal{A}|$ denotes the cardinality of the set $\mathcal{A}$. That is, $\{\Tt_k\}$ belongs to a \textit{finite} family \cite{couillet2009dcm}. In particular, if $\Tt_k=\Tt~\forall k$, then Assumption \ref{as:1a} is satisfied, since $\frac1M\|\Ht^\herm\Ht\|\leq\|\Tt\|\|\hat{\Z}^\herm\hat{\Z}\|$, where $\hat{\Z}=[\hat{\z}_1,\dots,\hat{\z}_K]^\herm$ and both $\|\Tt\|$ and $\|\hat{\Z}^\herm\hat{\Z}\|$ are uniformly bounded for all large $M$ with probability one \cite{bai1998neo}.

\end{remark}

A deterministic equivalent $\dSINRkrzf$ of $\SINRkrzf$ is provided in the following theorem.

\begin{theorem}
  \label{th:2}
Let Assumptions \ref{as:0}, \ref{as:0ap}, and \ref{as:1a} hold true and let $\alpha>0$ and $\SINRkrzf$ be the SINR of user $k$ defined in \eqref{eq:sinr}. Then
  \begin{equation}
    \label{eq:theorem-rzf-sinr-1}
    \SINRkrzf - \SINRkrzf^\circ\tom 0,
  \end{equation}
  almost surely, where $\SINRkrzf^\circ$ is given by
  \begin{equation}
    \label{eq:sinr-rzf}
    \SINRkrzf^\circ = \frac{p_k(1-\tau_k^2)\left(m_k^{\circ}\right)^2}
    {\Upsilon_k^\circ(1 - \tau_k^2[ 1 - (1 + m_k^\circ)^2])+\frac{\Psi^\circ}{\SNR}(1+m_k^{\circ})^2},
  \end{equation}
  with $m_k^{\circ}=e_k$, where the $e_1,\dots,e_K$ form the unique positive solutions of
  \begin{align}
    \label{eq:110}
    e_i &= \frac1M\tr\Tt_i\T \\
    \label{eq:tt}
    \T &= \left(\frac1M\sum_{j=1}^K\frac{\Tt_j}{1 + e_j} + \alpha\I_M\right)^{-1}
  \end{align}
  and $\Psi^\circ$ and $\Upsilon_k^\circ$ read
  \begin{align}
    \label{psi_rzf}
     \Psi^\circ &= \frac1M\sum_{j=1}^K\frac{p_je_j^\prime}{(1 + e_j)^2},\\
    \label{ups_rzf}
     \Upsilon_k^\circ &= \frac1M\sum_{j=1,j\neq k}^K\frac{p_j e_{j,k}^\prime}{(1 + e_j)^2},
  \end{align}
  with $\e^\prime=[e_1^\prime,\dots,e_K^\prime]^\trans$ and $\e_k^\prime=[e_{1,k}^\prime,\dots,e_{K,k}^\prime]^\trans$ given by
\begin{align}
  \label{eq:166a}
  \e^\prime &= \left(\I_K - \J\right)^{-1}\v,\\
  \label{eq:166b}
  \e_k^\prime &= \left(\I_K - \J\right)^{-1}\v_k,
\end{align}
where $\J$, $\v$ and $\v_k$ take the form
\begin{align*}
  [\J]_{ij} &= \frac{\frac1M\tr\Tt_i\T\Tt_j\T}{M(1+e_j)^2}, \\
  \v &= \left[\frac1M\tr\Tt_1\T^2,\dots,\frac1M\tr\Tt_K\T^2 \right]^\trans,\\
  \v_k &= \left[\frac1M\tr\Tt_1\T\Tt_k\T,\dots,\frac1M\tr\Tt_K\T\Tt_k\T \right]^\trans.
\end{align*}
\end{theorem}
\begin{IEEEproof}
  The proof of Theorem \ref{th:2} is given in Appendix \ref{sec:proof-theorem:2}.
\end{IEEEproof}

\begin{corollary}
\label{sec:regul-zero-forc-1}
Let Assumptions \ref{as:0} and \ref{as:0ap} hold true and let $\alpha>0$ and $\Tt_k=\Tt~\forall k$, then $\SINRkrzf^\circ$ takes the form
  \begin{equation}
    \label{eq:9}
    \SINRkrzf^\circ = \frac{\frac{p_k}{P/K} m^\circ(1-\tau_k^2)\left[e_{22} + \alpha\beta(1+m^\circ)^2e_{12}\right]}{e_{22}(1-\frac{p_k}{P})\left[1-\tau_k^2(1\!-\!(1+m^\circ)^2)\right]+\frac{e_{12}}{\SNR}(1+m^\circ)^2},
  \end{equation}
  where $m^\circ$ is the unique positive solution of
  \begin{align}
    \label{eq:32}
    m^\circ &= \frac1M\tr\Tt\T \\
    \label{eq:32t}
    \T &= \left(\frac{\Tt/\beta}{1+m^{\circ}}
      + \alpha\I_M\right)^{-1}
  \end{align}
  and $e_{ij}$ is given by
  \begin{equation}
    \label{eq:69}
    e_{ij} = \frac{1}{(1+m^\circ)^j}\frac1M \tr\Tt^i\T^j.
  \end{equation}
\end{corollary}
\begin{IEEEproof}
  Substituting $\Tt_k=\Tt~\forall k$ into Theorem \ref{th:2}, we have $e_i=m_k^\circ=m^\circ$ given in \eqref{eq:32}, $e^\prime_i=e^\prime=[\beta(1+m^\circ)^2e_{12}]/(\beta-e_{22})$ and $e_{i,k}^\prime=\tilde{e}^\prime=[\beta(1+m^\circ)^2e_{22}]/(\beta-e_{22})$. Therefore, the terms $\Psi^\circ$ and $\Upsilon_k^\circ$ become $(P/K)e_{12}/(\beta-e_{22})$ and $(P/K[1-p_k/P])e_{22}/(\beta-e_{22})$, respectively. Furthermore, $m^\circ$ can be written as
  \begin{align}
    m^\circ &= \frac1M\tr\Tt\T\left(\frac{\Tt/\beta}{1+m^\circ} + \alpha\I_M\right)\T\nonumber\\
    \label{eq:59}
    &= \alpha(1+m^\circ)^2e_{12}+\frac1\beta(1+m^\circ)e_{22}.
  \end{align}
  Substituting these terms into \eqref{eq:sinr-rzf} yields \eqref{eq:9} which completes the proof.
\end{IEEEproof}
Note that under Assumption \ref{as:0ap}, the term $\frac{p_k}{P}$ in \eqref{eq:9} can be omitted since the convergence in \eqref{eq:theorem-rzf-sinr-1} still holds true. We will make use of this simplification when studying different applications of the SINR approximations.


\begin{corollary}
  \label{sec:regul-zero-forc-11}
  Let Assumption \ref{as:0ap} hold true and let $\alpha>0$ and $\Tt_k=\I_M~\forall k$, then $\dSINRkrzf$ takes the form
  \begin{equation}
    \label{eq:139a}
    \dSINRkrzf=\frac{\frac{p_k}{P/K}m^\circ(1-\tau_k^2)\left[1+\alpha\beta(1+m^\circ)^2\right]}{(1-\frac{p_k}{P})\left[1-\tau_k^2(1-(1+m^\circ)^2)\right] + \frac1\SNR(1+m^\circ)^2},
  \end{equation}
  where $m^\circ$ is given as
  \begin{align}
    \label{eq:32a}
    m^\circ = \frac{\beta-1-\beta\alpha+\sqrt{(\beta-1)^2+2(1+\beta)\alpha\beta + \alpha^2\beta^2}}{2\alpha\beta}.
  \end{align}  
\end{corollary}
\begin{IEEEproof}
  Substituting $\Tt=\I_M$ into Corollary \ref{sec:regul-zero-forc-1}, we have $e_{12}=e_{22}$ which yields \eqref{eq:139a}. Moreover, \eqref{eq:32} becomes a quadratic equation in $m^\circ$ with unique positive solution \eqref{eq:32a}, which completes the proof.
\end{IEEEproof}

In particular, we will consider two different RZF precoders. The first RZF precoder is defined by $\alpha=\frac{1}{\beta\SNR}$ and is referred to as RZF \textit{channel distortion unaware} (RZF-CDU) precoder. Under imperfect CSIT the RZF-CDU precoder is mismatched to the true channel. The second RZF precoder is called RZF \textit{channel distortion aware} (RZF-CDA) precoder and does account for imperfect CSIT. The optimal regularization parameter for the RZF-CDA precoder is derived in Section \ref{sec:sum-rate-maximizing}. 

Moreover, there are two limiting cases of the RZF precoder corresponding to $\alpha\to\infty$ and $\alpha\to 0$. For $\alpha\to\infty$ the RZF precoder converges to the matched filter (MF) precoder $\G_{\rm mf} = \xi\Ht^\herm$. A deterministic equivalent $\SINR_{k,\rm mf}^\circ$ for the MF precoder can be derived by taking the limit $\SINR_{k,\rm mf}^\circ = \lim_{\alpha\to\infty}\dSINRkrzf$. However, since the performance of the MF precoder is rather poor and $\SINR_{k,\rm mf}^\circ$ does not involve Stieltjes transforms anymore, we will not discuss this precoding scheme in the present work. The reader is referred to \cite{wagner2011phd} or \cite{hoydis2011mmm} for a detailed large system analysis of the MF precoder. In the case of $\alpha\to 0$, the RZF precoder converges to the ZF precoder, which is discussed in the next section.   

\subsection{Zero-forcing Precoding}
\label{sec:zero-forc-prec}

For $\alpha\eqq 0$, the RZF precoding matrix in \eqref{eq:precoding-matrix-rci} reduces to the ZF precoding matrix $\G_{\rm zf}$ which reads
\begin{equation*}
  \label{eq:precoding-matrix-ci}
  \G_{\rm zf} = \xi\Ht^{\herm}\left(\Ht\Ht^\herm\right)^{-1},
\end{equation*}
where $\xi$ is a scaling factor to fulfill the power constraint
\eqref{eq:transmit-power-constraint} and is given by
\begin{equation*}
  \xi^2 = \frac{P}{\tr\P(\Ht\Ht^\herm)^{-1}} = \frac{P}{\Psib},
\end{equation*}
where $\Psib\eqdef \tr\P(\Ht\Ht^\herm)^{-1}$. Defining $\underline{\hat{\W}}\!\eqdef\!\Ht^\herm(\Ht\Ht^\herm)^{-2}\Ht$, the SINR $\SINRkzf$ of user $k$ in \eqref{eq:sinr-linear-precoding} under ZF precoding reads
\begin{equation}
  \label{eq:sinr-zf}
  \SINRkzf = \frac{p_k|\h_k^{\herm}\underline{\hat{\W}}\hht_k|^2}{\h_k^{\herm}\underline{\hat{\W}}\Ht_{[k]}^{\herm}\P_{[k]}\Ht_{[k]}\underline{\hat{\W}}\h_k
    + \frac{\Psib}{\SNR}}.
\end{equation}

To obtain a deterministic equivalent of the SINR in \eqref{eq:sinr-zf}, we need to ensure that the minimum eigenvalue of $\Ht\Ht^\herm$ is bounded away from zero for all large $M$, almost surely. Therefore, the following assumption is required.
\begin{assumption}
  \label{as:1}
  There exists $\varepsilon>0$ such that, for all large $M$, we have $\lambda_{\min}(\frac1M\Ht\Ht^\herm)>\varepsilon$ with probability one.
\end{assumption}
\begin{remark}
  \label{re:1}
  If $\Tt_k=\Tt~\forall k$ and $\lambda_{\min}(\Tt)>\varepsilon>0$ (i.e., in contrast to Theorem \ref{th:2}, $\Tt$ must be invertible), for all $M$, then Assumption \ref{as:1} holds true if $\beta>1$. Indeed, for $\beta>1$, from \cite{bai1998neo}, there exists $\zeta>0$ such that, for all large $M$, $\lambda_{\min}(\hat{\Z}\hat{\Z}^\herm)>\zeta$, where $\hat{\Z}=[\hat{\z}_1,\dots,\hat{\z}_K]^\herm$, with probability one. Therefore, for all large $M$, $\lambda_{\min}(\frac1M\Ht\Ht^\herm)\geq\lambda_{\min}(\hat{\Z}\hat{\Z}^\herm)\lambda_{\min}(\Tt)>\zeta\varepsilon>0$ almost surely.
\end{remark}

Furthermore, we require the following assumption for the channel model with per-user correlation.
\begin{assumption}
  \label{as:1b}
  Assume that $\underline{e}_{i} = \lim_{\alpha\to 0}\alpha e_i(\alpha)$ exists for all $i$ and $\underline{e}_{i}>\varepsilon$ $\forall i$ for some $\varepsilon>0$, for all $M$.
\end{assumption}
\begin{remark}
  Under these conditions, the $\underline{e}_1,\dots,\underline{e}_K$ are the unique positive solutions of \eqref{eq:50}. In particular, Assumption \ref{as:1b} holds true if $\Tt_k=\Tt~\forall k$, $\beta>1$ and $\lambda_{\min}(\Tt)>\varepsilon>0$. This is detailed in the proof of Corollary \ref{co:1}.
\end{remark}

\begin{theorem}
\label{th:3}
Let Assumptions \ref{as:0}, \ref{as:0ap}, \ref{as:1a}, \ref{as:1} and \ref{as:1b} hold true and let $\SINRkzf$ be the SINR of user $k$ under ZF precoding defined in \eqref{eq:sinr-zf}. Then 
\begin{equation*}
  \label{eq:theorem-zf-sinr-1}
  \SINRkzf - \SINRkzf^\circ\tom 0,
\end{equation*}
almost surely, where $\SINRkzf^\circ$ is given by
\begin{equation}
  \label{eq:de-zf-puc}
  \SINRkzf^\circ = p_k\frac{1-\tau_k^2}{\tau_k^2\underline{\Upsilon}_k^\circ + \frac{\Psib^\circ}{\SNR}},
\end{equation}
where $\Psib^\circ$ and $\underline{\Upsilon}_k^\circ$ read
\begin{align}
  \Psib^\circ &= \frac1M\sum_{j=1}^K\frac{p_j}{\underline{e}_j},\nonumber\\
  \label{eq:zf-upsilon}
  \underline{\Upsilon}_k^\circ &= \frac1M\sum_{j=1,j\neq k}^Kp_j\frac{\underline{e}_{j,k}^\prime}{\underline{e}_j^2}.
\end{align}
The functions $\underline{e}_{1},\dots,\underline{e}_K$ form the unique positive solution of
\begin{align}
  \label{eq:50}
  \underline{e}_{i} &= \frac1M\tr\Tt_i\underline{\T} \\
  \label{eq:6tzf}
  \underline{\T} &= \left(\frac1M\sum_{j=1}^K\frac{\Tt_j}{\underline{e}_j} + \I_M\right)^{-1}.
\end{align}
Further, define $\underline{\e}_k^\prime=[\underline{e}_{1,k}^\prime,\dots,\underline{e}_{K,k}^\prime]^\trans$, which is given as
\begin{align}
  \label{eq:28df}
  \underline{\e}_k^\prime = \left(\I_K - \underline{\J}\right)^{-1}\underline{\v}_k,
\end{align}
where $\underline{\J}$ and $\underline{\v}_k$ take the form
\begin{align*}
  [\underline{\J}]_{ij} &= \frac{\frac1M\tr\Tt_i\underline{\T}\Tt_j\underline{\T}}{M\,\underline{e}_j^2}, \\
  \underline{\v}_k &= \left[\frac1M\tr\Tt_1\underline{\T}\Tt_k\underline{\T},\dots,\frac1M\tr\Tt_K\underline{\T}\Tt_k\underline{\T} \right]^\trans.
\end{align*}
\end{theorem}
\begin{IEEEproof}
  The proof of Theorem \ref{th:3} is given in Appendix \ref{sec:proof-theorem:3}.
\end{IEEEproof}

\begin{corollary}
  \label{co:1}
  Let Assumptions \ref{as:0} and \ref{as:0ap} hold true. Further, let $\beta>1$, $\Tt_k\eqq\Tt~\forall k$ with $\lambda_{\min}(\Tt)>\varepsilon$, $\varepsilon>0$, for all $M$, then Theorem \ref{th:3} holds true and $\SINRkzf^\circ$ takes the form
  \begin{equation*}
    \label{eq:sinr-zf-1}
    \SINRkzf^\circ = \frac{p_k}{P/K}\frac{1-\tau_k^2}{\tau_k^2\underline{\Upsilon}^\circ\left[1 - \frac{p_k}{P} \right] + \frac{\Psib^\circ}{\SNR}}
  \end{equation*}
  with
  \begin{align}
    \label{eq:64}
    \Psib^\circ &= \frac1{\beta\underline{e}},\\
    \label{eq:66}
    \underline{\Upsilon}^\circ  &=\frac{\underline{e}_2/\underline{e}^2}{\beta-\underline{e}_2/\underline{e}^2},\\
    \underline{e}_2 &= \frac1M\tr\Tt^2\underline{\T}^2\nonumber
  \end{align}
  where $\underline{e}$ is the unique positive solution of
  \begin{align}
    \label{eq:zf-ca}  
    \underline{e} &= \frac1M\tr\Tt\underline{\T},\\
    \underline{\T} &= \left(\I_M+\frac1{\underline{e}\beta}\Tt\right)^{-1}.
  \end{align}
\end{corollary}
\begin{IEEEproof}
  For $\Tt_k\eqq\Tt~\forall k$, we obtain from \eqref{eq:110}
  \begin{align}
    \underline{e}_{i} &= \lim_{\alpha\to 0}\alpha e_i(\alpha)=\underline{e} \nonumber \\
     &= \lim_{\alpha\to 0}\left\{\frac1M\tr\Tt\left(\frac{1}{\beta}\frac{\Tt}{\alpha + \alpha e(\alpha)} + \I_M \right)^{-1} \right\}\nonumber\\
    \label{eq:37}
    &= \frac1M\tr\Tt\left(\frac{\Tt}{\beta \underline{e}} + \I_M \right)^{-1}.
  \end{align}
A lower bounded of \eqref{eq:37} is given as $\underline{e} \geq \lambda_{\min}(\Tt)(1-1/\beta)$ which is uniformly bounded away from zero if $\Tt$ is invertible and $\beta>1$. Thus, under these conditions, Assumption \ref{as:1b} is satisfied. Moreover, the $\underline{e}_{j,k}^\prime$ in \eqref{eq:28df} rewrite 
\begin{equation*}
  \underline{e}_{j,k}^\prime = \underline{e}^\prime =\frac{\beta\underline{e}_2}{\beta-\frac{\underline{e}_2}{\underline{e}^2}}
\end{equation*}
and therefore, 
\begin{equation*}
  \underline{\Upsilon}^\circ_k = \frac{\underline{e}_2/\underline{e}^2}{\beta-\frac{\underline{e}_2}{\underline{e}^2}}\frac{P}{K}\left[1-\frac{p_k}{P}\right].
\end{equation*}
Dividing $\underline{\Upsilon}^\circ_k$ by $\frac{P}{K}\left[1-\frac{p_k}{P}\right]$ and $\underline{\Psi}^\circ=\frac{P}{\underline{e}M}$ by $P/K$, we obtain $\underline{\Upsilon}^\circ$ given in \eqref{eq:66} and $\underline{\Psi}^\circ$ given in \eqref{eq:64}, respectively, which completes the proof.
\end{IEEEproof}

\begin{corollary}
  \label{co:2}
   Let Assumption \ref{as:0ap} hold true and let $\beta>1$ and $\Tt_k\eqq\I_M~\forall k$, then $\SINRkzf^\circ$ takes the explicit form
  \begin{equation}
    \label{eq:cor-zf-sinr}
    \SINRkzf^\circ = \frac{p_k}{P/K}\frac{1-\tau_k^2}{\tau_k^2[1-\frac{p_k}{P}] + \frac1\SNR}(\beta-1).
  \end{equation}
\end{corollary}
\begin{IEEEproof}
 By substituting $\Tt=\I_M$ into \eqref{eq:zf-ca}, $\underline{e}$ is explicitly given by $\underline{e}\eqq(\beta-1)/\beta$. We further have $\frac{\underline{e}_2}{\underline{e}^2}=1$ and $\underline{\Psi}^\circ = \underline{\Upsilon}^\circ = (\beta-1)^{-1}$.
\end{IEEEproof}

\subsection{Rate Approximations}
\label{sec:appr-syst-sum}

We are interested in the individual rates $R_k$ of the users as well as the average system sum rate $\Rsum$. Since the logarithm is a continuous function, by applying the continuous mapping theorem \cite{billingsley2008probability}, it follows from the almost sure convergence $\SINR_k - \SINR_k^\circ\tom 0$, that 
\begin{equation}
  \label{eq:7}
  R_k - R_k^\circ \tom 0,
\end{equation}
almost surely, where $R_k^\circ=\log(1+\SINR_k^\circ)$. An approximation $\Rsumh$ of the ergodic sum rate $\Rsum$ is obtained by replacing the instantaneous (i.e., without averaging over the channel distribution) SINR $\SINR_k$ with its large system approximation $\SINR^\circ_k$, i.e.,
\begin{equation}
  \label{eq:sum-rate-de}
  \Rsumh = \sum_{k=1}^K\log\left(1 + \SINR^\circ_k \right).
\end{equation}
It follows that
\begin{equation}
  \label{eq:53}
  \frac1K\left(\Rsum - \Rsumh\right)\tom 0,
\end{equation}
holds true almost surely.

Another quantity of interest is the rate gap between the achievable rate under perfect and imperfect CSIT. We define the rate gap $\Delta R_k$ of user $k$ as 
\begin{equation}
  \label{eq:139b}
  \Delta R_k\eqdef \bar{R}_k-R_k,
\end{equation}
where $\bar{R}_k$ is the rate of user $k$ under perfect CSIT, i.e., for $\tau_k^2=0~\forall k$. Then, from \eqref{eq:7} it follows that a deterministic equivalent $\Delta R_k^\circ$ of the rate gap of user $k$ such that
\begin{equation*}
  \label{eq:58}
  \Delta R_k - \Delta R_k^\circ \tom 0,
\end{equation*}
almost surely, is given by
\begin{equation}
  \label{eq:42}
  \Delta R_k^\circ = \bar{R}_k^\circ - R_k^\circ,
\end{equation}
where $\bar{R}_k^\circ$ is a deterministic equivalent of the rate of user $k$ under perfect CSIT.

Since we will require the per-user rate gaps for uncorrelated channels ($\Tt_k=\I_M~\forall k$) in the limited feedback analysis in Sections \ref{sec:optim-feedb-large} and \ref{sec:optim-train-large}, we introduce hereafter $\Delta R_k^\circ$ for RZF-CDU and ZF precoding.

\begin{corollary}[RZF-CDU precoding]
  \label{lemma-rzf-lfb-2}
  Let $\Tt_k\eqq\I_M~\forall k$, $p_k=P/K~\forall k$, $\tau^2_k=\tau^2~\forall k$ and define $\Delta R_{k,\rm rzf-cdu}$ as the rate gap of user $k$ under RZF-CDU precoding. Then a deterministic equivalent $\Delta R_{k,\rm rzf-cdu}^\circ=\Delta R_{\rm rzf-cdu}^\circ$ such that
  \begin{equation*}
    \Delta R_{k,\rm rzf-cdu} - \Delta R^\circ_{\rm rzf-cdu}\tom 0
  \end{equation*}
  almost surely, is given by
  \begin{align*}
    \Delta R_{\rm rzf-cdu}^\circ = \log\left(\frac{1 + m^\circ}{1 +\frac{m^\circ(1-\tau^2)\left[1+\frac1\SNR(1+m^\circ)^2\right]}{1-\tau^2+(1+m^\circ)^2[\tau^2+\frac1\SNR]}}\right),
  \end{align*}
  where $m^\circ$ is given in \eqref{eq:32a}.
\end{corollary}
\begin{IEEEproof}
  With Corollary \ref{sec:regul-zero-forc-11}, compute $\Delta R^\circ_{\rm rzf-cdu}$ as defined in \eqref{eq:42}, where $\bar{R}_{\rm rzf-cdu}^\circ=\log(1+m^\circ)$.
\end{IEEEproof}

\begin{corollary}[ZF precoding]
  \label{lemma-zf-lfb}
  Let $\Tt_k\eqq\I_M~\forall k$, $p_k=P/K~\forall k$ and define $\Delta R_{k,\rm zf}$ to be the rate gap of user $k$ under ZF precoding. Then
  \begin{equation*}
    \Delta R_{k,\rm zf} - \Delta R^\circ_{k,\rm zf}\tom 0
  \end{equation*}
  almost surely, with $\Delta R^\circ_{k,\rm zf}$ given by
  \begin{equation*}
    \Delta R^\circ_{k,\rm zf} = 
      \log\left(\frac{1+\SNR(\beta-1)}{1 + \SNR\omega_k(\beta-1)}\right)
  \end{equation*}
  where $\omega_k$ is defined given by
  \begin{equation}
    \label{eq:omega-ak}
    \omega_k = \frac{1-\tau_k^2}{1 + \tau_k^2\SNR}.
  \end{equation}
\end{corollary}
\begin{IEEEproof}
  Substitute the SINR from Corollary \ref{co:2} into \eqref{eq:42}. 
\end{IEEEproof}

\begin{remark}
  \label{rem:dct}
  In practice, one is often interested in the average system performance, e.g., the ergodic SINR $\EE[\SINR_k]$ or ergodic rate $\EE[R_k]$. Since the SINR $\SINR_k$ is uniformly bounded on $M$ for the considered precoding schemes, we can apply the dominated convergence theorem \cite[Theorem 16.4]{billingsley2008probability} and obtain
\begin{equation*}
  \EE[\SINR_k] - \SINR^\circ_k \tom 0,
\end{equation*}
where the expectation is taken over the probability space generating the sequence $\{\H(\omega),~M\geq 1\}$ with $\H=[\h_1,\dots,\h_K]^\herm\inn\Cbb^{K \times M}$. The same holds true for the per-user rate $R_k$, i.e., $\EE[R_k] - R^\circ_k \tom 0$.
\end{remark}

\subsection{Numerical Results}
\label{sec:determ-equiv-sum}

We validate Theorem \ref{th:2} and Theorem \ref{th:3} by comparing the ergodic sum rate \eqref{eq:sum-rate}, obtained by Monte-Carlo (MC) simulations of i.i.d.\ Rayleigh block-fading channels, to the large system approximation $\Rsumh$, for finite system dimensions and equal power allocation $\P\eqq\frac1K\I_K$.

The correlation $\Tt_k$ of the $k$th user channel is modeled as in \cite{jakes1994microwave} by assuming a diffuse two-dimensional field of isotropic scatterers around the receivers. The waves impinge the receiver $k$ uniformly at an azimuth angle $\theta$ ranging from $\theta_{k,\rm min}$ to $\theta_{k,\rm max}$. Denoting $d_{ij}$ the distance between transmit antenna $i$ and $j$, the correlation is modeled as
\begin{equation}
  \label{eq:10c}
  [\Tt_k]_{ij} = \frac1{\theta_{k,\rm max} - \theta_{k,\rm
      min}}\int_{\theta_{k,\rm min}}^{\theta_{k,\rm
      max}}e^{~\mathbf{i}\frac{2\pi}{\lambda}d_{ij}\cos(\theta)}d\theta,
\end{equation}
where $\lambda$ denotes the signal wavelength. The users are assumed to be distributed uniformly around the transmitter at an angle $\varphi_k\eqq 2\pi k/K$ and as a simple example, we choose $\theta_{k,\rm min}\eqq -\pi$ and $\theta_{k,\rm max}\eqq\varphi_k-\pi$. Note that for small $\theta_{k,\rm max} - \theta_{k,\rm min}$ (in our example for small values of $k$), the corresponding signal of user $k$ is highly correlated since the signal arrives from a very narrow angle. Thus, the correlation model \eqref{eq:10c} yields rank-deficient correlating matrices for some users. The transmitter is equipped with a uniform linear array (ULA). To ensure that $\|\Tt_k\|$ is bounded as $M$ grows large, we assume that the distance between adjacent antennas is independent of $M$, i.e., the length of the ULA increases with $M$.

\begin{figure}[t]
  \centering
  \begin{tikzpicture}
  \tikzstyle{every pin}=[fill=white,draw=black]
    \pgfplotsset{every axis legend/.append style={
        cells={anchor=west}, at={(0.98,0.98)}, anchor=north east}}
    \pgfplotsset{every axis/.append style={line width=0.5pt}}
    \pgfplotsset{every axis/.append style={mark options=solid, mark size=2.5pt}}

    \begin{semilogyaxis}[xlabel={$M$}, ylabel={$(\Rsum-\Rsum^\circ)/\Rsum$},
      grid=major, xmin=3, xmax=40, xtick={3,5,10,...,40}, ymin=0,
      ymax=1,ytickten={-2,-1,0}]

      \addplot[red, solid, mark=o] plot coordinates { 
(3.000,0.090) (5.000,0.060) (10.000,0.032) (15.000,0.022)   (20.000,0.018) (25.000,0.015) (30.000,0.012) (35.000,0.011) (40.000,0.010) };

\addplot[blue, solid, mark=square] plot coordinates { (3.000,0.04429) (5.000,0.02598) (10.000,0.01480)   (15.000,0.00998) (20.000,0.00807) (25.000,0.00634) (30.000,0.00524) (35.000,0.00459) (40.000,0.00456)};

      \legend{ {$\Tt_k\neq\I_M$, $\tau_k^2=0.1$}\\
        {$\Tt_k=\I_M$, $\tau_k^2=0$}\\};
      
    \end{semilogyaxis}
  \end{tikzpicture}
  \caption{RZF, $(\Rsum-\Rsum^\circ)/\Rsum$ vs. $M$ for a fixed SNR of $\SNR=10$ dB with $M\eqq K$,     $\alpha=1/\SNR$.}
  \label{fig:de-rzf-sinr}
\end{figure}

\begin{figure}[t]
  \centering
  \begin{tikzpicture}
  \tikzstyle{every pin}=[fill=white,draw=black]
    \pgfplotsset{every axis legend/.append style={
        cells={anchor=west}, at={(0.02,0.98)}, anchor=north west}}
    \pgfplotsset{every axis/.append style={line width=0.5pt}}
    \pgfplotsset{every axis/.append style={mark options=solid, mark size=0.5pt}}

    \begin{axis}[xlabel={$\SNR$ [dB]}, ylabel={sum rate [bits/s/Hz]},
      grid=major, xmin=0, xmax=30, xtick={0,5,...,30}, ymin=0,
      ymax=180, ytick={0,20,...,180},cycle list name=\defListORZF]

      \addplot plot coordinates { (0.000,20.981) (5.000,37.234) (10.000,57.093) (15.000,79.145)         (20.000,102.451) (25.000,126.463) (30.000,150.871) };

      \addplot plot coordinates { (0.000,18.866) (5.000,33.529) (10.000,51.988) (15.000,73.000)         (20.000,95.493) (25.000,118.638) (30.000,141.825) };

      \addplot plot coordinates { (0.000,18.630) (5.000,31.223) (10.000,41.633) (15.000,44.224)         (20.000,38.320) (25.000,28.561) (30.000,19.245) };

      \addplot plot coordinates { (0.000,16.778) (5.000,28.439) (10.000,39.182) (15.000,43.611)         (20.000,39.832) (25.000,31.450) (30.000,22.741) };

      \addplot plot[error bars/.cd,y dir=both,y explicit, error bar
      style={mark size=2.5pt}] coordinates { (0.000,21.074)
        +-(0.464,0.464) (5.000,37.359) +-(0.758,0.758) (10.000,57.405)
        +-(1.307,1.307) (15.000,79.918) +-(2.247,2.247)
        (20.000,103.871) +-(3.723,3.723) (25.000,129.395)
        +-(6.445,6.445) (30.000,157.366) +-(11.948,11.948)};

      \addplot plot[error bars/.cd,y dir=both,y explicit, error bar style={mark size=2.5pt}] coordinates {         (0.000,19.021) +-(0.504,0.504) (5.000,33.698) +-(0.779,0.779) (10.000,52.203) +-(1.261,1.261)         (15.000,73.293) +-(2.030,2.030) (20.000,96.239) +-(3.387,3.387) (25.000,119.845) +-(4.977,4.977)         (30.000,143.725) +-(7.927,7.927)};

      \addplot plot[error bars/.cd,y dir=both,y explicit, error bar style={mark size=2.5pt}] coordinates {         (0.000,18.767) +-(0.479,0.479) (5.000,31.428) +-(0.821,0.821) (10.000,42.046) +-(1.316,1.316)         (15.000,45.124) +-(1.800,1.800) (20.000,40.116) +-(2.402,2.402) (25.000,31.276) +-(2.978,2.978)         (30.000,22.872) +-(3.342,3.342)};

      \addplot plot[error bars/.cd,y dir=both,y explicit, error bar style={mark size=2.5pt}] coordinates {         (0.000,17.001) +-(0.521,0.521) (5.000,28.643) +-(0.810,0.810) (10.000,39.742) +-(1.221,1.221)         (15.000,44.574) +-(1.819,1.819) (20.000,41.617) +-(2.356,2.356) (25.000,34.222) +-(2.893,2.893)         (30.000,26.679) +-(3.370,3.370)};

      \draw (axis cs:15,120) node[fill=white,draw=black] (pint0) {$\tau_k^2\eqq 0$}; \draw (axis cs:20,100)       node[draw,black,ellipse,minimum height=1.0cm] (ell0) {}; \draw[black] (pint0) -- (ell0);

      \draw (axis cs:15,20) node[fill=white,draw=black] (pint1) {$\tau_k^2\eqq 0.1$}; \draw (axis cs:20,40)       node[draw,black,ellipse,minimum height=0.5cm] (ell1) {}; \draw[black] (pint1) -- (ell1);

      \legend{ {$\Tt_k\eqq\I_M$}\\
               {$\Tt_k\!\neq\!\I_M$}\\};

    \end{axis}
  \end{tikzpicture}
  \caption{RZF, sum rate vs. SNR with $M\eqq K\eqq 30$ and $\alpha=1/\SNR$, simulation results are indicated     by circle marks with error bars indicating the standard deviation.}
  \label{fig:de-rzf-2}
\end{figure}

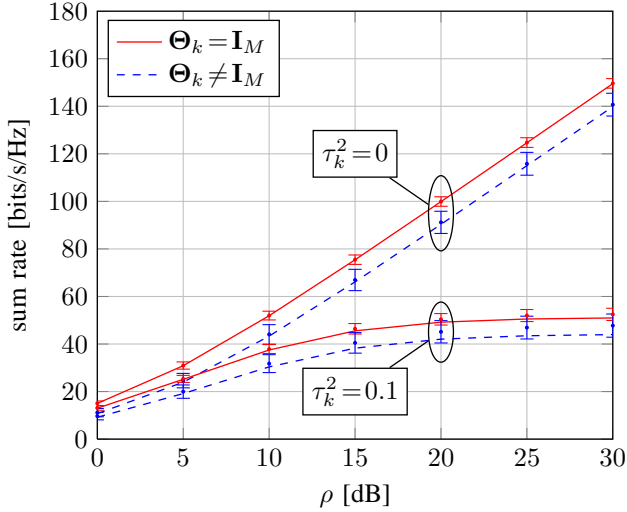
\begin{figure}[t]
  \centering
  \begin{tikzpicture}
  \tikzstyle{every pin}=[fill=white,draw=black]
    \pgfplotsset{every axis legend/.append style={
        cells={anchor=west}, at={(0.02,0.98)}, anchor=north west}}
    \pgfplotsset{every axis/.append style={line width=0.5pt}}
    \pgfplotsset{every axis/.append style={mark options=solid, mark size=0.5pt}}

    \begin{axis}[xlabel={$\SNR$ [dB]}, ylabel={sum rate [bits/s/Hz]},
      grid=major, xmin=0, xmax=30, xtick={0,5,...,30}, ymin=0,
      ymax=180, ytick={0,20,...,180},cycle list name=\defListORZF]

      \addplot plot coordinates { (0.000,15.005) (5.000,30.869) (10.000,51.901) (15.000,75.428)         (20.000,99.884) (25.000,124.652) (30.000,149.519) };

      \addplot plot coordinates { (0.000,10.706) (5.000,23.946)
        (10.000,43.314) (15.000,66.130) (20.000,90.336)
        (25.000,115.022) (30.000,139.864) };

      \addplot plot coordinates { (0.000,12.997) (5.000,25.155)
        (10.000,37.495) (15.000,45.542) (20.000,49.193)
        (25.000,50.527) (30.000,50.970) };

      \addplot plot coordinates { (0.000,9.192) (5.000,19.164)
        (10.000,30.373) (15.000,38.259) (20.000,42.045)
        (25.000,43.479) (30.000,43.963) };

      \addplot plot[error bars/.cd,y dir=both,y explicit, error bar style={mark size=2.5pt}] coordinates {         (0.000,15.058) +-(1.001,1.001) (5.000,30.952) +-(1.520,1.520) (10.000,51.971) +-(1.838,1.838)         (15.000,75.477) +-(1.980,1.980) (20.000,99.939) +-(1.990,1.990) (25.000,124.738) +-(2.025,2.025)         (30.000,149.631) +-(1.999,1.999)};

      \addplot plot[error bars/.cd,y dir=both,y explicit, error bar style={mark size=2.5pt}] coordinates {         (0.000,11.143) +-(1.751,1.751) (5.000,24.608) +-(3.037,3.037) (10.000,44.038) +-(4.112,4.112)         (15.000,66.941) +-(4.485,4.485) (20.000,91.175) +-(4.667,4.667) (25.000,115.810) +-(4.798,4.798)         (30.000,140.684) +-(4.793,4.793)};

      \addplot plot[error bars/.cd,y dir=both,y explicit, error bar style={mark size=2.5pt}] coordinates {         (0.000,13.107) +-(0.948,0.948) (5.000,25.333) +-(1.483,1.483) (10.000,37.893) +-(1.883,1.883)         (15.000,46.412) +-(2.200,2.200) (20.000,50.393) +-(2.412,2.412) (25.000,51.976) +-(2.534,2.534)         (30.000,52.473) +-(2.570,2.570)};

      \addplot plot[error bars/.cd,y dir=both,y explicit, error bar style={mark size=2.5pt}] coordinates {         (0.000,9.697) +-(1.597,1.597) (5.000,19.976) +-(2.804,2.804) (10.000,31.753) +-(3.773,3.773)         (15.000,40.489) +-(4.324,4.324) (20.000,45.145) +-(4.701,4.701) (25.000,46.915) +-(4.791,4.791)         (30.000,47.743) +-(4.871,4.871)};

      \draw (axis cs:15,120) node[fill=white,draw=black] (pint0) {$\tau_k^2\eqq 0$}; \draw (axis cs:20,95)       node[draw,black,ellipse,minimum height=1.0cm] (ell0) {}; \draw[black] (pint0) -- (ell0);

      \draw (axis cs:15,20) node[fill=white,draw=black] (pint1) {$\tau_k^2\eqq 0.1$}; \draw (axis cs:20,45)       node[draw,black,ellipse,minimum height=0.8cm] (ell1) {}; \draw[black] (pint1) -- (ell1);

      \legend{ {$\Tt_k\eqq\I_M$}\\
               {$\Tt_k\!\neq\!\I_M$}\\};

    \end{axis}
  \end{tikzpicture}
  \caption{ZF, sum rate vs. SNR with $M\eqq 30$, $K\eqq 15$, simulation results are indicated by circle marks with error bars indicating the standard deviation.}
  \label{fig:de-zf-2}
\end{figure}

The simulation results presented in Figure \ref{fig:de-rzf-sinr} depict the absolute error of the sum rate approximation $\Rsumh$ compared to the ergodic sum rate $\Rsum$, averaged over $10\,000$ independent channel realizations. The notation ``$\Tt_k\!\neq\!\I_M$'' indicates that $\Tt_k$ is modeled according to \eqref{eq:10c} with $d_{ij}/\lambda\eqq 0.5$. From Figure \ref{fig:de-rzf-sinr}, we observe that the approximated sum rate $\Rsumh$ becomes more accurate with increasing $M$.

Figures \ref{fig:de-rzf-2} and \ref{fig:de-zf-2} compare the ergodic sum rate to the deterministic approximation \eqref{eq:sum-rate-de} under RZF and ZF precoding, respectively. The error bars indicate the
standard deviation of the MC results. It can be observed that the approximation lies roughly within one standard deviation of the MC simulations. From Figure \ref{fig:de-rzf-2}, under imperfect CSIT ($\tau^2_k=0.1$), the sum rate is decreasing for high SNR, because the regularization parameter $\alpha$ does not account for $\tau^2_k$ and thus the matrix $\Ht^\herm\Ht+M\alpha\I_M$ in the RZF precoder becomes ill-conditioned. Figure \ref{fig:de-zf-2} shows that, for $M>K$, the sum rate is not decreasing at high SNR, because the CSIT $\Ht$ is much better conditioned. The optimal regularization is discussed in Section~\ref{sec:sum-rate-maximizing-1}. Further observe that in Figure \ref{fig:de-rzf-2} the deterministic approximation becomes less accurate for high SNR. The reason is that in the derivation of the approximated SINR, we apply Theorem~\ref{th:1} in $z=-\alpha=-1/\SNR$ and thus the bounds in  Proposition \ref{prop:Edp} (Appendix \ref{sec:proof-convergence}) are proportional to the SNR. Therefore, to increase the accuracy of the approximated SINR, larger dimensions are required in the high SNR regime.

We conclude that the approximations in Theorems \ref{th:2} and \ref{th:3} are accurate even for small dimensions and can be applied to various optimization problems discussed in the sequel.

\section{Sum Rate Maximizing Regularization}
\label{sec:sum-rate-maximizing}

The optimal regularization parameter $\alpha^{\star\circ}$ maximizing \eqref{eq:sum-rate-de} is defined as
\begin{equation}
  \label{eq:optimization-alpha}
  \alpha^{\star\circ} = \underset{\alpha>0}{\arg\max}\sum_{k=1}^K\log\left(1 + \SINRkrzf^\circ \right).
\end{equation}
In general, the optimization problem \eqref{eq:optimization-alpha} is not convex in $\alpha$ and the solution has to be computed via a one-dimensional line search. 

In the following, we confine ourselves to the case of common correlation $\Tt_k=\Tt~\forall k$, since for per-user correlation a common regularization parameter is not optimal anymore \cite{christensen2008weighted,wagner2011wsr}. Under common transmit correlation, we subsequently assume that the distortions $\tau_k^2$ of the CSIT $\hht_k$ are identical for all users, since the users' channels are statistically equivalent. Under these conditions $\P=\frac1K\I_K$ maximizes \eqref{eq:sum-rate-de} and the optimization problem \eqref{eq:optimization-alpha} has the following solution.


\begin{proposition}
  \label{pr:1}
  Let $\Tt_k\eqq\Tt$, $0\leq\tau_k\eqq\tau <1$ $\forall k$ and $\P=\frac1K\I_K$. The approximated SINR $\dSINRkrzf$ of user $k$ under RZF precoding (equivalently, the approximated per-user rate and the sum rate) is maximized for a regularization parameter $\alpha\eqdef\alpha^{\star\circ}$, given as a positive solution to the fixed-point equation
  \begin{equation}
    \label{eq:alpha-optimal}
    \alpha^{\star\circ} = \frac{\left[1+\nu(\alpha^{\star\circ})+\tau^2\SNR \frac{e_{22}(\alpha^{\star\circ})}{e_{12}(\alpha^{\star\circ})}\right]\frac{1}{\beta\SNR}}{(1-\tau^2)[1+\nu(\alpha^{\star\circ})] + \tau^2\nu(\alpha^{\star\circ})[1+m^\circ(\alpha^{\star\circ})]^2}
  \end{equation}
  where $m^\circ(\alpha)$ is defined in \eqref{eq:32} and $\nu(\alpha)$ is given by
  \begin{equation}
    \label{eq:12a}
    \nu(\alpha) = \frac{1}{(1+m^\circ)e_{22}}\frac{e_{13}}{e_{12}}\left[\frac{e_{22}}{e_{12}} - \frac{e_{23}}{e_{13}}\right]
  \end{equation}
  with $e_{ij}$ defined in \eqref{eq:69}.
\end{proposition}
\begin{IEEEproof}
  The proof is provided in Appendix \ref{sec:proof-proposition}.
\end{IEEEproof}

Note that the solution in Proposition \ref{pr:1} assumes a \textit{fixed} distortion $\tau^2$. Later in Section \ref{sec:optim-feedb-large} the distortion becomes a function of the quantization codebook size and in Section \ref{sec:optim-train-large} it depends on the uplink SNR as well as on the amount of channel training.

Under perfect CSIT ($\tau^2=0$), Proposition \ref{pr:1} simplifies to the well-known solution $\alpha^{\star\circ} = \frac{1}{\beta\SNR}$, \textit{independent} of $\Tt$, which has previously been derived in \cite{peel2005vpt,nguyen2008mtb,muharar2009dbt}. As mentioned in \cite{peel2005vpt}, for \textit{large} $M$ the RZF-CDA precoder is identical to the MMSE precoder in \cite{dabbagh2008multiple,joham2002transmit}. The authors in \cite{muharar2009dbt} showed that, under \textit{perfect} CSIT, $\alpha^{\star\circ}$ is independent of the correlation $\Tt$. However, for imperfect CSIT ($\tau^2\neq 0$), the optimal regularization parameter \eqref{eq:alpha-optimal} depends on the transmit correlation through $m^\circ(\alpha)$ and $e_{ij}(\alpha)$. For uncorrelated channels ($\Tt=\I_M$), we have $e_{12}=e_{22}$ and $\nu(\alpha)=0$ and therefore the explicit solution
\begin{equation}
  \label{eq:alpha-optimal-2}
  \alpha^{\star\circ} = \left(\frac{1+\tau^2\SNR}{1-\tau^2}\right)\frac{1}{\beta\SNR}.
\end{equation}
Note that in this case, it can be shown that $\alpha^{\star\circ}$ in \eqref{eq:alpha-optimal-2} is the \textit{unique} positive solution to \eqref{eq:optimization-alpha}.

For imperfect CSIT ($\tau^2>0$), the RZF-CDA precoder and the MMSE precoder with regularization parameter $\alpha_{\rm MMSE}=\tau^2\beta^{-1}+1/(\beta\SNR)$ \cite{dabbagh2008multiple} are not identical anymore, even in the large $M,K$ limit. Unlike the case of perfect CSIT, $\alpha^{\star\circ}$ now depends on the correlation matrix $\Tt$ through $m^\circ(\alpha^{\star\circ})$ and $e_{ij}(\alpha^{\star\circ})$. The impact of $m^\circ$ and $e_{ij}$ on the sum rate of RZF-CDA precoding is evaluated through numerical simulations in Figure \ref{fig:de-rzf-a-2}. Further note that since $m^\circ(\alpha)$ and $e_{ij}$ are bounded from above under the conditions explained in Remark \ref{re:snr} below, at asymptotically high SNR the regularization parameter $\alpha^{\star\circ}$ in \eqref{eq:alpha-optimal} converges to $\alpha_\infty^{\star\circ}\eqdef\lim_{\SNR\to\infty}\alpha^{\star\circ}$, where $\alpha_\infty^{\star\circ}$ is a positive solution of
\begin{equation}
  \label{eq:55}
  \alpha_\infty^{\star\circ} = \frac{\frac{\tau^2}{\beta}\frac{e_{22}(\alpha_\infty^{\star\circ})}{e_{12}(\alpha_\infty^{\star\circ})}}{(1-\tau^2)[1+\nu(\alpha_\infty^{\star\circ})] + \tau^2\nu(\alpha_\infty^{\star\circ})[1+m^\circ(\alpha_\infty^{\star\circ})]^2}.
\end{equation}
For uncorrelated channels, the limit in \eqref{eq:55} takes the form
\begin{equation*}
  \alpha_\infty^{\star\circ} = \frac{\tau^2}{(1-\tau^2)\beta}.
\end{equation*}
Thus, for asymptotically high SNR, RZF-CDA precoding is \textit{not} the same as ZF precoding, since the regularization parameter $\alpha^{\star\circ}$ is non-zero due to the residual interference caused by the imperfect CSIT. Similar observations have been made in \cite{dabbagh2008multiple} for the MMSE precoder.

\begin{remark}
\label{re:snr}
  Note that in \eqref{eq:55} we apply the limit $\SNR\to\infty$ on a result obtained from an SINR approximation which is almost surely exact as $M,K\to\infty$. This is correct if $\Psi=\tr\P\Ht(\Ht^\herm\Ht + M\alpha\I_M)^{-2}\Ht^{\herm}$ in \eqref{eq:sinr} is bounded for asymptotically high SNR as $M,K\to\infty$. For $\tau^2>0$ it is clear that $\Psi^\circ$ is bounded since $\alpha^{\star\circ}>0$ for all SNR. In the case where $\tau^2=0$, we have $\lim_{\SNR\to\infty}\alpha^{\star\circ}=0$ and thus for $\beta=1$ the support of the limiting eigenvalue distribution of $\frac1M\Ht\Ht^\herm$ includes zero resulting in an unbounded $\Psi^\circ$. From Remark \ref{re:1}, for $\beta>1$, $\Tt_k=\Tt~\forall k$ and $\lambda_{\min}(\Tt)>\varepsilon>0$ there exists $\xi>0$ such that $\lambda_{\min}(\frac1M\Ht\Ht^\herm)>\xi$ for all large $M$. Thus, $\Psi^\circ$ is bounded. On the contrary, for $\Tt_k\neq\Tt_j$ ($k\neq j$), $\beta>1$ and $\lambda_{\min}(\Tt_k)>\varepsilon>0~\forall k$, it has not been proved that $\lambda_{\min}(\frac1M\Ht\Ht^\herm)>\xi$ and we have to evoke Assumption \ref{as:1} to ensure that $\Psi^\circ$ is bounded. Thus, for $\tau^2=0$, the limit \eqref{eq:55} is only well defined for $\beta>1$. Further note that if $\Psi^\circ$ is bounded as $M,K\to\infty$ the limits $M,K\to\infty$ and $\SNR\to\infty$ can be inverted without affecting the result.
\end{remark}

For various special cases, substituting \eqref{eq:alpha-optimal} into the deterministic equivalent of the SINR $\dSINRkrzf$ in \eqref{eq:9} yields the following simplified expressions.

\begin{corollary}
  \label{cor:theorem-rzf-sinr}
  Let Assumptions \ref{as:0} and \ref{as:0ap} hold true and let $\Tt_k\eqq\Tt$, $\tau_k^2\eqq 0$, $p_k=P/K$ $\forall k$, $\alpha^{\star\circ}=\frac{1}{\beta\SNR}$ and $\SINR_{k,\rm rzf-cda}$ be the sum rate maximizing SINR of user $k$ under RZF precoding. Then $$\SINR_{k,\rm rzf-cda} - \SINR^\circ_{k,\rm rzf-cda}\tom 0,$$ almost surely, where $\SINR^\circ_{k,\rm rzf-cda}$ is given by
  \begin{align}
    \label{eq:asymtotic-sinr-mc}
    \SINR^\circ_{k,\rm rzf-cda} \eqdef \SINR^\circ_{\rm rzf-cda} = m^{\circ}(-\alpha^{\star\circ}), 
  \end{align}
  where $ m^{\circ}(-\alpha^{\star\circ})$ is the unique positive solution to 
  \begin{equation*}
     m^{\circ}(-\alpha^{\star\circ}) = \frac1M\tr\Tt\left(\frac{\Tt/\beta}{1+m^{\circ}(-\alpha^{\star\circ})} + \alpha^{\star\circ}\I_M\right)^{-1}.
  \end{equation*}
\end{corollary}
\begin{IEEEproof}
  Substituting $\alpha^{\star\circ}=\frac{1}{\beta\SNR}$ into \eqref{eq:9} together with $\tau^2=0$, we obtain \eqref{eq:asymtotic-sinr-mc} which completes the proof.
\end{IEEEproof}

For uncorrelated channels $\Tt_k=\I_M~\forall k$, the solution to \eqref{eq:asymtotic-sinr-mc} is explicit and summarized in the following corollary.

\begin{corollary}
  \label{cor:theorem-rzf-sinr-2}
  Let $\Tt_k\eqq\I_M$, $\tau_k^2\eqq\tau^2$, $p_k=P/K$ $\forall k$ and $\SINR_{k,\rm rzf-cda}$ be the sum rate maximizing SINR of user $k$ under RZF precoding. Then $\SINR_{k,\rm rzf-cda} - \SINR^\circ_{k,\rm rzf-cda}\tom 0$, almost surely, where $\SINR^\circ_{k,\rm rzf-cda}$ is given by
  \begin{align}
    \label{eq:asymtotic-sinr-mc-2}
    \SINR^\circ_{k,\rm rzf-cda} \eqdef \SINR^\circ_{\rm rzf-cda} =
    \frac{\omega}{2}\SNR(\beta-1)+\frac{\chi}{2}-\frac12,
  \end{align}
  where $\omega\inn[0,1]$ and $\chi$ are given by
  \begin{align}
    \label{eq:omega}
    \omega &= \frac{1-\tau^2}{1 + \tau^2\SNR}, \\
    \label{eq:chi}
    \chi(\omega) &= \sqrt{(\beta-1)^2\omega^2\SNR^2+2(1+\beta)\omega\SNR+1}.
  \end{align}
\end{corollary}
\begin{IEEEproof}
  Substituting $\Tt=\I_M$ into Corollary \ref{cor:theorem-rzf-sinr} leads to a quadratic equation in $m^{\circ}(-\alpha^{\star\circ})$ for which the unique positive solution is given by \eqref{eq:asymtotic-sinr-mc-2}, which completes the proof.
\end{IEEEproof}

A deterministic equivalent $\Delta R^\circ_{\rm rzf-cda}$ of the rate gap $\Delta R_{k,\rm rzf-cda}$ under RZF-CDA precoding is provided in the following corollary.

\begin{corollary}[RZF-CDA precoding]
  \label{lemma-rzf-lfb}
  Let $\Tt_k\eqq\I_M~\forall k$, $p_k=P/K~\forall k$, $\tau^2_k=\tau^2~\forall k$ and define $\Delta R_{k,\rm rzf-cda}$ as the rate gap of user $k$ under RZF-CDA precoding. Then,
  \begin{equation*}
    \label{eq:57}
    \Delta R_{k,\rm rzf-cda} - \Delta R^\circ_{\rm rzf-cda}\tom 0
  \end{equation*}
  almost surely, with
  \begin{align*}
    \Delta R^\circ_{\rm rzf-cda} = \log\left(\frac{1 + \SNR(\beta-1) + \chi(1)}{1 + \omega\SNR(\beta-1) + \chi(\omega)}\right),
  \end{align*}
  where $\omega$ and $\chi$ are defined in \eqref{eq:omega} and \eqref{eq:chi}, respectively.
\end{corollary}
\begin{IEEEproof}
  With Corollary \ref{cor:theorem-rzf-sinr-2}, compute $\Delta R^\circ_{\rm rzf-cda}$ as defined in \eqref{eq:42}.
\end{IEEEproof}

The impact of the regularization parameter on the ergodic sum rate is depicted in Figures \ref{fig:de-rzf-a} and \ref{fig:de-rzf-a-2}.

\begin{figure}[t]
  \centering
  \begin{tikzpicture}
  \tikzstyle{every pin}=[fill=white,draw=black]
    \pgfplotsset{every axis legend/.append style={
        cells={anchor=west}, at={(0.02,0.98)}, anchor=north west}}
    \pgfplotsset{every axis/.append style={line width=0.5pt}}
    \pgfplotsset{every axis/.append style={mark options=solid, mark size=2.5pt}}

    \begin{axis}[xlabel={$\SNR$ [dB]}, ylabel={ergodic sum rate [bits/s/Hz]},
      grid=major, xmin=0, xmax=30, xtick={0,5,...,30}, ymin=0,
      ymax=16, ytick={0,2,...,16}, legend columns=3]

      \addplot[red, dashed] plot coordinates { (0.000,3.374)
        (5.000,5.806) (10.000,8.250) (15.000,10.193) (20.000,11.169)
        (25.000,11.796) (30.000,11.964) };


      \addplot[blue, mark=o] plot coordinates { (0.000,3.409)
        (5.000,5.822) (10.000,8.323) (15.000,10.090) (20.000,11.049)
        (25.000,11.467) (30.000,11.681) };

      \addplot[green, mark=asterisk] plot coordinates { (0.000,3.351)
        (5.000,5.763) (10.000,8.254) (15.000,9.947) (20.000,10.911)
        (25.000,11.392) (30.000,11.536) };

      \addplot[black, mark=square] plot coordinates { (0.000,3.347)
        (5.000,5.706) (10.000,7.904) (15.000,9.010) (20.000,8.997)
        (25.000,8.625) (30.000,8.405) };

        \addplot[green, mark=triangle] plot coordinates {
          (0.000,0.870) (5.000,2.080) (10.000,3.975) (15.000,5.901)
          (20.000,7.250) (25.000,7.853) (30.000,8.155) };

      \legend{ 
        {$\alpha\eqq\alpha^\star$}\\
        {$\alpha\eqq\bar{\alpha}^{\star}$}\\
        {$\alpha\eqq\alpha^{\star\circ}$}\\
        {$\alpha\eqq \frac{1}{\beta\SNR}$}\\
        {$\alpha\eqq 0$ (ZF)}\\};

    \end{axis}
  \end{tikzpicture}
  \caption{RZF, ergodic sum rate vs. SNR with $M\eqq K \eqq 5$, $\Tt_k\eqq\I_M~\forall k$, $\P=\frac1K\I_K$ and $\tau^2\eqq 0.1$.}
  \label{fig:de-rzf-a}
\end{figure}

In Figure \ref{fig:de-rzf-a}, we compare the ergodic sum rate performance for different regularization parameters $\alpha$ with CSIT distortion $\tau_k^2=\tau^2= 0.1~\forall k$. The upper bound $\alpha\eqq\alpha^\star$ is obtained by optimizing $\alpha$ for every channel realization, whereas $\bar{\alpha}^{\star}$ maximizes the ergodic sum rate. It can be observed that both $\bar{\alpha}^{\star}$ and $\alpha^{\star\circ}$ perform close to the optimal $\alpha^\star$. Furthermore, if the channel quality $\tau^2$ is unknown at the transmitter (and hence assumed to be equal to zero), the performance is decreasing as soon as $\tau^2$ dominates (i.e. the inter-user interference limits the performance) the noise power $\sigma^2$ and approaches the sum rate of ZF precoding for high SNR. We conclude that (i) adapting the regularization parameter yields a significant performance increase and (ii) that the proposed RZF-CDA precoder with $\alpha^{\star\circ}$ performs close to optimal even for small system dimensions.

\begin{figure}[t]
  \centering
  \begin{tikzpicture}
  \tikzstyle{every pin}=[fill=white,draw=black]
    \pgfplotsset{every axis legend/.append style={
        cells={anchor=west}, at={(0.02,0.98)}, anchor=north west}}
    \pgfplotsset{every axis plot/.append style={smooth}}
    \pgfplotsset{every axis/.append style={line width=0.5pt}}
    \pgfplotsset{every axis/.append style={mark options=solid, mark size=2.5pt}}

    \begin{axis}[xlabel={$\SNR$ [dB]}, ylabel={ergodic sum rate [bits/s/Hz]},
      grid=major, xmin=0, xmax=30, xtick={0,5,...,30}, ymin=0,
      ymax=16, ytick={0,2,...,16}]


      \addplot[red, solid, mark=o] plot coordinates { (0.000,3.5238)
        (5.000,6.1132) (10.000,9.0063) (15.000,11.5693) (20.000,13.2027)
        (25.000,14.0111) (30.000,14.3222) };

      \addplot[blue, dashed, mark=square] plot coordinates {
        (0.000,3.515) (5.000,6.109) (10.000,9.012) (15.000,11.518)
        (20.000,13.206) (25.000,13.981) (30.000,14.297) };


      \addplot[red, solid, mark=o] plot coordinates { (0.000,3.2126)
        (5.000,5.5140) (10.000,8.2338) (15.000,10.7969) (20.000,12.6583)
        (25.000,13.6328) (30.000,14.0923)};

      \addplot[blue, dashed, mark=square] plot coordinates {
        (0.000,3.220) (5.000,5.525) (10.000,8.234) (15.000,10.747)
        (20.000,12.490) (25.000,13.365) (30.000,13.747) };


      \addplot[red, solid, mark=o] plot coordinates { (0.000,2.1358)
        (5.000,3.4126) (10.000,5.2174) (15.000,7.4810) (20.000,9.8976)
        (25.000,11.9390) (30.000,13.1573) };

      \addplot[blue, dashed, mark=square] plot coordinates {
        (0.000,2.140) (5.000,3.411) (10.000,5.203) (15.000,7.256)
        (20.000,8.909) (25.000,9.831) (30.000,10.236) };

      \draw (axis cs:16,14) node[fill=white,draw=black] (pint0)
      {$v=0.1$};
      \draw (axis cs:15,11.515) node[draw,black,thick,ellipse,minimum height=0.2cm] (ell0) {};
      \draw[black,thick] (pint0) -- (ell0);

      \draw (axis cs:5,10) node[fill=white,draw=black] (pint1)
      {$v=0.5$};
      \draw (axis cs:10,8.2) node[draw,black,thick,ellipse,minimum height=0.2cm] (ell1) {};
      \draw[black,thick] (pint1) -- (ell1);

      \draw (axis cs:25,6) node[fill=white,draw=black] (pint2)
      {$v=0.9$};
      \draw (axis cs:20,9.3) node[draw,black,thick,ellipse,minimum height=0.8cm] (ell2) {};
      \draw[black,thick] (pint2) -- (ell2);

      \legend{ 
        {RZF-CCA}\\
        {RZF-CCU}\\};

    \end{axis}
  \end{tikzpicture}
  \caption{RZF, ergodic sum rate vs. SNR with $M\eqq K \eqq 5$, $\P=\frac1K\I_K$ and $\tau^2\eqq 0.05$.}
  \label{fig:de-rzf-a-2}
\end{figure}

In Figure \ref{fig:de-rzf-a-2}, we simulate the impact of transmit correlation in the computation of $\alpha^{\star\circ}$ on the sum rate. For this purpose, we use the standard exponential correlation model, i.e.,
\begin{equation*}
  \label{eq:45}
  [\Tt]_{ij} = v^{|i-j|}.
\end{equation*}
We compare two different RZF precoders: A first precoder coined RZF common correlation aware (RZF-CCA) that takes the channel correlation into account and computes $\alpha$ according to \eqref{eq:alpha-optimal}, and a second precoder, called RZF common correlation unaware (RZF-CCU) that does not take $\Tt$ into account and computes $\alpha$ as in \eqref{eq:alpha-optimal-2}. We observe that for high correlation, i.e., $v=0.9$, the RZF-CCA precoder significantly outperforms the RZF-CCU precoder at medium to high SNR, whereas both precoders perform
equally well at low SNR. Therefore, we conclude that it is beneficial to account for transmit correlation, especially in highly correlated channels. Further simulations (not provided here) suggest that the sum rate gain of RZF-CCA over RZF-CCU precoding is less pronounced for lower CSIT qualities (i.e., increasing $\tau^2$), because in this case the impact of the CSIT quality $\tau^2$ is more significant than the impact of $\Tt$ on the sum rate.

\section{Optimal Number of Users and Power Allocation}
\label{sec:sum-rate-maximizing-1}

In this section, we address two problems: (i) the determination of the sum rate maximizing number of users per transmit antenna for a fixed $M$ and (ii) the optimization of the power distribution among a \textit{given} set of users with unequal CSIT qualities.

Consider problem (i). Intuitively, an optimal number of users $K^{\star}$ exists because serving more users creates more interference which in turn reduces the rates of the users. At some point the accumulated rate loss, due to the additional interference caused by scheduling another user, will outweigh the sum rate gain and hence the system sum rate will decrease. In particular, we consider a fair scenario where the SINR approximation of all users are equal. Here, the (approximated) optimal solution can be expressed under a closed form for ZF precoding.

In problem (ii), we optimize the power allocation matrix $\P$ for a given $K$. More precisely, we focus on common correlation $\Tt_k=\Tt~\forall k$ with \textit{different} CSIT qualities $\tau_k^2$, since in this case the (approximated) optimal power distribution $\P^{\star\circ}$ is the solution of a classical water-filling algorithm.

\subsection{Sum Rate Maximizing Number of Users}
\label{sec:sum-rate-maximizing-2}

Consider the problem of finding the system loading $\beta^{\star\circ}$ maximizing the approximated sum rate per transmit antenna for a \textit{fixed} $M$, i.e.,
\begin{equation}
  \label{eq:zf-max-sr}
  \beta^{\star\circ} = \underset{\beta}{\arg\max} \frac1{\beta}\frac1K\sum_{k=1}^K\log\left(1+\SINR_k^\circ\right),
\end{equation}
where $\SINR_k^\circ$ denotes either $\SINRkzf^\circ$ with $\beta>1$ or $\SINRkrzf^\circ$ with $\beta\geq 1$. In general \eqref{eq:zf-max-sr} has to be solved by a one-dimensional line search. However, in case of ZF precoding and uncorrelated antennas, the optimization problem \eqref{eq:zf-max-sr} has a closed-form solution given in the following proposition.

\begin{proposition}
  Let $\Tt_k=\I_M$, $\tau_k=\tau$ $\forall k$ and $\P=\frac{P}{K}\I_K$, the sum rate maximizing system loading per transmit antenna $\beta^{\star\circ}$ is given by
  \begin{equation}
    \label{eq:zf-beta-optimal-solution}
    \beta^{\star\circ} = \left(1-\frac1 a\right)\left(1 + \frac{1}{\WW(x)} \right),
  \end{equation}
  where $a\eqq\frac{1-\tau^2}{\tau^2+\frac1\SNR}$, $x=\frac{a-1}{e}$ and $\WW(x)$ is the Lambert W-function defined as $z\!=\!\WW(z)e^{\WW(z)}$, $z\inn\Cbb$.
\end{proposition}
\begin{IEEEproof}
  Substituting the SINR in Corollary \ref{co:2} into \eqref{eq:zf-max-sr} and differentiating along $\beta$ leads to
  \begin{equation}
    \label{eq:zf-beta-implicit-eq-2}
    \frac{a\beta}{1+a(\beta-1)}=\log\left(1+a(\beta-1) \right)
  \end{equation}
  Denoting $w(\beta)=\frac{a-1}{a(\beta-1)+1}$, we can rewrite \eqref{eq:zf-beta-implicit-eq-2} as
  \begin{equation*}
    \label{eq:zf-beta-3}
    w(\beta) e^{w(\beta)} = x.
  \end{equation*}
  Noticing that $w(\beta)=\WW(x)$ and solving for $\beta$ yields \eqref{eq:zf-beta-optimal-solution}, which completes the proof.
\end{IEEEproof}

For $\tau\inn[0,1]$, $\beta>1$ we have $w\!\geq\!-1$ and $x\geq -e^{-1}$. In this case $\WW(x)$ is a well-defined function. If $\tau^2=0$, we obtain the results in \cite{hochwald2002space}, although in \cite{hochwald2002space} they are not given in closed form. Note that for $\tau^2=0$, we have $\lim_{\SNR\to\infty}\beta^{\star\circ}=1$, i.e., the optimal system loading tends to one. Further note that only integer values of $M/\beta^{\star\circ}$ are meaningful in practice.

\subsection{Power Optimization under Common Correlation}
\label{sec:power-optim-tt_k=tt}

From Corollaries \ref{sec:regul-zero-forc-1} and \ref{co:1}, the approximated sum rate \eqref{eq:sum-rate-de} for both RZF and ZF precoding takes the form
\begin{equation}
  \label{eq:38}
  \Rsumh = \sum_{k=1}^K\log\left[1 + p_k\nu_k^\circ(\tau_k) \right],
\end{equation}
with $\nu_k^\circ(\tau_k)=\SINR_k^\circ/p_k$, where the only dependence on user $k$ stems from $\tau_k$. The user powers $p^{\star\circ}_k$ that maximize \eqref{eq:38}, subject to $\sum_{k=1}^Kp_k\leq P$, $p_k\geq 0$, are thus given by the classical water-filling solution \cite{palomar2005practical}
\begin{equation}
  \label{eq:54}
  p^{\star\circ}_k = \left[\mu - \frac{1}{\nu_k^\circ(\tau_k)}\right]^+,
\end{equation}
where $[x]^+\eqdef\max(0,x)$ and $\mu$ is the water level chosen to satisfy $\sum_{k=1}^Kp_k= P$. For $\tau_k^2 = \tau^2 ~forall k$, the optimal user powers \eqref{eq:54} are all equal, i.e., $p^{\star\circ}_k=p^{\star\circ}=P/K$ and $\P^{\star\circ}\eqdef\diag(p^{\star\circ}_1,\dots,p^{\star\circ}_K)=\frac{P}{K}\I_K$. In this case though, it could still be beneficial to adapt the number of users as discussed in Section \ref{sec:sum-rate-maximizing-2}.

\subsection{Numerical Results}
\label{sec:numerical-results-1}

Figure \ref{fig:zfbf-optimal-beta-high-snr-cor-k} compares the optimal number of users $K^{\star\circ}=M/\beta^{\star\circ}$ in \eqref{eq:zf-beta-optimal-solution} to $K^\star$ obtained by choosing the $K\inn\{1,2,\dots,M\}$ such that the ergodic sum rate is maximized, whereas Figure \ref{fig:zf-sr-op-beta} depicts the impact of a suboptimal number of users on the ergodic sum rate of the system.

From Figure \ref{fig:zfbf-optimal-beta-high-snr-cor-k}, it can be observed that (i) the approximated results $K^{\star\circ}$ do fit well with the simulation results even for small dimensions, (ii) $(K^\star,K^{\star\circ})$ increase with the SNR and (iii), for $\tau^2\!\neq\!0$, $(K^\star,K^{\star\circ})$ saturate for high SNR at a value lower than $M$. Therefore, under imperfect CSIT, it is not optimal anymore to serve the maximum number of users $K=M$ for asymptotically high SNR. Instead, depending on $\tau^2$, a lower number of users $K<M$ should be served even at high SNR which implies a reduced multiplexing gain of the system. The impact of different numbers of users on the sum rate is depicted in Figure \ref{fig:zf-sr-op-beta}.

\begin{figure}[t]
  \centering
  \begin{tikzpicture}
    \tikzstyle{every pin}=[fill=white,draw=black]
    \pgfplotsset{every axis legend/.append style={
        cells={anchor=west}, at={(0.02,0.98)}, anchor=north west}} 
    \pgfplotsset{every axis/.append style={line width=0.5pt}} 
    \begin{axis}[ title={}, xlabel={$\SNR$ [dB]}, ylabel={optimal
        number of users},
      grid={both}, ymin={0}, ymax={24}, xmin={0}, xmax={30},
      xtick={0,5,...,30}, ytick={0,2,...,24}, no markers]

      \addplot[red, solid] plot coordinates { (0.000,2.7877289076)
        (0.303,2.8364844413) (0.606,2.8850820906) (0.909,2.9334790169)
        (1.212,2.9816321422) (1.515,3.0294982698) (1.818,3.0770342096)
        (2.121,3.1241969094) (2.424,3.1709435894) (2.727,3.2172318816)
        (3.030,3.2630199727) (3.333,3.3082667494) (3.636,3.3529319466)
        (3.939,3.3969762971) (4.242,3.4403616813) (4.545,3.4830512779)
        (4.848,3.5250097119) (5.152,3.5662032012) (5.455,3.6065996989)
        (5.758,3.6461690312) (6.061,3.6848830282) (6.364,3.7227156473)
        (6.667,3.7596430873) (6.970,3.7956438918) (7.273,3.8306990400)
        (7.576,3.8647920244) (7.879,3.8979089133) (8.182,3.9300383977)
        (8.485,3.9611718209) (8.788,3.9913031904) (9.091,4.0204291716)
        (9.394,4.0485490632) (9.697,4.0756647538)
        (10.000,4.1017806607) (10.303,4.1269036506)
        (10.606,4.1510429443) (10.909,4.1742100054)
        (11.212,4.1964184154) (11.515,4.2176837365)
        (11.818,4.2380233631) (12.121,4.2574563659)
        (12.424,4.2760033287) (12.727,4.2936861812)
        (13.030,4.3105280290) (13.333,4.3265529843)
        (13.636,4.3417859970) (13.939,4.3562526905)
        (14.242,4.3699792014) (14.545,4.3829920268)
        (14.848,4.3953178782) (15.152,4.4069835451)
        (15.455,4.4180157678) (15.758,4.4284411202)
        (16.061,4.4382859027) (16.364,4.4475760471)
        (16.667,4.4563370305) (16.970,4.4645938007)
        (17.273,4.4723707123) (17.576,4.4796914710)
        (17.879,4.4865790892) (18.182,4.4930558492)
        (18.485,4.4991432749) (18.788,4.5048621114)
        (19.091,4.5102323113) (19.394,4.5152730269)
        (19.697,4.5200026089) (20.000,4.5244386092)
        (20.303,4.5285977886) (20.606,4.5324961280)
        (20.909,4.5361488432) (21.212,4.5395704019)
        (21.515,4.5427745431) (21.818,4.5457742992)
        (22.121,4.5485820176) (22.424,4.5512093859)
        (22.727,4.5536674553) (23.030,4.5559666667)
        (23.333,4.5581168756) (23.636,4.5601273776)
        (23.939,4.5620069337) (24.242,4.5637637955)
        (24.545,4.5654057295) (24.848,4.5669400414)
        (25.152,4.5683735994) (25.455,4.5697128570)
        (25.758,4.5709638752) (26.061,4.5721323433)
        (26.364,4.5732235998) (26.667,4.5742426519)
        (26.970,4.5751941943) (27.273,4.5760826267)
        (27.576,4.5769120718) (27.879,4.5776863909)
        (28.182,4.5784091995) (28.485,4.5790838825)
        (28.788,4.5797136073) (29.091,4.5803013381)
        (29.394,4.5808498473) (29.697,4.5813617281)
        (30.000,4.5818394054) };

      \addplot[blue, dashed] plot coordinates { (0.000,3.000)
        (0.303,3.000) (0.606,3.000) (0.909,3.000) (1.212,3.000)
        (1.515,3.000) (1.818,3.000) (2.121,3.000) (2.424,3.000)
        (2.727,3.000) (3.030,3.000) (3.333,3.000) (3.636,4.000)
        (3.939,4.000) (4.242,4.000) (4.545,4.000) (4.848,4.000)
        (5.152,4.000) (5.455,4.000) (5.758,4.000) (6.061,4.000)
        (6.364,4.000) (6.667,4.000) (6.970,4.000) (7.273,4.000)
        (7.576,4.000) (7.879,4.000) (8.182,4.000) (8.485,4.000)
        (8.788,4.000) (9.091,4.000) (9.394,4.000) (9.697,4.000)
        (10.000,4.000) (10.303,4.000) (10.606,4.000) (10.909,4.000)
        (11.212,4.000) (11.515,4.000) (11.818,4.000) (12.121,4.000)
        (12.424,4.000) (12.727,4.000) (13.030,4.000) (13.333,4.000)
        (13.636,5.000) (13.939,5.000) (14.242,5.000) (14.545,5.000)
        (14.848,5.000) (15.152,5.000) (15.455,5.000) (15.758,5.000)
        (16.061,5.000) (16.364,5.000) (16.667,5.000) (16.970,5.000)
        (17.273,5.000) (17.576,5.000) (17.879,5.000) (18.182,5.000)
        (18.485,5.000) (18.788,5.000) (19.091,5.000) (19.394,5.000)
        (19.697,5.000) (20.000,5.000) (20.303,5.000) (20.606,5.000)
        (20.909,5.000) (21.212,5.000) (21.515,5.000) (21.818,5.000)
        (22.121,5.000) (22.424,5.000) (22.727,5.000) (23.030,5.000)
        (23.333,5.000) (23.636,5.000) (23.939,5.000) (24.242,5.000)
        (24.545,5.000) (24.848,5.000) (25.152,5.000) (25.455,5.000)
        (25.758,5.000) (26.061,5.000) (26.364,5.000) (26.667,5.000)
        (26.970,5.000) (27.273,5.000) (27.576,5.000) (27.879,5.000)
        (28.182,5.000) (28.485,5.000) (28.788,5.000) (29.091,5.000)
        (29.394,5.000) (29.697,5.000) (30.000,5.000) };

      \addplot[red, solid] plot coordinates { (0.000,5.5754578151)
        (0.303,5.6729688827) (0.606,5.7701641812) (0.909,5.8669580337)
        (1.212,5.9632642844) (1.515,6.0589965395) (1.818,6.1540684192)
        (2.121,6.2483938188) (2.424,6.3418871788) (2.727,6.4344637633)
        (3.030,6.5260399454) (3.333,6.6165334988) (3.636,6.7058638932)
        (3.939,6.7939525941) (4.242,6.8807233627) (4.545,6.9661025559)
        (4.848,7.0500194239) (5.152,7.1324064024) (5.455,7.2131993979)
        (5.758,7.2923380625) (6.061,7.3697660564) (6.364,7.4454312946)
        (6.667,7.5192861746) (6.970,7.5912877836) (7.273,7.6613980800)
        (7.576,7.7295840487) (7.879,7.7958178266) (8.182,7.8600767955)
        (8.485,7.9223436419) (8.788,7.9826063808) (9.091,8.0408583431)
        (9.394,8.0970981263) (9.697,8.1513295076)
        (10.000,8.2035613214) (10.303,8.2538073012)
        (10.606,8.3020858885) (10.909,8.3484200107)
        (11.212,8.3928368308) (11.515,8.4353674729)
        (11.818,8.4760467262) (12.121,8.5149127318)
        (12.424,8.5520066575) (12.727,8.5873723623)
        (13.030,8.6210560580) (13.333,8.6531059685)
        (13.636,8.6835719940) (13.939,8.7125053810)
        (14.242,8.7399584029) (14.545,8.7659840536)
        (14.848,8.7906357563) (15.152,8.8139670902)
        (15.455,8.8360315357) (15.758,8.8568822404)
        (16.061,8.8765718055) (16.364,8.8951520942)
        (16.667,8.9126740609) (16.970,8.9291876015)
        (17.273,8.9447414245) (17.576,8.9593829419)
        (17.879,8.9731581784) (18.182,8.9861116983)
        (18.485,8.9982865498) (18.788,9.0097242229)
        (19.091,9.0204646226) (19.394,9.0305460538)
        (19.697,9.0400052178) (20.000,9.0488772184)
        (20.303,9.0571955772) (20.606,9.0649922561)
        (20.909,9.0722976865) (21.212,9.0791408037)
        (21.515,9.0855490863) (21.818,9.0915485983)
        (22.121,9.0971640353) (22.424,9.1024187717)
        (22.727,9.1073349106) (23.030,9.1119333335)
        (23.333,9.1162337512) (23.636,9.1202547552)
        (23.939,9.1240138674) (24.242,9.1275275910)
        (24.545,9.1308114590) (24.848,9.1338800828)
        (25.152,9.1367471988) (25.455,9.1394257141)
        (25.758,9.1419277503) (26.061,9.1442646865)
        (26.364,9.1464471996) (26.667,9.1484853039)
        (26.970,9.1503883885) (27.273,9.1521652534)
        (27.576,9.1538241436) (27.879,9.1553727818)
        (28.182,9.1568183991) (28.485,9.1581677649)
        (28.788,9.1594272147) (29.091,9.1606026762)
        (29.394,9.1616996946) (29.697,9.1627234562)
        (30.000,9.1636788108) };

      \addplot[blue, dashed] plot coordinates { (0.000,6.000)
        (0.303,6.000) (0.606,6.000) (0.909,6.000) (1.212,6.000)
        (1.515,6.000) (1.818,6.000) (2.121,6.000) (2.424,6.000)
        (2.727,7.000) (3.030,7.000) (3.333,7.000) (3.636,7.000)
        (3.939,7.000) (4.242,7.000) (4.545,7.000) (4.848,7.000)
        (5.152,7.000) (5.455,7.000) (5.758,7.000) (6.061,7.000)
        (6.364,8.000) (6.667,8.000) (6.970,8.000) (7.273,8.000)
        (7.576,8.000) (7.879,8.000) (8.182,8.000) (8.485,8.000)
        (8.788,8.000) (9.091,8.000) (9.394,8.000) (9.697,8.000)
        (10.000,8.000) (10.303,8.000) (10.606,8.000) (10.909,8.000)
        (11.212,9.000) (11.515,9.000) (11.818,9.000) (12.121,9.000)
        (12.424,9.000) (12.727,9.000) (13.030,9.000) (13.333,9.000)
        (13.636,9.000) (13.939,9.000) (14.242,9.000) (14.545,9.000)
        (14.848,9.000) (15.152,9.000) (15.455,9.000) (15.758,9.000)
        (16.061,9.000) (16.364,9.000) (16.667,9.000) (16.970,9.000)
        (17.273,9.000) (17.576,9.000) (17.879,9.000) (18.182,9.000)
        (18.485,9.000) (18.788,9.000) (19.091,9.000) (19.394,9.000)
        (19.697,9.000) (20.000,9.000) (20.303,9.000) (20.606,9.000)
        (20.909,9.000) (21.212,9.000) (21.515,9.000) (21.818,9.000)
        (22.121,9.000) (22.424,9.000) (22.727,9.000) (23.030,9.000)
        (23.333,9.000) (23.636,9.000) (23.939,9.000) (24.242,9.000)
        (24.545,9.000) (24.848,9.000) (25.152,9.000) (25.455,9.000)
        (25.758,9.000) (26.061,9.000) (26.364,9.000) (26.667,9.000)
        (26.970,9.000) (27.273,9.000) (27.576,9.000) (27.879,9.000)
        (28.182,9.000) (28.485,9.000) (28.788,9.000) (29.091,9.000)
        (29.394,9.000) (29.697,9.000) (30.000,9.000) };

      \addplot[red, solid] plot coordinates { (0.000,11.1509156302)
        (0.303,11.3459377653) (0.606,11.5403283624)
        (0.909,11.7339160675) (1.212,11.9265285688)
        (1.515,12.1179930790) (1.818,12.3081368385)
        (2.121,12.4967876377) (2.424,12.6837743576)
        (2.727,12.8689275266) (3.030,13.0520798908)
        (3.333,13.2330669975) (3.636,13.4117277864)
        (3.939,13.5879051882) (4.242,13.7614467253)
        (4.545,13.9322051117) (4.848,14.1000388477)
        (5.152,14.2648128048) (5.455,14.4263987958)
        (5.758,14.5846761250) (6.061,14.7395321128)
        (6.364,14.8908625892) (6.667,15.0385723493)
        (6.970,15.1825755673) (7.273,15.3227961600)
        (7.576,15.4591680974) (7.879,15.5916356531)
        (8.182,15.7201535909) (8.485,15.8446872838)
        (8.788,15.9652127616) (9.091,16.0817166863)
        (9.394,16.1941962526) (9.697,16.3026590152)
        (10.000,16.4071226428) (10.303,16.5076146024)
        (10.606,16.6041717770) (10.909,16.6968400214)
        (11.212,16.7856736617) (11.515,16.8707349459)
        (11.818,16.9520934523) (12.121,17.0298254636)
        (12.424,17.1040133149) (12.727,17.1747447247)
        (13.030,17.2421121160) (13.333,17.3062119370)
        (13.636,17.3671439880) (13.939,17.4250107620)
        (14.242,17.4799168058) (14.545,17.5319681072)
        (14.848,17.5812715126) (15.152,17.6279341804)
        (15.455,17.6720630714) (15.758,17.7137644807)
        (16.061,17.7531436110) (16.364,17.7903041885)
        (16.667,17.8253481218) (16.970,17.8583752030)
        (17.273,17.8894828491) (17.576,17.9187658839)
        (17.879,17.9463163567) (18.182,17.9722233966)
        (18.485,17.9965730995) (18.788,18.0194484457)
        (19.091,18.0409292452) (19.394,18.0610921077)
        (19.697,18.0800104357) (20.000,18.0977544368)
        (20.303,18.1143911544) (20.606,18.1299845121)
        (20.909,18.1445953729) (21.212,18.1582816075)
        (21.515,18.1710981726) (21.818,18.1830971966)
        (22.121,18.1943280705) (22.424,18.2048375435)
        (22.727,18.2146698213) (23.030,18.2238666670)
        (23.333,18.2324675025) (23.636,18.2405095104)
        (23.939,18.2480277348) (24.242,18.2550551820)
        (24.545,18.2616229180) (24.848,18.2677601656)
        (25.152,18.2734943976) (25.455,18.2788514281)
        (25.758,18.2838555007) (26.061,18.2885293730)
        (26.364,18.2928943992) (26.667,18.2969706078)
        (26.970,18.3007767770) (27.273,18.3043305069)
        (27.576,18.3076482873) (27.879,18.3107455635)
        (28.182,18.3136367981) (28.485,18.3163355298)
        (28.788,18.3188544294) (29.091,18.3212053523)
        (29.394,18.3233993892) (29.697,18.3254469125)
        (30.000,18.3273576216)};

      \addplot[blue, dashed] plot coordinates { (0.000,11.000)
        (0.303,11.000) (0.606,12.000) (0.909,12.000) (1.212,12.000)
        (1.515,12.000) (1.818,12.000) (2.121,13.000) (2.424,13.000)
        (2.727,13.000) (3.030,13.000) (3.333,13.000) (3.636,13.000)
        (3.939,14.000) (4.242,14.000) (4.545,14.000) (4.848,14.000)
        (5.152,14.000) (5.455,15.000) (5.758,15.000) (6.061,15.000)
        (6.364,15.000) (6.667,15.000) (6.970,15.000) (7.273,15.000)
        (7.576,15.000) (7.879,16.000) (8.182,16.000) (8.485,16.000)
        (8.788,16.000) (9.091,16.000) (9.394,16.000) (9.697,16.000)
        (10.000,17.000) (10.303,17.000) (10.606,17.000)
        (10.909,17.000) (11.212,17.000) (11.515,17.000)
        (11.818,17.000) (12.121,17.000) (12.424,17.000)
        (12.727,17.000) (13.030,17.000) (13.333,17.000)
        (13.636,18.000) (13.939,18.000) (14.242,18.000)
        (14.545,18.000) (14.848,18.000) (15.152,18.000)
        (15.455,18.000) (15.758,18.000) (16.061,18.000)
        (16.364,18.000) (16.667,18.000) (16.970,18.000)
        (17.273,18.000) (17.576,18.000) (17.879,18.000)
        (18.182,18.000) (18.485,18.000) (18.788,18.000)
        (19.091,18.000) (19.394,18.000) (19.697,18.000)
        (20.000,18.000) (20.303,18.000) (20.606,18.000)
        (20.909,18.000) (21.212,18.000) (21.515,18.000)
        (21.818,18.000) (22.121,19.000) (22.424,19.000)
        (22.727,18.000) (23.030,19.000) (23.333,18.000)
        (23.636,19.000) (23.939,18.000) (24.242,19.000)
        (24.545,19.000) (24.848,19.000) (25.152,18.000)
        (25.455,19.000) (25.758,19.000) (26.061,19.000)
        (26.364,19.000) (26.667,19.000) (26.970,19.000)
        (27.273,19.000) (27.576,19.000) (27.879,19.000)
        (28.182,19.000) (28.485,19.000) (28.788,19.000)
        (29.091,19.000) (29.394,19.000) (29.697,19.000)
        (30.000,19.000) };

      \draw (axis cs:15,2) node[fill=white,draw=black] (pin0) {$M=8$};
      \draw[black,thick] (axis cs:10.000,4) -- (pin0);

      \draw (axis cs:25,12) node[fill=white,draw=black] (pin1) {$M=16$};
      \draw [black,thick] (axis cs:20.000,9) -- (pin1);

      \draw (axis cs:15,14) node[fill=white,draw=black] (pin2) {$M\!=\!32$};
      \draw[black,thick] (axis cs:10.000,16.4071226428) -- (pin2);

      \legend {{$K^{\star\circ}\!=\!M/\beta^{\star\circ}$,
          \eqref{eq:zf-beta-optimal-solution}}, {$K^{\star}$
          from exhaustive search} };
    \end{axis}

  \end{tikzpicture}

  \caption{ZF, sum rate maximizing number of users vs. SNR with $\Tt_k=\I_M~\forall k$, $\tau^2=0.1$ and $\P=\frac1K\I_K$.}
  \label{fig:zfbf-optimal-beta-high-snr-cor-k}
\end{figure}

\begin{figure}[t]
  \centering
  \begin{tikzpicture}
    \tikzstyle{every pin}=[fill=white,draw=black]
    \pgfplotsset{every axis legend/.append style={
        cells={anchor=west}, at={(0.02,0.98)}, anchor=north west}}
    \pgfplotsset{every axis plot/.append style={smooth}}
    \pgfplotsset{every axis/.append style={line width=0.5pt}}
    \pgfplotsset{every axis/.append style={mark options=solid, mark size=2.5pt}}

    \begin{axis}[xlabel={$\SNR$ [dB]}, ylabel={ergodic sum rate [bits/s/Hz]},
      grid=major, xmin=0, xmax=30, xtick={0,5,...,30}, ymin=0,
      ymax=35, ytick={0,5,...,35}]


      \addplot[red, solid, mark=o] plot coordinates
      {(0.000,7.6470427971) (5.000,13.8943612077)
        (10.000,20.7284137839) (15.000,25.9526134804)
        (20.000,28.5320650565) (25.000,29.6095277914)
        (30.000,29.9354590042)};

      \addplot[blue, dashed, mark=square] plot coordinates {
        (0.000,7.639) (5.000,13.913) (10.000,20.773) (15.000,25.861)
        (20.000,28.563) (25.000,29.544) (30.000,29.884) };

      \addplot[black, dotted, mark=asterisk] plot coordinates {
        (0.000,7.1358288751) (5.000,13.7491153254)
        (10.000,20.7646624447) (15.000,25.6389974077)
        (20.000,28.1009313617) (25.000,29.0471376906)
        (30.000,29.3467212055)};

      \addplot[green, solid, mark=triangle] plot coordinates {
        (0.000,7.336) (5.000,12.051) (10.000,16.401) (15.000,19.452)
        (20.000,21.052) (25.000,21.760) (30.000,21.993) };

      \legend{
        {$K\eqq K^\star$}\\
        {$K\eqq K^{\star\circ}$}\\ 
        {$K\eqq 8$}\\
        {$K\eqq 4$}\\};
    \end{axis}
  \end{tikzpicture}
  \caption{ZF, $\Rsum$ vs. SNR with $M\!=\!16$, $\Tt_k=\I_M~\forall k$, $\P=\frac1K\I_K$ and $\tau^2\!=\!0.1$.}
  \label{fig:zf-sr-op-beta}
\end{figure}

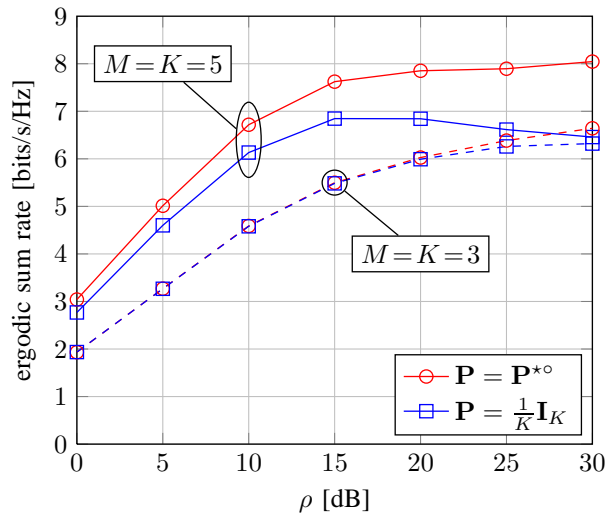
\begin{figure}[t]
  \centering
  \begin{tikzpicture}
  \tikzstyle{every pin}=[fill=white,draw=black]
    \pgfplotsset{every axis legend/.append style={
        cells={anchor=west}, at={(0.98,0.02)}, anchor=south east}}
    \pgfplotsset{every axis/.append style={line width=0.5pt}}
    \pgfplotsset{every axis/.append style={mark options=solid, mark size=2.5pt}}

    \begin{axis}[xlabel={$\SNR$ [dB]}, ylabel={ergodic sum rate [bits/s/Hz]},
      grid=major, xmin=0, xmax=30, xtick={0,5,...,30}, ymin=0,
      ymax=9, ytick={0,1,...,9}]

      \addplot[red, mark=o] plot coordinates { (0.000,3.041)
        (5.000,5.013) (10.000,6.717) (15.000,7.623) (20.000,7.852)
        (25.000,7.895) (30.000,8.045) };

      \addplot[blue, mark=square] plot coordinates { (0.000,2.769)
        (5.000,4.601) (10.000,6.132) (15.000,6.846) (20.000,6.843)
        (25.000,6.615) (30.000,6.458) };

      \addplot[dashed, red, mark=o] plot coordinates { (0.000,1.933)
        (5.000,3.270) (10.000,4.588) (15.000,5.497) (20.000,6.027)
        (25.000,6.384) (30.000,6.642) };

      \addplot[dashed, blue, mark=square] plot coordinates {
        (0.000,1.936) (5.000,3.267) (10.000,4.581) (15.000,5.484)
        (20.000,5.992) (25.000,6.261) (30.000,6.321) };

      \draw (axis cs:5,8) node[fill=white,draw=black] (pint0)
      {$M\eqq K\eqq 5$};
      \draw (axis cs:10,6.4) node[draw,black,ellipse,minimum height=1.0cm] (ell0) {};
      \draw[black] (pint0) -- (ell0);

      \draw (axis cs:20,4) node[fill=white,draw=black] (pint1)
      {$M\eqq K\eqq 3$};
      \draw (axis cs:15,5.5) node[draw,black,ellipse,minimum height=0.2cm] (ell1) {};
      \draw[black] (pint1) -- (ell1);

      \legend{ {$\P=\P^{\star\circ}$}\\
               {$\P=\frac1K\I_K$}\\};

    \end{axis}
  \end{tikzpicture}
  \caption{RZF-CDU, $\Rsum$ vs. $\SNR$ with $\alpha\!=\!1/\SNR$, $\Tt_k\!=\!\I_M~\forall k$, $P\!=\!1$ and $\tau_k^2\inn\mathcal{T}_1,\cup_{k=1}^3\tau_k^2=\mathcal{T}_1$ ($M\eqq 5$) and $\tau_k^2\inn\mathcal{T}_2,\cup_{k=1}^3\tau_k^2=\mathcal{T}_2$ ($M\eqq 3$).}
  \label{fig:de-rzf-op}
\end{figure}

From Figure \ref{fig:zf-sr-op-beta} we observe that (i) the approximate solution $K^{\star\circ}$ achieves most of the sum rate and (ii) adapting the number of users with the SNR is beneficial compared to a fixed $K$. Moreover, from Figure \ref{fig:zfbf-optimal-beta-high-snr-cor-k}, we identify $K\eqq 8$ as an optimal choice (for $M=16$) for medium SNR and, as expected, the performance is optimal in the medium SNR regime and suboptimal at low and high SNR. From Figure \ref{fig:zfbf-optimal-beta-high-snr-cor-k} it is clear that $K=4$ is highly suboptimal in the medium and high SNR range and we observe a significant loss in sum rate. Consequently, the number of users must be adapted to the channel conditions and the approximate result $K^{\star\circ}$ is a good choice to determine the optimal number of users.

In Figure \ref{fig:de-rzf-op}, under RZF-CDU precoding, we compare the ergodic sum rate performance with power allocation $\P=\P^{\star\circ}$ from \eqref{eq:54} to equal power allocation $\P=\frac1K\I_K$. We consider a system with $M\eqq K \eqq 5$, where the CSIT qualities vary significantly among the users, i.e., $\tau_k^2\inn\mathcal{T}_1$ with $\mathcal{T}_1=\{0.8, 0.3, 0.2, 0.1, 0.05\}$, $\cup_{k=1}^5\tau_k^2=\mathcal{T}_1$. We observe a significant gain over the whole SNR range when optimal power allocation is applied. In contrast, if the CSIT distortion of the users' channels with $M\eqq K \eqq 3$ does not differ considerably ($\tau_k^2\inn\mathcal{T}_2$, $\cup_{k=1}^3\tau_k^2=\mathcal{T}_2$ with $\mathcal{T}_2=\{0.2,0.15,0.1\}$), we only observe a small gain at high SNR. For increasing SNR, the SINRs become increasingly distinct depending on the $\tau_k^2$. Therefore, it might be optimal to turn off the users with lowest CSIT accuracy as the SNR increases, which explains why the sum rate gain is larger at high SNR than at low SNR. However, recall that the water-filling solution is optimal under Assumption \ref{as:0ap} ($\|\P\| = O(1/K)$) and large $M$. We thus conclude that the optimal power allocation proposed in \eqref{eq:54} achieves significant performance gains, especially at high SNR, when the quality of the available CSIT varies considerably among the users' channels.

\section{Optimal Feedback in Large FDD Multi-user Systems}
\label{sec:optim-feedb-large}

Consider a frequency-division duplex (FDD) system, where the users quantize their perfectly estimated channel vectors and send the codebook quantization index back to the transmitter over an independent feedback channel of limited rate. The feedback channels are assumed to be error-free and of zero delay. The quantization codebooks are generated prior to transmission and are known to both transmitter and respective receiver. Due to the finite rate feedback link, imposing a finite codebook size, the transmitter has only access to an imperfect estimate of the true downlink channel. To obtain tractable expressions, we restrict the subsequent analysis to i.i.d.\ Gaussian channels $\h_k\sim\mathcal{CN}(\0,\I_M)~\forall k$.

In the sequel, we follow the limited feedback analysis in \cite{jindal2006mbclf}, where each user's \textit{channel direction} $\tilde{\h}_k\eqdef\frac{\h_k}{\|\h_k\|_2}$ is quantized using $B$ bits which are subsequently fed back to the transmitter. Under Rayleigh fading, the channel $\h_k$ can be decomposed as $\h_k=\|\h_k\|_2\cdot\tilde{\h}_k$, where we suppose that the channel magnitude $\|\h_k\|_2$ is perfectly known to the transmitter since it can be efficiently quantized with only a few bits \cite{jindal2006mbclf}. Without loss of generality,\footnote{The derived scaling results hold for \textit{any} quantization codebook \cite{jindal2006mbclf}.} we assume random vector quantization (RVQ), where each user \textit{independently} generates a random codebook $\mathcal{C}_k\eqdef\{\w_{ki},\dots,\w_{k2^B}\}$ containing $2^B$ vectors $\w_{ki}\inn\Cbb^M$ that are isotropically distributed on the $M$-dimensional unit sphere. Subsequently, user $k$ quantizes its channel direction $\tilde{\h}_k$ to the closest $\w_{ki}$ according to
\begin{equation*}
  \hat{\tilde{\h}}_k = \arg\underset{\w_{ki}\;\inn\;\mathcal{C}_k}{\max}\|\tilde{\h}_k^\herm\w_{ki}\|.
\end{equation*}
Under RVQ, the quantized channel direction $\hat{\tilde{\h}}_k\inn\mathcal{C}_k$ is isotropically distributed on the $M$-dimensional unit sphere due to the statistical properties of both, the random codebook $\mathcal{C}_k$ and the channels $\h_k$. Thus, for \textit{fine quantization} with small errors, the entries of both $\tilde{\h}_k$ and $\hat{\h}_k=\|\h_k\|_2\cdot\hat{\tilde{\h}}_k$ can be modeled with good approximation as i.i.d.\ Gaussian of zero mean and unit variance. The quantization error vector $\e_k$ can be approximated as $\e_k\sim\mathcal{CN}(\0,\I_M)$ \cite{marco2005van} and we can write
\begin{equation}
  \label{eq:40a}
  \hat{\h}_k = \sqrt{1-\tau_k^2}\h_k + \tau_k\e_k,
\end{equation}
where $\tau_k^2$ is the quantization error variance. The scaling in \eqref{eq:40a} is required to ensure that the elements of $\hat{\h}_k$ have unit variance. Therefore, the effect of imperfect CSIT under RVQ in \eqref{eq:40a} is captured by the channel model \eqref{eq:3}. For RVQ, the quantization error $\tau_k^2\eqdef\|\tilde{\h}_k^\herm\hat{\tilde{\h}}_k\|$ can be upper bounded as \cite[Lemma 1]{jindal2006mbclf}
\begin{equation}
  \label{eq:41}
  \tau_k^2< 2^{-\frac{B}{M-1}}.
\end{equation}
The bound in \eqref{eq:41} is tight for large $B$ \cite{jindal2006mbclf}. Moreover, since the quantization codebooks of the users are supposed to be of equal size, the resulting CSIT distortions can be assumed identical, i.e., $\tau^2_k=\tau^2~\forall k$. Under this assumption and equal power allocation, for large $M$, the SINR $\SINR^\circ$ is identical for all users and, hence, optimizing $\SINR^\circ$ is equivalent to optimizing the per-user rate $R^\circ=\log_2(1+\SINR^\circ)$ bits/s/Hz and the sum rate $\Rsumh=KR^\circ$. 


In the following, in particular under RVQ, we will derive the necessary scaling of the distortion $\tau^2$ to ensure that
\begin{equation*}
  \label{eq:60}
  \Delta R_k - \log_2b \tom 0,
\end{equation*}
almost surely, where $\Delta R_k$ is defined in \eqref{eq:139b} and $b\geq 1$. That is, a constant rate gap of $\log_2b$ is maintained \textit{exactly} as $M,K\to\infty$. A constant rate gap ensures that the full multiplexing gain of $K$ is achieved. Thus, the proposed scaling also guarantees a larger but constant rate gap to the optimal DPC solution with perfect CSIT. The choice of a rate offset $\log_2b$ is motivated by mere mathematical convenience to avoid terms of the form $2^b$ and to be compliant with \cite{jindal2006mbclf}. 

With this strategy we closely follow \cite{jindal2006mbclf}. In \cite[Theorem 1]{jindal2006mbclf}, the author derived an upper bound of the \textit{ergodic} per-user gap $\Delta \tilde{R}_{\rm zf}$ for ZF precoding with $M=K$ and \textit{unit norm} precoding vectors under RVQ, which is given by 
\begin{equation}
  \label{eq:170jj}
  \Delta \tilde{R}_{\rm zf} < \log_2\left(1 + \SNR\cdot 2^{-\frac{B}{M-1}} \right).
\end{equation}
We cannot directly compare the deterministic equivalents to the upper bound in \eqref{eq:170jj} for two reasons, (i) under ZF precoding and $M=K$, a deterministic equivalent for the per-user rate gap does not exist and (ii) \cite{jindal2006mbclf} considers unit norm precoding vectors, whereas in this paper we only impose a total power constraint \eqref{eq:transmit-power-constraint}. Concerning (i), at high SNR, we can use the deterministic equivalent for RZF-CDU precoding given in Corollary \ref{lemma-rzf-lfb-2} as a good approximation for ZF precoding, since for high SNR the rates of RZF-CDU and ZF precoding converge. Regarding (ii), deriving a deterministic equivalent of the SINR under linear precoding with a unit norm power constraint on the precoding vectors is difficult, since it introduces an additional non-trivial dependence on the channel. However, it is useful to compare the accuracy of the upper bound in \eqref{eq:170jj} and the deterministic equivalent $\Delta R^\circ_{k,\rm rzf-cdu}$ in Corollary \ref{lemma-rzf-lfb-2} at high SNR.

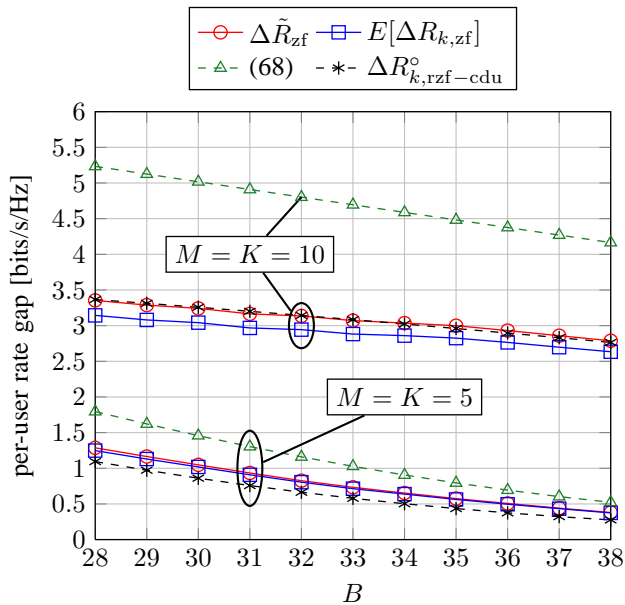
\begin{figure}[ht]
  \centering
  \begin{tikzpicture}
  \tikzstyle{every pin}=[fill=white,draw=black]
    \pgfplotsset{every axis legend/.append style={
        cells={anchor=west}, at={(0.5,1.05)}, anchor=south}}
    \pgfplotsset{every axis/.append style={line width=0.5pt}}
    \pgfplotsset{every axis/.append style={mark options=solid, mark size=2.5pt}}

    \begin{axis}[xlabel={$B$}, ylabel={per-user rate gap [bits/s/Hz]},
      grid=major, xmin=28, xmax=38, xtick={28,29,...,38}, ymin=0,
      ymax=6, ytick={0,0.5,...,6}, legend columns=2]

      \addplot[red, solid, mark=o] plot coordinates { (28.000,1.287) (29.000,1.165) (30.000,1.048)         (31.000,0.934) (32.000,0.826) (33.000,0.733) (34.000,0.653) (35.000,0.575) (36.000,0.507)         (37.000,0.441) (38.000,0.380) };

      \addplot[blue, solid, mark=square] plot coordinates { (28.000,1.248) (29.000,1.130) (30.000,1.019)         (31.000,0.909) (32.000,0.805) (33.000,0.714) (34.000,0.638) (35.000,0.563) (36.000,0.496)         (37.000,0.434) (38.000,0.374) };

      \addplot[green, dashed, mark=triangle] plot coordinates { (28.000,1.795) (29.000,1.622) (30.000,1.458)         (31.000,1.304) (32.000,1.160) (33.000,1.028) (34.000,0.906) (35.000,0.795) (36.000,0.694)         (37.000,0.603) (38.000,0.523) };

      \addplot[black, dashed, mark=asterisk] plot coordinates { (28.000,1.094) (29.000,0.973) (30.000,0.861)         (31.000,0.758) (32.000,0.664) (33.000,0.579) (34.000,0.503) (35.000,0.435) (36.000,0.375)         (37.000,0.322) (38.000,0.276) };

      \addplot[red, solid, mark=o] plot coordinates { (28.000,3.354) (29.000,3.288) (30.000,3.242)         (31.000,3.165) (32.000,3.136) (33.000,3.071) (34.000,3.036) (35.000,2.999) (36.000,2.931)         (37.000,2.859) (38.000,2.788) };

      \addplot[blue, solid, mark=square] plot coordinates { (28.000,3.144) (29.000,3.080) (30.000,3.042)         (31.000,2.968) (32.000,2.944) (33.000,2.882) (34.000,2.860) (35.000,2.825) (36.000,2.765)         (37.000,2.698) (38.000,2.634) };

      \addplot[green, dashed, mark=triangle] plot coordinates { (28.000,5.233) (29.000,5.125) (30.000,5.017)         (31.000,4.909) (32.000,4.802) (33.000,4.695) (34.000,4.588) (35.000,4.482) (36.000,4.376)         (37.000,4.270) (38.000,4.165) };

      \addplot[black, dashed, mark=asterisk] plot coordinates { (28.000,3.366) (29.000,3.313) (30.000,3.258)         (31.000,3.201) (32.000,3.143) (33.000,3.084) (34.000,3.023) (35.000,2.960) (36.000,2.896)         (37.000,2.831) (38.000,2.765) };

      \draw (axis cs:34,2) node[fill=white,draw=black] (pint0) {$M=K=5$}; 
      \draw (axis cs:31,1) node[draw,black,thick,ellipse,minimum height=1.0cm] (ell0) {}; 
      \draw[black,thick] (pint0) -- (ell0);
      
      \draw (axis cs:31,4) node[fill=white,draw=black] (pint1) {$M=K=10$}; \draw (axis cs:32,3)       node[draw,black,thick,ellipse,minimum height=0.6cm] (ell1) {}; \draw[black,thick] (pint1) -- (ell1);
      \draw[black,thick] (pint1) -- (axis cs:32,4.802);

      \legend{ 
        {$\Delta \tilde{R}_{\rm zf}$}\\
        {$\EE[\Delta R_{k,\rm zf}]$}\\
        {\eqref{eq:170jj}}\\
        {$\Delta R^\circ_{k,\rm rzf-cdu}$}\\};

    \end{axis}
  \end{tikzpicture}

  \caption{ZF, per-user rate gap vs. number of bits per user with $\SNR=25$ dB, $\Tt_k\eqq\I_M~\forall k$.}
  \label{fig:zf-comp}
\end{figure}

Figure \ref{fig:zf-comp}, depicts the per-user rate gap as a function of the feedback bits $B$ per user under ZF precoding at a SNR of 25 dB. We simulated the ergodic per-user rate gap $\Delta \tilde{R}_{\rm zf}$ and $\EE[\Delta R_{k,\rm zf}]$ of ZF precoding with unit norm precoding vectors and total power constraint, respectively. We compare the numerical results to the upper bound \eqref{eq:170jj} and to the deterministic equivalent $\Delta R^\circ_{k,\rm rzf-cdu}$ for $M=K=5$ and $M=K=10$. For both system dimensions $\Delta \tilde{R}_{\rm zf}$ and $\EE[\Delta R_{k,\rm zf}]$ are close, suggesting that our results derived under the total power constraint may be good approximations for the case of unit norm precoding vectors as well. As mentioned in \cite{jindal2006mbclf}, the accuracy of the upper bound increases with increasing $B$ but the deterministic equivalent $\Delta R^\circ_{k,\rm rzf-cdu}$ appears to be more accurate for both $M=K=5$ and $M=K=10$. In fact, for $M=K=10$, $\Delta R^\circ_{k,\rm rzf-cdu}$ approximates the per-user rate gap significantly more accurately than the upper bound \eqref{eq:170jj} for the given SNR. We conclude that the proposed deterministic equivalent $\Delta R^\circ_{k,\rm rzf-cdu}$ is sufficiently accurate and can be used to derive scaling laws for the optimal feedback rate. 

In the following, we compare the scaling of $\tau^2$ under RZF-CDA, RZF-CDU and ZF ($M>K$) precoding to the upper bound given for ZF ($M=K$) precoding in \cite[Theorem 3]{jindal2006mbclf}. For the sake of comparison, we restate \cite[Theorem 3]{jindal2006mbclf}.

\begin{theorem}
  \label{sec:optim-feedb-large-1}
  \cite[Theorem 3]{jindal2006mbclf}. In order to maintain a rate
  offset no larger than $\log_2b$ (per user) between zero-forcing with
  perfect CSIT and with finite-rate feedback (i.e., $\Delta
  R(\SNR)\leq\log_2b$ $\forall\SNR$), it is sufficient to scale the
  number of feedback bits per mobile according to
  \begin{align*}
    \label{eq:44}
    B_{\rm zf} &= (M-1)\log_2\SNR - (M-1)\log_2(b-1)\\
    &\approx \frac{M-1}{3}\SNR_{\rm dB} - (M-1)\log_2(b-1).
  \end{align*}
\end{theorem}
where $\SNR_{\rm dB}=10\log_{10}\SNR$. It is also mentioned that the result in \cite[Theorem 3]{jindal2006mbclf} holds true for RZF-CDU precoding for high SNR, since ZF and RZF-CDU precoding converge for asymptotically high SNR. Furthermore, it is claimed, corroborated by simulation results, that \cite[Theorem 3]{jindal2006mbclf} is true under RZF-CDU precoding for \textit{all} SNR.

In order to correctly interpret the subsequent results, it is important to understand the differences between our approach and the approach in \cite{jindal2006mbclf}. The scaling given in \cite[Theorem 3]{jindal2006mbclf} is a strict upper bound on the \textit{ergodic} per-user rate gap $\EE_{\H}[\Delta R_k]$ for all SNR and all $M=K$ under a unit norm constraint on the precoding vectors. In contrast, our approach yields a necessary scaling of $\tau^2$ that maintains a given \textit{instantaneous} target rate gap $\log_2b$ \textit{exactly} as $M,K\to\infty$ under a total power constraint. Therefore, our results are \textit{not} upper bounds for small $M$, i.e., we cannot guarantee that $\Delta R_k < \log_2b$ for small dimensions. But since for asymptotically large $M$, the rate gap is maintained exactly and we apply an upper bound on the CSIT distortion under RVQ \eqref{eq:41}, it follows that our results become indeed upper bounds for large $M$. Simulations reveal that under the derived scaling of $\tau^2$, the per-user rate gap is very close to $\log_2b$ even for small dimension, e.g., $M=10$. Concerning the ergodic and instantaneous per-user rate gap, the reader is reminded that our results hold also for ergodic per-user rates as a consequence of the dominated convergence theorem, see Remark \ref{rem:dct}.

Consequently, a comparison of the results in \cite{jindal2006mbclf} to our solutions is meaningful, especially for larger values of $M$ where our results become upper bounds.

In the following section, we apply the deterministic equivalents of the per-user rate gap under RZF-CDA, RZF-CDU and ZF precoding provided in Corollaries \ref{lemma-rzf-lfb}, \ref{lemma-rzf-lfb-2} and \ref{lemma-zf-lfb}, respectively, to derive scaling laws for the amount of feedback necessary to achieve full multiplexing gain.

\subsection{Channel Distortion Aware Regularized Zero-forcing Precoding}

\begin{proposition}
  \label{theorem:rzf-tau}
  Let $\Tt_k\eqq\I_M~\forall k$. Then the CSIT distortion $\tau^2$, such that the rate gap $\Delta R_{k,\rm     rzf-cda}$ of user $k$ between RZF-CDA precoding with perfect CSIT and imperfect CSIT satisfies
  \begin{equation*}
    \Delta R_{k,\rm rzf-cda} - \log_2b \tom 0
  \end{equation*}
  almost surely, has to scale as
  \begin{align}
    \label{eq:theorem:rzf-tau-1}
    \tau^2 &= \frac{\phi^\circ_{\rm rzf-cda}(\SNR,b)}{\SNR}, \\
    \label{eq:theorem:rzf-tau-phi}
    \phi^\circ_{\rm rzf-cda}(\SNR,b) & = \frac{\SNR\left[(1+\beta)b + \delta(\beta-1)\right] -
      \frac1{2b}(\delta^2-b^2)}{(1+\beta)b + \delta(\beta-1) + \frac1{2b}(\delta^2 - b^2)},\\
    \delta &= 1 - b +\chi(1) + \SNR(\beta-1)\nonumber,
  \end{align}
  where $\chi$ is defined in \eqref{eq:chi}. With $\beta=1$, the distortion $\tau^2$ has to scale as
  \begin{align*}
    \tau^2 = \frac{1 + 4\SNR - \frac{\delta^2}{b^2}}{3 + \frac{\delta^2}{b^2}}\frac1\SNR.
  \end{align*}
\end{proposition}
\begin{IEEEproof}
  Set $\Delta \bar{R}_{\rm rzf-cda}$ given in Corollary \ref{lemma-rzf-lfb} equal to $\log_2b$ and solve for $\tau^2$.
\end{IEEEproof}
Although the proposed scaling of $\tau^2$ in \eqref{eq:theorem:rzf-tau-1} converges to zero for asymptotically high SNR, we can approximate the term $\phi^\circ_{\rm rzf-cda}(\SNR,b)$ in the high SNR regime.
\begin{proposition}
  \label{pr:fdd-rzf-1}
  For asymptotically high SNR, the term $\phi^\circ_{\rm rzf-cda}(\SNR,b)$ defined in \eqref{eq:theorem:rzf-tau-phi} converges to the following limits,
\begin{equation}
  \label{eq:phi-high-snr-rzf}
  \lim_{\SNR\to\infty} \phi^\circ_{\rm rzf-cda}(\SNR,b) = 
  \begin{cases} 
    b^2-1 & \textrm{if}~\beta = 1  \\
    b - 1 & \textrm{if}~\beta > 1.
  \end{cases}
\end{equation}
\end{proposition}
\begin{IEEEproof}
  For $\beta\eqq 1$ observe that $\delta$ scales as $2\sqrt{\SNR}$. Thus, for $\SNR\to\infty$, \eqref{eq:theorem:rzf-tau-phi} converges to $b^2-1$. If $\beta>1$, the term $\delta$ takes the form
\begin{equation*}
  \delta = 1 - b + (\beta-1)\SNR  + |1-\beta|\SNR\left(1 + o(1) \right) \overset{\SNR\to\infty}{\longrightarrow}     2\SNR(\beta-1) + 1-b.
\end{equation*}
Therefore, for $\SNR\to\infty$, \eqref{eq:theorem:rzf-tau-phi} converges to $b-1$, which completes the proof.
\end{IEEEproof}

\begin{remark}
  Note that $\lim_{\SNR\to\infty}\frac{\phi^\circ_{\rm rzf-cda}(\SNR,b)}{\SNR}=0$ and thus, we require $\beta>1$ to ensure that the limit $\SNR\to\infty$ of the deterministic equivalent is well defined, see Remark \ref{re:snr}. However, for finite SNR with the approximation in Proposition~\ref{pr:fdd-rzf-1}, we have $\tau^2>0$ and the scaling result holds true.
\end{remark}

To compare Proposition \ref{theorem:rzf-tau} to \cite[Theorem 3]{jindal2006mbclf}, we use the upper bound on the quantization distortion \eqref{eq:41}, i.e., $\tau^2\eqq 2^{-\frac{B^\circ_{\rm rzf-cda}}{M-1}}$, where $B^\circ_{\rm rzf-cda}$ is the number of feedback bits per user under RZF-CDA precoding. Thus, \eqref{eq:theorem:rzf-tau-1} can be rewritten as
\begin{equation}
  \label{eq:rzf-B}
  B^\circ_{\rm rzf-cda} = (M-1)\log_2\SNR - (M-1)\log_2\phi^\circ_{\rm rzf-cda}(\SNR,b).
\end{equation}

\subsection{Channel Distortion Unaware Regularized Zero-forcing Precoding}

Although the RZF-CDU precoder is suboptimal under imperfect CSIT, the results are useful to compare to the work in \cite{jindal2006mbclf}. 


\begin{proposition}
\label{th:fdd-rzf}
  Let $\Tt_k\eqq\I_M~\forall k$. Then the CSIT distortion $\tau^2$, such that the rate gap $\Delta R_{k,\rm rzf-cdu}$ with $\alpha=1/(\beta\SNR)$ of user $k$ between RZF-CDU precoding with perfect CSIT and imperfect CSIT satisfies
  \begin{equation*}
    \Delta R_{k,\rm rzf-cdu} - \log_2b \tom 0
  \end{equation*}
  almost surely, has to scale as
  \begin{align*}
    \tau^2 &= \frac{\phi^\circ_{\rm rzf-cdu}(\SNR,b)}{\SNR}, \\
    \phi^\circ_{\rm rzf-cdu}(\SNR,b) &=
    \frac{(b-1)(1+m^\circ)(\SNR + \bar{m}^\circ)}
    {(b-1-m^\circ)[1-\bar{m}^\circ] +
      bm^\circ[1+\frac1\SNR \bar{m}^\circ]},
  \end{align*}
  where $m^\circ$ is defined in \eqref{eq:32a} and $\bar{m}^\circ\eqdef (1+m^\circ)^2$.
\end{proposition}
\begin{IEEEproof}
  Set $\Delta R_{k,\rm rzf-cdu}$ from Corollary \ref{lemma-rzf-lfb-2} equal to $\log_2b$ and solve for $\tau^2$.
\end{IEEEproof}

An approximation of the term $\phi^\circ_{\rm rzf-cdu}(\SNR,b)$ at high SNR is given in the following proposition.

\begin{proposition}
\label{pr:fdd-rzf-cdu}
  For asymptotically high SNR, $\phi^\circ_{\rm rzf-cdu}(\SNR,b)$ converges to the following limits,
  \begin{equation}
    \label{eq:phi-high-snr-rzf-2}
    \lim_{\SNR\to\infty} \phi^\circ_{\rm rzf-cdu}(\SNR,b) = 
    \begin{cases} 
      2(b-1) & \textrm{if}~\beta = 1  \\
      b-1 & \textrm{if}~\beta > 1.
    \end{cases}
  \end{equation}
\end{proposition}
\begin{IEEEproof}[Proof of Proposition \ref{pr:fdd-rzf-cdu}]
  For $\beta\eqq 1$ and $\SNR$ large, $m^\circ$ scales as $\sqrt{\SNR}$. Therefore, $\lim_{\SNR\to\infty}\phi^\circ_{\rm rzf-cdu}(\SNR,b)=2(b-1)$. If $\beta >1$, for large $\SNR$, the term $m^\circ$ scales as $\SNR(\beta-1)$. With this approximation we obtain $\lim_{\SNR\to\infty}\phi^\circ_{\rm rzf-cdu}(\SNR,b) = b-1$, which completes the proof.
\end{IEEEproof}

Applying the upper bound on the CSIT distortion under RVQ \eqref{eq:41} with $B^\circ_{\rm rzf-cdu}$ bits per user, we obtain
\begin{align}
  \label{eq:prop-rzf}
  B^\circ_{\rm rzf-cdu} = (M-1)\log_2\SNR - (M-1)\log_2\phi^\circ_{\rm rzf-cdu}(\SNR,b).
\end{align}

\subsection{Zero-forcing Precoding}

The following results are only valid for $\beta>1$ and thus, they cannot be compared to \cite[Theorem 3]{jindal2006mbclf} which are derived under the assumption $M=K$. However, for high SNR the results for the RZF-CDU precoder are a good approximation for the ZF precoder as well, even for $\beta=1$.

\begin{corollary}
  \label{theorem:zf-tau}
  Let $\beta>1$ and $\Tt_k\eqq\I_M~\forall k$. To maintain a rate offset $\Delta R_{k,\rm zf}$ such that
  \begin{equation*}
    \Delta R_{k,\rm zf} - \log_2b \tom 0
  \end{equation*}
  almost surely, the distortion $\tau^2$ has to scale according to
  \begin{align}
    \tau^2 &= \frac{\phi^\circ_{\rm zf}(\SNR,b)}{\SNR}, \nonumber\\
    \label{eq:theorem:zf-tau-phi}
    \phi^\circ_{\rm zf}(\SNR,b) & = \frac{(b-1)[1 +
      \SNR(\beta-1)]}{1-b+(\beta-1)[\SNR+b]}.
  \end{align}
\end{corollary}
\begin{IEEEproof}
  From Corollary \ref{lemma-zf-lfb}, set $\Delta R^\circ_{\rm zf} =\log_2b$ and solve for $\tau^2$.
\end{IEEEproof}

\begin{proposition}
\label{pr:fdd-zf}
For asymptotically high SNR, $\phi^\circ_{\rm zf}(\SNR,b)$ in \eqref{eq:theorem:zf-tau-phi} converges to
\begin{equation}
  \label{eq:phi-high-snr}
  \lim_{\SNR\to\infty} \phi^\circ_{\rm zf}(\SNR,b) = b-1.
\end{equation}  
\end{proposition}
\begin{IEEEproof}
  From \eqref{eq:theorem:zf-tau-phi}, the result is immediate.
\end{IEEEproof}

Under RVQ with $B^\circ_{\rm zf}$ feedback bits per user, we have
\begin{align}
  \label{eq:prop-zf}
  B^\circ_{\rm zf} = (M-1)\log_2\SNR - (M-1)\log_2\phi^\circ_{\rm zf}(\SNR,b).
\end{align}

\subsection{Discussion and Numerical Results}
\label{sec:discussion}

At this point, we can draw the following conclusions. The optimal scaling of the CSIT distortion $\tau^2$ is lower for $\beta=1$ compared to $\beta>1$. For $\beta=1$, the optimal scaling of the feedback bits $B^\circ_{\rm rzf-cda}$, $B^\circ_{\rm rzf-cdu}$ and $B$ for ZF in \cite[Theorem 3]{jindal2006mbclf} are different, even at high SNR. In fact, for large $M$, under RZF-CDU precoding and ZF precoding, the upper bound in \cite[Theorem 3]{jindal2006mbclf} appears to be too pessimistic in the scaling of the feedback bits. From \eqref{eq:prop-rzf} and \eqref{eq:phi-high-snr-rzf-2}, a more accurate choice may be
\begin{equation}
  \label{eq:rzf-bits}
  B^\circ_{\rm rzf-cdu} = (M-1)\log_2\SNR - (M-1)\log_2(2(b-1)),
\end{equation}
i.e., $M-1$ bits less than proposed in \cite[Theorem 3]{jindal2006mbclf}. However, recall that \eqref{eq:rzf-bits} becomes an upper bound for large $M$ and a rate gap of at least $\log_2b$ bits/s/Hz cannot be guaranteed for small values of $M$. Moreover, for high SNR, $\beta=1$ and large $M$, to maintain a rate offset of $\log_2b$, the RZF-CDA precoder requires $(M-1)\log_2(\frac{b+1}{2})$ bits \textit{less} than the RZF-CDU and ZF precoder and $(M-1)\log_2(b+1)$ bits \textit{less} than the scaling proposed in \cite[Theorem 3]{jindal2006mbclf}.

In contrast, for $\beta>1$ and high SNR, we have $B^\circ_{\rm rzf-cda} \eqq B^\circ_{\rm rzf-cdu}\eqq B^\circ_{\rm zf}$. Intuitively, the reason is that, for $\beta>1$, the channel matrix is well conditioned and the RZF and ZF precoders perform similarly. Therefore, both schemes are equally sensitive to imperfect CSIT and thus the scaling of $\tau^2$ is the same for high SNR.

Note that our model comprises a generic distortion of the CSIT. That is, the distortion can be a combination of different additional factors, e.g., channel estimation at the receivers, channel mismatch due to feedback delay or feedback errors (see \cite{caire2010mma}) as long as they can be modeled as additive noise \eqref{eq:3}. Moreover, we consider i.i.d.\ block-fading channels, which can be seen as a worst case scenario in terms of feedback overhead. It is possible to exploit channel correlation in time, frequency and space to refine the CSIT or to reduce the amount of feedback.

\begin{figure}[t]
  \centering
  \begin{tikzpicture}
      \pgfplotsset{every axis legend/.append style={
          cells={anchor=west}, at={(0.5,1.05)}, anchor=south, font=\scriptsize}}
      \pgfplotsset{every axis/.append style={line width=0.5pt}}
      \pgfplotsset{every axis/.append style={mark options=solid, mark size=2.5pt}}
      
      \begin{axis}[xlabel={$\SNR$ [dB]}, ylabel={ergodic sum rate [bits/s/Hz]},
        grid=major, xmin=0, xmax=30, xtick={0,5,...,30}, ymin=0,
      ymax=70, ytick={0,10,...,70}, legend columns=2]
      

      \addplot[red, solid] plot coordinates { (0.000,7.179)
        (5.000,12.724) (10.000,19.626) (15.000,27.518) (20.000,36.134)
        (25.000,46.305) (30.000,58.635) };
      
      \addplot[red, dashed] plot coordinates { (0.000,7.031) (5.000,12.597) (10.000,19.574) (15.000,28.058)         (20.000,37.714) (25.000,48.635) (30.000,61.518) };

      \addplot[blue,solid,mark=o] plot coordinates { (0.000,1.202)
        (5.000,3.273) (10.000,10.054) (15.000,17.755) (20.000,26.259)
        (25.000,35.578) (30.000,46.157) };


      \addplot[black,solid,mark=square] plot coordinates {
        (0.000,1.158) (5.000,2.844) (10.000,8.654) (15.000,15.690)
        (20.000,23.635) (25.000,32.622) (30.000,43.842) };


      \addplot[green,solid,mark=triangle] plot coordinates {
        (0.000,1.162) (5.000,7.635) (10.000,14.748) (15.000,22.408)
        (20.000,30.740) (25.000,40.045) (30.000,51.590) };



      \addplot[red,solid,mark=asterisk] plot coordinates { 
        (0.000,1.1560) (5.000,4.1353) (10.000,11.5205) (15.000,18.9808)
        (20.000,27.0027) (25.000,36.0402) (30.000,47.0551) };


      \draw[<->,>=stealth] (axis cs:17.500,31.6875) -- (axis
      cs:17.500,22.1416); 

      \draw (axis cs:10,50) node[fill=white,draw=black] (pin) {$\Delta
        \Rsum\approx 10$ bits/s/Hz}; \draw[black,thick] (axis
      cs:17.5,26.9145) -- (pin);

      \legend{ {RZF, $\tau^2=0$ (total power)}\\
        {RZF, $\tau^2=0$ (unit norm) }\\
        {RZF-CDA, $B^\circ_{\rm rzf-cda}$ \eqref{eq:rzf-B}}\\
        {RZF-CDU, $B^\circ_{\rm rzf-cda}$ \eqref{eq:rzf-B}}\\
        {RZF-CDU, $B\eqq\frac{M-1}{3}\SNR_{\rm dB}$}\\
        {RZF-CDU, $B^\circ_{\rm rzf-cdu}$} \eqref{eq:rzf-bits}\\};
    \end{axis}
  \end{tikzpicture}
  \caption{RZF, ergodic sum rate vs. SNR under RZF precoding and RVQ with $B$ feedback bits per user, where $B$ is chosen to maintain a sum rate offset of $K\log_2b\eqq 10$, $\Tt_k=\I_M~\forall k$ and $M=K=10$.}
  \label{fig:lim-fb}
\end{figure}
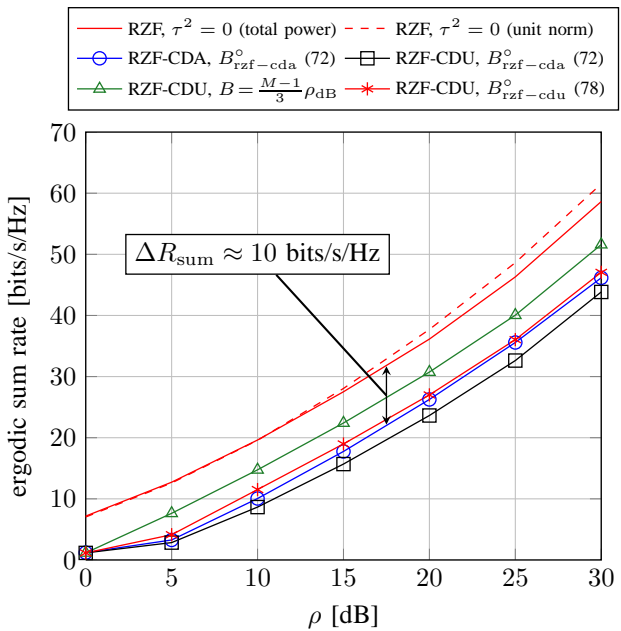

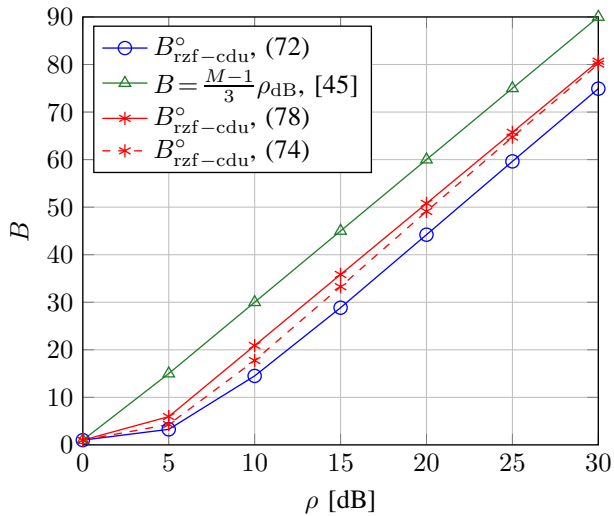
\begin{figure}[t]
  \centering
  \begin{tikzpicture}
    \pgfplotsset{every axis legend/.append style={
        cells={anchor=west}, at={(0.02,0.98)}, anchor=north west}}
    \pgfplotsset{every axis/.append style={line width=0.5pt}}
    \pgfplotsset{every axis/.append style={mark options=solid, mark size=2.5pt}}

    \begin{axis}[xlabel={$\SNR$ [dB]}, ylabel={$B$},
      grid=major, xmin=0, xmax=30, xtick={0,5,...,30}, ymin=0,
      ymax=90, ytick={0,10,...,90}]


      \addplot[blue,solid,mark=o] plot coordinates { (0.000,1.000)
        (5.000,3.273) (10.000,14.504) (15.000,28.854) (20.000,44.208)
        (25.000,59.630) (30.000,74.919) };


      \addplot[green,solid,mark=triangle] plot coordinates {
        (0.000,1.000) (5.000,15.000) (10.000,30.000) (15.000,45.000)
        (20.000,60.000) (25.000,75.000) (30.000,90.000) };

      \addplot[red,solid,mark=asterisk] plot coordinates {
        (0.000,1.000) (5.000,5.9487) (10.000,20.8974) (15.000,35.8460)
        (20.000,50.7947) (25.000,65.7434) (30.000,80.6921) };

      \addplot[red,dashed,mark=asterisk] plot coordinates {
        (0.000,1.000) (5.000,4.287) (10.000,17.748) (15.000,33.282)
        (20.000,49.116) (25.000,64.730) (30.000,80.101) };


      \legend{{$B_{\rm rzf-cdu}^{\circ}$, \eqref{eq:rzf-B}}\\
        {$B\eqq\frac{M-1}{3}\SNR_{\rm dB}$, \cite{jindal2006mbclf}}\\
        {$B_{\rm rzf-cdu}^{\circ}$, \eqref{eq:rzf-bits}}\\
        {$B_{\rm rzf-cdu}^{\circ}$, \eqref{eq:prop-rzf}}\\};

    \end{axis}
  \end{tikzpicture}
  \caption{RZF, $B$ feedback bits per user vs. SNR, with $B$ to maintain a sum rate offset of $K\log_2b\eqq 10$ and $\Tt_k\eqq\I_M~\forall k$, $M\eqq K \eqq 10$.}
  \label{fig:lim-fb-bits}
\end{figure}

Figures \ref{fig:lim-fb} and \ref{fig:lim-fb-bits} depict the ergodic sum rate of RZF precoding under RVQ and the corresponding number of feedback bits per user $B$, respectively. To avoid an infinitely high regularization parameter $\alpha^{\star\circ}$, the minimum number of feedback bits is set to one.

In Figure \ref{fig:lim-fb}, we plot the ergodic sum rate for RZF precoding under perfect CSIT with total power constraint (red solid lines) and unit norm constraint on the precoding vectors (red dashed line). We observe, that the sum rate under unit norm constraint is slightly larger at high SNR, suggesting that our scaling results for RZF precoding derived under a total power constraint become inaccurate under the unit norm constraint at high SNR. Hence, one has to be cautious when comparing the scaling in \cite[Theorem 3]{jindal2006mbclf} directly to the scaling derived with the large system approximations at high SNR. From Figure \ref{fig:lim-fb}, we further observe that (i) the desired sum rate offset of $10$ bits/s/Hz is approximately maintained over the given SNR range when $B$ is chosen according to \eqref{eq:rzf-B} and the high SNR approximation in \eqref{eq:rzf-bits} under RZF-CDA and RZF-CDU precoding, respectively, (ii) given an equal number of feedback bits \eqref{eq:rzf-B}, the RZF-CDA precoder achieves a significantly higher sum rate compared to RZF-CDU for medium and high SNR, e.g., about 2.5 bits/s/Hz at $20$ dB and (iii) to maintain a sum rate offset of $K$ bits/s/Hz, the proposed feedback scaling of $B\eqq\frac{M-1}{3}\SNR_{\rm dB}$ for unit norm precoding vectors \cite{jindal2006mbclf} is very pessimistic, since the sum rate offset to RZF with total power constraint and unit norm constraint is about $6$ bits/s/Hz and $7$ bits/s/Hz at $20$ dB, respectively.

We conclude that the proposed RZF-CDA precoder significantly increases the sum rate for a given feedback rate or equivalently significantly reduces the amount of feedback given a target rate. Moreover, the scaling of the number of feedback bits under RZF-CDU precoding proposed in \cite[Theorem 3]{jindal2006mbclf} appears to be less accurate under a total power constraint than our large system approximation in \eqref{eq:rzf-B}.

\section{Optimal Training in Large TDD Multi-user Systems}
\label{sec:optim-train-large}

Consider a time-division duplex (TDD) system where uplink (UL) and downlink (DL) share the \textit{same} channel at different times. Therefore, the transmitter estimates the channel from known pilot signaling of the receivers. The channel coherence interval $T$, i.e., the amount of channel uses for which the channel is approximately constant, is divided into $T_t$ channel uses for UL training and $T-T_t$ channel uses for coherent transmission in the DL. Note that in order to coherently decode the information symbols, the users need to know their effective (precoded) channels. This is usually accomplished by a dedicated training phase (using precoded pilots) in the DL prior to the data transmission. As shown in \cite{jindal2009wvj}, a minimal amount of training (at most one pilot symbol) is sufficient when data and pilots are processed jointly. Therefore, we assume that the users have \textit{perfect} knowledge of their effective channels and we neglect the overhead associated with the DL training.

In the considered TDD system, the imperfections in the CSIT are caused by (i) channel estimation errors in the UL, (ii) imperfect channel reciprocity due to different hardware in the transmitter and receiver and (iii) the channel coherence interval $T$. In what follows, we assume that the channel is perfectly reciprocal and we study the joint impact of (i) and (iii) for uncorrelated channels ($\Tt_k = \I_M~\forall k$).

\subsection{Uplink Training Phase}
\label{sec:uplink-train-phase}

In our setup, the distortion $\tau^2$ of the CSIT is solely caused by an imperfect channel estimation at the transmitter and is identical for all entries of $\H$. To acquire CSIT, each user transmits the same amount $T_t\geq K$ of \textit{orthogonal} pilot symbols over the UL channel to the transmitter. Subsequently, the transmitter estimates all $K$ channels simultaneously. At the transmitter, the signal $\r_k$ received from user $k$ is given by
\begin{equation*}
  \label{eq:tdd:2}
  \r_k = \sqrt{T_t P_{ul}}\h_k + \n_k,
\end{equation*}
where we assumed perfect reciprocity of UL and DL channels and $P_{ul}$ is the average available transmit power at the receivers. That is, the UL and DL channel coefficients are equal and the UL noise $\n_k\eqq[n_1,n_2,\dots,n_M]^\trans$ is assumed identical for all users and statistically equivalent to its DL analog. Subsequently, the transmitter performs an MMSE estimation of each channel coefficient $h_{ij}\sim\mathcal{CN}(0,1)$ ($i\eqq 1,\dots, K$, $j\eqq 1,\dots, M$). Due to the orthogonality property of the MMSE estimation \cite{poor1994introduction}, the estimates $\hat{h}_{ij}$ of $h_{ij}$ and the corresponding estimation errors $\tilde{h}_{ij}=h_{ij}-\hat{h}_{ij}$ are uncorrelated and i.i.d.\ complex Gaussian distributed. Hence, we can write 
\begin{equation*}
  \label{eq:tdd:3}
  \hat{h}_{ij} = h_{ij} + \tilde{h}_{ij},
\end{equation*}
where $h_{ij}$ and $\tilde{h}_{ij}$ are independent with zero mean and variance $1-\tau^2$ and $\tau^2$, respectively. The variance $\tau^2$ of the estimation error $\tilde{h}_{ij}$ is given by \cite{caire2010mma}
\begin{equation}
  \label{eq:tdd:4}
  \tau^2 = \frac1{1 + T_t\SNR_{ul}},
\end{equation}
where we defined the uplink SNR $\SNR_{ul}$ as $\SNR_{ul}\eqdef P_{ul}/\sigma^2$.

\subsection{Optimization of Channel Training}
\label{sec:optim-amount-chann}

We focus on equal power allocation among the users, i.e., $p_k=P/K~\forall k$, because it is optimal for large $M$ and $\tau_k^2=\tau^2~\forall k$, see Section \ref{sec:power-optim-tt_k=tt}. Since $T_t$ channel uses have already been consumed to train the transmitter about the user channels, there remains an interval of length $T-T_t$ for DL data transmission and thus we have the pre-log factor $1-T_t/T$. The net sum rate approximation reads
\begin{equation}
  \label{eq:140}
  \Rsumh = K\left(1 - \frac{T_t}{T}\right)\log\left(1+\SINR^\circ_k\right).
\end{equation}

To compute the training length $T_t$ that maximizes the net sum rate approximation \eqref{eq:140}, we substitute $\SINRkzf^\circ$ from Corollary \ref{co:2} into \eqref{eq:140} and the approximated net sum rate $\Rsumh^{\rm zf}$ under ZF precoding takes the form
\begin{equation}
  \label{eq:tdd:1}
   \Rsumh^{\rm zf} = K\left(1 - \frac{T_{t,\rm zf}}{T}\right)\log\left(1 + \frac{1-\tau^2}{\tau^2 + \frac{1}{\SNR_{dl}}}(\beta-1)\right),
\end{equation}
where $\SNR_{dl}\eqdef P/\sigma^2$. Similarly, for RZF-CDA precoding the approximated net sum rate $\Rsumh^{\rm rzf}$ reads
\begin{equation}
  \label{eq:tdd:1-1}
   \Rsumh^{\rm rzf} = K\left(1 - \frac{T_{t,\rm rzf}}{T}\right)\log\left(1 +  \SINRrzf^{\circ}\right),
\end{equation}
where $\SINRrzf^{\circ}$ is given in Corollary \ref{cor:theorem-rzf-sinr-2}.

Substituting \eqref{eq:tdd:4} into \eqref{eq:tdd:1} and \eqref{eq:tdd:1-1}, we obtain
\begin{align}
  \label{eq:tdd:5}
  \Rsumh^{\rm zf} &= K\left(1 -
    \frac{T_{t,{\rm zf}}}{T}\right)\log\left(1 +
    \frac{T_{t,{\rm zf}}\SNR_{ul}(\beta-1)}{1+T_{t,{\rm zf}}\frac{\SNR_{ul}}{\SNR_{dl}}+\frac1{\SNR_{dl}}}\right),\\
  \label{eq:tdd:5-1}
  \Rsumh^{\rm rzf} &=
  K\left(1-\frac{T_{t,{\rm rzf}}}{T}\right)\log\left(\frac12 +\frac12
    \omega\SNR_{dl}(\beta-1) + \frac{\chi(\omega)}{2}\right),\\
  \label{eq:tdd:5-d}
  \chi(\omega) &= \sqrt{(\beta-1)^2\omega^2\SNR_{dl}^2 + 2\omega\SNR_{dl}(1+\beta) + 1},\\
  \omega &= \frac{T_{t,{\rm rzf}}\SNR_{ul}}{1+T_{t,{\rm rzf}}\SNR_{ul} + \SNR_{dl}}\nonumber.
\end{align}
For $\beta\!>\!1$ under ZF precoding and $\beta\!\geq\!1$ for RZF-CDA precoding, it is easy to verify that the functions $\Rsumh^{\rm zf}$ and $\Rsumh^{\rm rzf}$ are strictly concave in $T_{t,{\rm zf}}$ and $T_{t,{\rm rzf}}$ in the interval $[K,T]$, respectively, where $K$ is the minimum amount of training required, due to the orthogonality constraint of the pilot sequences. Therefore, we can apply standard convex optimization algorithms \cite{boyd2004convex} to evaluate
\begin{align}
  \label{eq:tdd:6}
  T_{t,{\rm zf}}^{\star\circ} = \underset{K\leq T_{t,{\rm
        zf}}\leq
    T}{\arg\max} \Rsumh^{\rm zf} ,\\
  \label{eq:tdd:6-1}
  T_{t,{\rm rzf}}^{\star\circ} = \underset{K\leq T_{t,{\rm
        rzf}}\leq T}{\arg\max}\Rsumh^{\rm rzf}.
\end{align}
In the following, we derive approximate explicit solutions to \eqref{eq:tdd:6} and \eqref{eq:tdd:6-1} for high SNR. We distinguish two cases, (i) the UL and DL SNR vary with finite ratio $c\!\eqdef\!\SNR_{dl}/\SNR_{ul}$ and (ii) $\SNR_{dl}$ varies, while $\SNR_{ul}$ remains finite. In contrast to case (i), the system in case (ii) is interference-limited due to the finite transmit power of the users.

\subsubsection{Case 1: finite ratio $\SNR_{dl}/\SNR_{ul}$}

We derive approximate, but explicit, solutions for the optimal training intervals $T_{t,{\rm zf}}^{\star\circ},T_{t,{\rm  rzf}}^{\star\circ}$ in the high SNR regime and derive their limiting values for asymptotically low SNR.

\paragraph{High SNR Regime}

An approximate closed form solution to \eqref{eq:tdd:6} and \eqref{eq:tdd:6-1} is summarized in the following proposition.

\begin{proposition}
  \label{sec:high-snr-regime}
  Let $\SNR_{dl},\SNR_{ul}$ be large with $c\eqq\SNR_{dl}/\SNR_{ul}$ constant. Then, an approximation of the sum rate maximizing amount of channel training $T_{t,{\rm zf}}^{\star\circ}$ and $T_{t,{\rm  rzf}}^{\star\circ}$ under ZF and RZF-CDA precoding is given by
  \begin{align}
    \label{eq:12}
    T_{t,{\rm zf}}^{\star\circ} &=
    \max\left[\frac{c}{2}\sqrt{1+2\frac{2T+c}{c\bar{R}^\circ_{\rm zf}}}
      -\frac{c}{2},K\right],\\
    \label{eq:13}
    T_{t,{\rm rzf}}^{\star\circ} &=
    \begin{cases}
      \max\left[\frac{c}{2}\sqrt{1+\frac{2T+c}{c\bar{R}^\circ_{\rm rzf}}} -\frac{c}{2},K\right]
      & \textrm{if}~\beta = 1, \\
      \max\left[\frac{c}{2}\sqrt{1+2\frac{2T+c}{c\bar{R}^\circ_{\rm rzf}}} -\frac{c}{2},K\right]
      & \textrm{if}~\beta > 1,
    \end{cases}
  \end{align}
  where $\bar{R}^\circ_{\rm zf}\eqq\log(1+\SNR_{dl}(\beta-1))$ and $\bar{R}^\circ_{\rm rzf}\eqq\log(\frac12 +\frac12 \SNR_{dl}(\beta-1) + \frac{\chi(1)}{2})$.
\end{proposition}
\begin{IEEEproof}
  The proof is presented in Appendix \ref{sec:proof-proposition-1}.
\end{IEEEproof}

Thus, for a fixed DL SNR $\SNR_{dl}$, the optimal training intervals scale as $T_{t,{\rm  zf}}^{\star\circ},T_{t,{\rm rzf}}^{\star\circ}\!\sim\!\sqrt{T}$.  Likewise, for a constant $T$, the optimal training intervals scale as $T_{t,{\rm zf}}^{\star\circ},T_{t,{\rm    rzf}}^{\star\circ}\!\sim\!1/\sqrt{\log(\SNR_{dl})}$. Under ZF precoding the same scaling has been reported in \cite{kobayashi2008much,kobayashi2009otf,salim2010fma}. From this scaling it is clear that, as $\SNR_{dl}\to\infty$, $T_t^{\star\circ}$ tends to $K$, the minimum amount of training.

Moreover, for $\beta\!>\!1$, $\bar{R}^\circ_{\rm rzf}\!\geq\! \bar{R}^\circ_{\rm zf}$ with equality if $\SNR_{dl}\!\to\!\infty$. Therefore, RZF-CDA requires less training than ZF, but the training interval of both schemes is equal for asymptotically high SNR. In case of full system loading ($\beta\eqq 1$), RZF-CDA requires less training compared to the scenario where $\beta\!>\!1$.

\paragraph{Low SNR Regime}

For asymptotically low SNR $\SNR_{dl},\SNR_{ul}\!\to\!0$ with constant ratio $c\eqq\SNR_{dl}/\SNR_{ul}$ the optimal amount of training is given in the subsequent proposition.

\begin{proposition}
  Let $\SNR_{dl},\SNR_{ul}\!\to\!0$ with constant ratio $c\eqq\SNR_{dl}/\SNR_{ul}$ and $T\geq 2K$. Then, the sum rate maximizing amount of channel training $T_{t,{\rm zf}}^{\star\circ}$ and $T_{t,{\rm rzf}}^{\star\circ}$ under ZF and RZF-CDA precoding converges to
\begin{align}
  \label{eq:34}
  \lim_{\SNR_{dl}\to 0}T_{t,{\rm zf}}^{\star\circ} =
  \lim_{\SNR_{dl}\to 0}T_{t,{\rm rzf}}^{\star\circ} = \frac{T}{2}.
\end{align}
\end{proposition}
\begin{IEEEproof}
  Applying $\log(1+x) = x + O(x^2)$ and $\SNR_{ul}=\SNR_{dl}/c$, equations \eqref{eq:tdd:5} and \eqref{eq:tdd:5-1} take the form
  \begin{align}
    \label{eq:17a}
    \Rsumh^{\rm zf} &= K\left(1 -
      \frac{T_{t,{\rm zf}}}{T}\right)\frac{T_{t,{\rm
          zf}}(\beta-1)}{c}\SNR_{dl}^2 + O(\SNR_{dl}^4), \\
    \label{eq:18}
    \Rsumh^{\rm rzf} &= K\left(1-\frac{T_{t,{\rm
          rzf}}}{T}\right)\frac{T_{t,{\rm rzf}}\beta}{c}\SNR_{dl}^2 + O(\SNR_{dl}^4).
\end{align}
Maximizing equations \eqref{eq:17a} and \eqref{eq:18} with respect to $T_{t,{\rm zf}}$ and $T_{t,{\rm rzf}}$, respectively, yields \eqref{eq:34}. Since, by definition, we assume orthogonal pilot sequences, hence $T_t\geq K$, the result \eqref{eq:34} implies that $T\geq 2K$, which completes the proof.
\end{IEEEproof}

For ZF precoding, the limit has also been reported in \cite{jose2010lpm}.

\subsubsection{Case 2: $\SNR_{dl}\gg\SNR_{ul}$ with finite $\SNR_{ul}$}

This scenario models a high capacity DL channel where the primary sum rate loss stems from the inaccurate CSIT estimate due to limited-rate UL signaling caused, e.g., by a finite transmit power of the users. Thus, the system becomes interference-limited and the optimal amount of channel training under ZF precoding is given in the following proposition.

\begin{proposition}
  Let $\SNR_{dl}\to\infty$ and $\SNR_{ul}$ finite. Then the (approximated) sum rate maximizing amount of channel training $T_{t,{\rm zf}}^{\star\circ}$ is given by
  \begin{equation}
    \label{eq:15}
    T_{t,{\rm zf}}^{\star\circ} = \frac1{\SNR_{ul}(\beta-1)}\left(\frac{a}{\WW(ae)}-1 \right),
  \end{equation}
  where $\WW(z)$ is the Lambert W-function.
\end{proposition}
\begin{IEEEproof}
  For ZF precoding and $\SNR_{dl}\!\to\!\infty$, the sum rate \eqref{eq:tdd:5} can be approximated as
  \begin{equation}
    \label{eq:14}
    \Rsumh^{\rm zf} \approx  K\left(1 - \frac{T_{t,{\rm
            zf}}}{T}\right)\log\left(1 + T_{t,{\rm zf}}\SNR_{ul}(\beta-1)\right).
  \end{equation}
  Setting the derivative of \eqref{eq:14} with respect to $T_{t,{\rm zf}}$ to zero, yields
  \begin{equation}
    \label{eq:18a}
    \log(a/\omega(T_{t,{\rm zf}})) = \omega(T_{t,{\rm zf}}) - 1,
  \end{equation}
  where $a\!\eqdef\!\SNR_{ul}T(\beta-1)+1$ and $\omega(T_{t,{\rm  zf}})\!\eqdef\!(Ta)/[T + T_{t,{\rm zf}}(a-1)]$. Equation \eqref{eq:18a} can be written as
  \begin{equation*}
    \label{eq:19}
    \omega(T_{t,{\rm zf}})e^{\omega(T_{t,{\rm zf}})} = ae.
  \end{equation*}
  Notice that $\omega(T_{t,{\rm zf}})\!=\!\WW(ae)$. Thus, solving $\omega(T_{t,{\rm zf}})\eqq \WW(ae)$ for $T_{t,{\rm zf}}$ yields \eqref{eq:15}.
\end{IEEEproof}

For asymptotically low $\SNR_{ul}$ we obtain $\lim_{\SNR_{ul}\to 0} T_{t,{\rm zf}}^{\star\circ} \eqq T/2$, implying that $T\geq 2K$.

For RZF-CDA precoding, no accurate closed-form solution to \eqref{eq:tdd:6-1} has yet been found.

\subsection{Numerical Results}
\label{sec:numerical-results}

\begin{figure}[t]
  \centering
  \begin{tikzpicture}
    \pgfplotsset{every axis legend/.append style={
        cells={anchor=west}, at={(0.02,0.98)}, anchor=north west}}
    \pgfplotsset{every axis/.append style={line width=0.5pt}}
    \pgfplotsset{every axis/.append style={mark options=solid, mark
        size=2.5pt}}
    \begin{axis}[xlabel={$T$}, ylabel={training interval $T_t$},
      grid=major, xmin=0, xmax=1000,
      xtick={0,200,...,1000}, ymin=0, ymax=50,
      ytick={0,5,...,50}]

      \addplot[red, solid, mark=none] plot coordinates { (4.000,4.000)
        (100.000,8.963) (200.000,13.854) (300.000,17.682)
        (400.000,20.887) (500.000,23.715) (600.000,26.378)
        (700.000,28.760) (800.000,31.093) (900.000,33.123)
        (1000.000,35.174) };

      \addplot[blue, dashed, mark=none] plot coordinates {
        (4.000,4.000) (20.000,4.000) (50.000,6.784) (100.000,10.724)
        (200.000,16.527) (300.000,21.067) (400.000,24.926)
        (500.000,28.342) (600.000,31.439) (700.000,34.294)
        (800.000,36.955) (900.000,39.457) (1000.000,41.826) };

      \addplot[green, solid, mark=triangle, mark size=2.5pt] plot
      coordinates { (4.000,4.000) (100.000,10.891) (200.000,16.661)
        (300.000,21.189) (400.000,25.042) (500.000,28.455)
        (600.000,31.550) (700.000,34.403) (800.000,37.063)
        (900.000,39.564) (1000.000,41.932) };

      \addplot[red, solid, mark=none] plot coordinates {
        (16.000,16.000) (20.000,16.000) (50.000,16.000)
        (100.000,16.000) (200.000,16.147) (300.000,20.119)
        (400.000,23.988) (500.000,27.216) (600.000,29.989)
        (700.000,32.891) (800.000,35.769) (900.000,37.981)
        (1000.000,40.172) };

      \addplot[green, solid, mark=triangle, mark size=2.5pt] plot
      coordinates { (16.000,16.000) (100.000,16.000) (200.000,16.661)
        (300.000,21.189) (400.000,25.042) (500.000,28.455)
        (600.000,31.550) (700.000,34.403) (800.000,37.063)
        (900.000,39.564) (1000.000,41.932) };

      \addplot[red, dashed, only marks, mark=o] plot coordinates {
        (4.000,4.000) (100.000,9.144) (200.000,14.025)
        (300.000,17.832) (400.000,21.166) (500.000,23.849)
        (600.000,26.495) (700.000,28.760) (800.000,31.599)
        (900.000,32.911) (1000.000,35.288) };

      \addplot[blue, dashed, only marks, mark=o] plot coordinates {
        (4.000,4.000) (100.000,10.588) (200.000,16.357)
        (300.000,20.871) (400.000,24.707) (500.000,28.104)
        (600.000,31.183) (700.000,34.021) (800.000,36.667)
        (900.000,39.155) (1000.000,41.510) };

      \addplot[red, dashed, only marks, mark=o] plot coordinates {
        (16.000,16.000) (100.000,16.000) (200.000,16.088)
        (300.000,20.007) (400.000,23.712) (500.000,26.912)
        (600.000,29.955) (700.000,32.712) (800.000,35.248)
        (900.000,37.523) (1000.000,39.805) };

      \draw (axis cs:800,15) node[fill=white,draw=black] (pin1) {$K\eqq 4$};
      \draw [black,thick] (pin1) -- (axis cs:600.000,26.150);

      \draw (axis cs:400,10) node[fill=white,draw=black] (pin2) {$K\eqq 16$};
      \draw [black,thick] (pin2) -- (axis cs:500.000,27.216);

      \legend{
        {$T_{t,\rm zf}^{\star}$, $T_{t,\rm rzf}^{\star}$}\\
        {$T_{t,\rm zf}^{\star\circ}$ \eqref{eq:tdd:6}, $T_{t,\rm
            rzf}^{\star\circ}$ \eqref{eq:tdd:6-1}}\\
        {$T_{t,\rm zf}^{\star\circ}$ \eqref{eq:12}, $T_{t,\rm
            rzf}^{\star\circ}$} \eqref{eq:13}\\};
    \end{axis}
  \end{tikzpicture}
  \caption{ZF and RZF-CDA, optimal amount of training with $\frac{M}{K}=2$, $\SNR_{dl}\eqq 20$ dB, $\SNR_{ul}\eqq 10$ dB, $\Tt_k=\I_M~\forall k$, RZF is indicated by circle marks.}
  \label{fig:accuracy}
\end{figure}
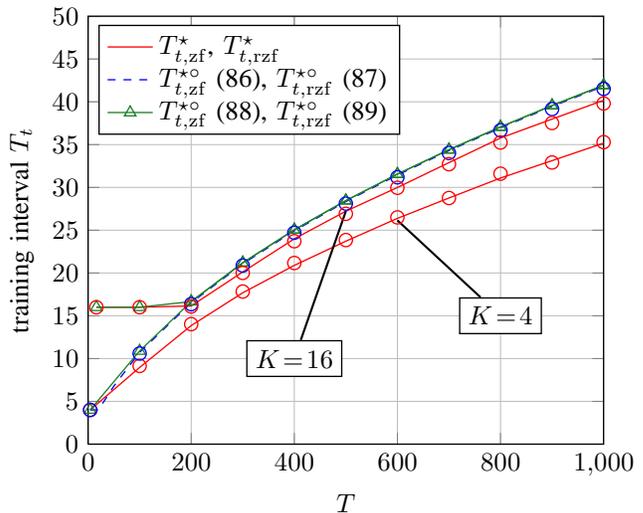

In Figure \ref{fig:accuracy}, we compare the approximated optimal training intervals $T_{t,{\rm zf}}^{\star\circ},T_{t,{\rm  rzf}}^{\star\circ}$ to $T_{t,{\rm zf}}^{\star},T_{t,{\rm rzf}}^{\star}$ computed via exhaustive search and averaged over $1\,000$ independent channel realizations. The regularization parameter $\alpha$ is computed using the large system approximation $\alpha^{\star\circ}$ in \eqref{eq:alpha-optimal-2}. Figure \ref{fig:accuracy} shows that the approximate solutions $T_{t,{\rm zf}}^{\star\circ},T_{t,{\rm rzf}}^{\star\circ}$ become very accurate for $K\eqq 16$. Moreover, it can be observed that the approximations in \eqref{eq:12} and \eqref{eq:13} match very well. Further note that for $\frac{M}{K}=2$, ZF and RZF-CDA need approximately the same amount of training, as predicted by equations \eqref{eq:12} and \eqref{eq:13}.

\begin{figure}[t]
  \centering
  \begin{tikzpicture}
    \pgfplotsset{every axis legend/.append style={
        cells={anchor=west}, at={(0.98,0.98)}, anchor=north east}}
    \pgfplotsset{every axis plot/.append style={smooth}}
    \pgfplotsset{every axis/.append style={line width=0.5pt}}
    \pgfplotsset{every axis/.append style={mark options=solid, mark
        size=2.5pt}}
    \begin{axis}[xlabel={$\SNR_{dl}$ [dB]}, ylabel={$T_t^{\star\circ}/T$},
      grid=major, xmin=-30, xmax=40,
      xtick={-30,-20,-10,0,10,20,30,40}, ymin=0, ymax=0.5,
      ytick={0,0.1,0.2,0.3,0.4,0.5}, extra y ticks={0.16,0.0533}, extra y
      tick style={grid=major}, extra y tick labels={$\frac{K}{100}$,$\frac{K}{300}$}]

      \addplot[red, solid, mark=none] plot coordinates {
        (-30.000,0.500) (-28.687,0.500) (-27.374,0.498)
        (-26.061,0.497) (-24.747,0.496) (-23.434,0.495)
        (-22.121,0.493) (-20.808,0.490) (-19.495,0.487)
        (-18.182,0.482) (-16.869,0.477) (-15.556,0.470)
        (-14.242,0.461) (-12.929,0.450) (-11.616,0.436)
        (-10.303,0.421) (-8.990,0.403) (-7.677,0.384) (-6.364,0.363)
        (-5.051,0.341) (-3.737,0.319) (-2.424,0.297) (-1.111,0.276)
        (0.202,0.257) (1.515,0.238) (2.828,0.222) (4.141,0.207)
        (5.455,0.193) (6.768,0.181) (8.081,0.170) (9.394,0.160)
        (10.707,0.160) (12.020,0.160) (13.333,0.160) (14.646,0.160)
        (15.960,0.160) (17.273,0.160) (18.586,0.160) (19.899,0.160)
        (21.212,0.160) (22.525,0.160) (23.838,0.160) (25.152,0.160)
        (26.465,0.160) (27.778,0.160) (29.091,0.160) (30.404,0.160)
        (31.717,0.160) (33.030,0.160) (34.343,0.160) (35.657,0.160)
        (36.970,0.160) (38.283,0.160) (39.596,0.160) (40.909,0.160)
        (42.222,0.160) (43.535,0.160) (44.848,0.160) (46.162,0.160)
        (47.475,0.160) (48.788,0.160) (50.101,0.160) (51.414,0.160)
        (52.727,0.160) (54.040,0.160) (55.354,0.160) (56.667,0.160)
        (57.980,0.160) (59.293,0.160) (60.606,0.160) (61.919,0.160)
        (63.232,0.160) (64.545,0.160) (65.859,0.160) (67.172,0.160)
        (68.485,0.160) (69.798,0.160) (71.111,0.160) (72.424,0.160)
        (73.737,0.160) (75.051,0.160) (76.364,0.160) (77.677,0.160)
        (78.990,0.160) (80.303,0.160) (81.616,0.160) (82.929,0.160)
        (84.242,0.160) (85.556,0.160) (86.869,0.160) (88.182,0.160)
        (89.495,0.160) (90.808,0.160) (92.121,0.160) (93.434,0.160)
        (94.747,0.160) (96.061,0.160) (97.374,0.160) (98.687,0.160)
        (100.000,0.160) };

      \addplot[blue, dashed, mark=none] plot coordinates {
        (-30.000,0.500) (-28.687,0.496) (-27.374,0.492)
        (-26.061,0.490) (-24.747,0.488) (-23.434,0.484)
        (-22.121,0.479) (-20.808,0.472) (-19.495,0.464)
        (-18.182,0.453) (-16.869,0.441) (-15.556,0.425)
        (-14.242,0.408) (-12.929,0.388) (-11.616,0.366)
        (-10.303,0.343) (-8.990,0.319) (-7.677,0.295) (-6.364,0.271)
        (-5.051,0.248) (-3.737,0.227) (-2.424,0.207) (-1.111,0.189)
        (0.202,0.173) (1.515,0.159) (2.828,0.146) (4.141,0.135)
        (5.455,0.125) (6.768,0.117) (8.081,0.109) (9.394,0.103)
        (10.707,0.097) (12.020,0.092) (13.333,0.087) (14.646,0.083)
        (15.960,0.080) (17.273,0.076) (18.586,0.073) (19.899,0.070)
        (21.212,0.068) (22.525,0.066) (23.838,0.063) (25.152,0.061)
        (26.465,0.060) (27.778,0.058) (29.091,0.056) (30.404,0.055)
        (31.717,0.053) (33.030,0.053) (34.343,0.053) (35.657,0.053)
        (36.970,0.053) (38.283,0.053) (39.596,0.053) (40.909,0.053)
        (42.222,0.053) (43.535,0.053) (44.848,0.053) (46.162,0.053)
        (47.475,0.053) (48.788,0.053) (50.101,0.053) (51.414,0.053)
        (52.727,0.053) (54.040,0.053) (55.354,0.053) (56.667,0.053)
        (57.980,0.053) (59.293,0.053) (60.606,0.053) (61.919,0.053)
        (63.232,0.053) (64.545,0.053) (65.859,0.053) (67.172,0.053)
        (68.485,0.053) (69.798,0.053) (71.111,0.053) (72.424,0.053)
        (73.737,0.053) (75.051,0.053) (76.364,0.053) (77.677,0.053)
        (78.990,0.053) (80.303,0.053) (81.616,0.053) (82.929,0.053)
        (84.242,0.053) (85.556,0.053) (86.869,0.053) (88.182,0.053)
        (89.495,0.053) (90.808,0.053) (92.121,0.053) (93.434,0.053)
        (94.747,0.053) (96.061,0.053) (97.374,0.053) (98.687,0.053)
        (100.000,0.053) };

      \addplot[black, dotted, mark=none] plot coordinates {
        (-30.000,0.484) (-28.687,0.484) (-27.374,0.480)
        (-26.061,0.473) (-24.747,0.464) (-23.434,0.454)
        (-22.121,0.441) (-20.808,0.425) (-19.495,0.407)
        (-18.182,0.387) (-16.869,0.365) (-15.556,0.341)
        (-14.242,0.316) (-12.929,0.290) (-11.616,0.265)
        (-10.303,0.241) (-8.990,0.218) (-7.677,0.196) (-6.364,0.176)
        (-5.051,0.158) (-3.737,0.142) (-2.424,0.128) (-1.111,0.115)
        (0.202,0.105) (1.515,0.095) (2.828,0.087) (4.141,0.080)
        (5.455,0.074) (6.768,0.069) (8.081,0.064) (9.394,0.061)
        (10.707,0.057) (12.020,0.054) (13.333,0.052) (14.646,0.049)
        (15.960,0.047) (17.273,0.045) (18.586,0.043) (19.899,0.042)
        (21.212,0.041) (22.525,0.039) (23.838,0.038) (25.152,0.037)
        (26.465,0.036) (27.778,0.035) (29.091,0.034) (30.404,0.033)
        (31.717,0.032) (33.030,0.032) (34.343,0.031) (35.657,0.030)
        (36.970,0.030) (38.283,0.029) (39.596,0.029) (40.909,0.028)
        (42.222,0.028) (43.535,0.027) (44.848,0.027) (46.162,0.026)
        (47.475,0.026) (48.788,0.025) (50.101,0.025) (51.414,0.025)
        (52.727,0.024) (54.040,0.024) (55.354,0.024) (56.667,0.023)
        (57.980,0.023) (59.293,0.023) (60.606,0.022) (61.919,0.022)
        (63.232,0.022) (64.545,0.021) (65.859,0.021) (67.172,0.021)
        (68.485,0.021) (69.798,0.020) (71.111,0.020) (72.424,0.020)
        (73.737,0.020) (75.051,0.020) (76.364,0.019) (77.677,0.019)
        (78.990,0.019) (80.303,0.019) (81.616,0.019) (82.929,0.018)
        (84.242,0.018) (85.556,0.018) (86.869,0.018) (88.182,0.018)
        (89.495,0.018) (90.808,0.017) (92.121,0.017) (93.434,0.017)
        (94.747,0.017) (96.061,0.017) (97.374,0.017) (98.687,0.017)
        (100.000,0.016)};

      \addplot[red, solid, mark=none] plot coordinates {
        (-30.000,0.500) (-28.687,0.498) (-27.374,0.498)
        (-26.061,0.497) (-24.747,0.496) (-23.434,0.494)
        (-22.121,0.493) (-20.808,0.490) (-19.495,0.487)
        (-18.182,0.482) (-16.869,0.476) (-15.556,0.469)
        (-14.242,0.459) (-12.929,0.447) (-11.616,0.432)
        (-10.303,0.414) (-8.990,0.393) (-7.677,0.370) (-6.364,0.345)
        (-5.051,0.319) (-3.737,0.294) (-2.424,0.270) (-1.111,0.248)
        (0.202,0.228) (1.515,0.211) (2.828,0.196) (4.141,0.183)
        (5.455,0.173) (6.768,0.163) (8.081,0.160) (9.394,0.160)
        (10.707,0.160) (12.020,0.160) (13.333,0.160) (14.646,0.160)
        (15.960,0.160) (17.273,0.160) (18.586,0.160) (19.899,0.160)
        (21.212,0.160) (22.525,0.160) (23.838,0.160) (25.152,0.160)
        (26.465,0.160) (27.778,0.160) (29.091,0.160) (30.404,0.160)
        (31.717,0.160) (33.030,0.160) (34.343,0.160) (35.657,0.160)
        (36.970,0.160) (38.283,0.160) (39.596,0.160) (40.909,0.160)
        (42.222,0.160) (43.535,0.160) (44.848,0.160) (46.162,0.160)
        (47.475,0.160) (48.788,0.160) (50.101,0.160) (51.414,0.160)
        (52.727,0.160) (54.040,0.160) (55.354,0.160) (56.667,0.160)
        (57.980,0.160) (59.293,0.160) (60.606,0.160) (61.919,0.160)
        (63.232,0.160) (64.545,0.160) (65.859,0.160) (67.172,0.160)
        (68.485,0.160) (69.798,0.160) (71.111,0.160) (72.424,0.160)
        (73.737,0.160) (75.051,0.160) (76.364,0.160) (77.677,0.160)
        (78.990,0.160) (80.303,0.160) (81.616,0.160) (82.929,0.160)
        (84.242,0.160) (85.556,0.160) (86.869,0.160) (88.182,0.160)
        (89.495,0.160) (90.808,0.160) (92.121,0.160) (93.434,0.160)
        (94.747,0.160) (96.061,0.160) (97.374,0.160) (98.687,0.160)
        (100.000,0.160) };

      \addplot[red, solid, only marks, mark=o] plot coordinates {
        (-30.000,0.500) (-20.000,0.488) (-10.000,0.409) (0.000,0.231)
        (10.000,0.160) (20.000,0.160) (30.000,0.160) (40.000,0.160)
        (50.000,0.160) (60.000,0.160) (70.000,0.160) (80.000,0.160)
        (90.000,0.160) (100.000,0.160) };

      \addplot[blue, dashed, mark=none] plot coordinates {
        (-30.000,0.496) (-28.687,0.496) (-27.374,0.493)
        (-26.061,0.491) (-24.747,0.488) (-23.434,0.484)
        (-22.121,0.479) (-20.808,0.472) (-19.495,0.463)
        (-18.182,0.453) (-16.869,0.439) (-15.556,0.423)
        (-14.242,0.404) (-12.929,0.383) (-11.616,0.359)
        (-10.303,0.333) (-8.990,0.307) (-7.677,0.280) (-6.364,0.254)
        (-5.051,0.229) (-3.737,0.206) (-2.424,0.186) (-1.111,0.168)
        (0.202,0.153) (1.515,0.140) (2.828,0.130) (4.141,0.121)
        (5.455,0.113) (6.768,0.107) (8.081,0.101) (9.394,0.096)
        (10.707,0.092) (12.020,0.088) (13.333,0.084) (14.646,0.081)
        (15.960,0.078) (17.273,0.075) (18.586,0.072) (19.899,0.070)
        (21.212,0.067) (22.525,0.065) (23.838,0.063) (25.152,0.061)
        (26.465,0.059) (27.778,0.058) (29.091,0.056) (30.404,0.055)
        (31.717,0.053) (33.030,0.053) (34.343,0.053) (35.657,0.053)
        (36.970,0.053) (38.283,0.053) (39.596,0.053) (40.909,0.053)
        (42.222,0.053) (43.535,0.053) (44.848,0.053) (46.162,0.053)
        (47.475,0.053) (48.788,0.053) (50.101,0.053) (51.414,0.053)
        (52.727,0.053) (54.040,0.053) (55.354,0.053) (56.667,0.053)
        (57.980,0.053) (59.293,0.053) (60.606,0.053) (61.919,0.053)
        (63.232,0.053) (64.545,0.053) (65.859,0.053) (67.172,0.053)
        (68.485,0.053) (69.798,0.053) (71.111,0.053) (72.424,0.053)
        (73.737,0.053) (75.051,0.053) (76.364,0.053) (77.677,0.053)
        (78.990,0.053) (80.303,0.053) (81.616,0.053) (82.929,0.053)
        (84.242,0.053) (85.556,0.053) (86.869,0.053) (88.182,0.053)
        (89.495,0.053) (90.808,0.053) (92.121,0.053) (93.434,0.053)
        (94.747,0.053) (96.061,0.053) (97.374,0.053) (98.687,0.053)
        (100.000,0.053) };

      \addplot[blue, dashed, only marks, mark=o] plot coordinates {
        (-30.000,0.496) (-20.000,0.467) (-10.000,0.327) (0.000,0.155)
        (10.000,0.094) (20.000,0.070) (30.000,0.055) (40.000,0.053)
        (50.000,0.053) (60.000,0.053) (70.000,0.053) (80.000,0.053)
        (90.000,0.053) (100.000,0.053) };

      \addplot[black, dotted, mark=none] plot coordinates {
        (-30.000,0.488) (-28.687,0.484) (-27.374,0.479)
        (-26.061,0.473) (-24.747,0.464) (-23.434,0.453)
        (-22.121,0.440) (-20.808,0.425) (-19.495,0.406)
        (-18.182,0.385) (-16.869,0.362) (-15.556,0.337)
        (-14.242,0.311) (-12.929,0.285) (-11.616,0.258)
        (-10.303,0.232) (-8.990,0.208) (-7.677,0.185) (-6.364,0.164)
        (-5.051,0.145) (-3.737,0.128) (-2.424,0.114) (-1.111,0.103)
        (0.202,0.093) (1.515,0.084) (2.828,0.078) (4.141,0.072)
        (5.455,0.067) (6.768,0.063) (8.081,0.060) (9.394,0.057)
        (10.707,0.054) (12.020,0.052) (13.333,0.050) (14.646,0.048)
        (15.960,0.046) (17.273,0.045) (18.586,0.043) (19.899,0.042)
        (21.212,0.040) (22.525,0.039) (23.838,0.038) (25.152,0.037)
        (26.465,0.036) (27.778,0.035) (29.091,0.034) (30.404,0.033)
        (31.717,0.032) (33.030,0.032) (34.343,0.031) (35.657,0.030)
        (36.970,0.030) (38.283,0.029) (39.596,0.029) (40.909,0.028)
        (42.222,0.028) (43.535,0.027) (44.848,0.027) (46.162,0.026)
        (47.475,0.026) (48.788,0.025) (50.101,0.025) (51.414,0.025)
        (52.727,0.024) (54.040,0.024) (55.354,0.024) (56.667,0.023)
        (57.980,0.023) (59.293,0.023) (60.606,0.022) (61.919,0.022)
        (63.232,0.022) (64.545,0.021) (65.859,0.021) (67.172,0.021)
        (68.485,0.021) (69.798,0.020) (71.111,0.020) (72.424,0.020)
        (73.737,0.020) (75.051,0.020) (76.364,0.019) (77.677,0.019)
        (78.990,0.019) (80.303,0.019) (81.616,0.019) (82.929,0.018)
        (84.242,0.018) (85.556,0.018) (86.869,0.018) (88.182,0.018)
        (89.495,0.018) (90.808,0.017) (92.121,0.017) (93.434,0.017)
        (94.747,0.017) (96.061,0.017) (97.374,0.017) (98.687,0.017)
        (100.000,0.016) };

      \addplot[black,dotted,only marks, mark=o] plot coordinates {
        (-30.000,0.488) (-20.000,0.414) (-10.000,0.226) (0.000,0.094)
        (10.000,0.056) (20.000,0.042) (30.000,0.033) (40.000,0.028)
        (50.000,0.025) (60.000,0.022) (70.000,0.020) (80.000,0.019)
        (90.000,0.018) (100.000,0.016) };

      \legend{
        {$T=100$}\\
        {$T=300$}\\
        {$T=1\,000$}\\};
    \end{axis}
  \end{tikzpicture}
  \caption{ZF and RZF-CDA, optimal relative amount of training $T_t^{\star\circ}/T$ vs. $\SNR_{dl}$ with $M\eqq 32$, $K\eqq 16$, $\SNR_{dl}/\SNR_{ul}\eqq 10$, $\Tt_k=\I_M~\forall k$, RZF is indicated by circle marks.}
  \label{fig:zf-topt}
\end{figure}

Figure \ref{fig:zf-topt} depicts the optimal relative amount of training $T_t^{\star\circ}/T$ for ZF and RZF-CDA precoding. We observe that $T_t^{\star\circ}/T$ decreases with increasing SNR as $1/\sqrt{\log(\SNR_{dl})}$. That is, for increasing SNR, the estimation becomes more accurate and resources for channel training are reallocated to data transmission. Furthermore, $T_t^{\star\circ}/T$ saturates at $K/T$ due to the orthogonality constraint on the pilot sequences. As expected from \eqref{eq:12} and \eqref{eq:13}, we observe that the optimal amount of training is less for RZF-CDA than for ZF precoding. Moreover, the relative amount of training $T_t^{\star\circ}/T$ for both ZF and RZF-CDA converges at low SNR to $1/2$ and at high SNR to the minimum amount of training $K$, as predicted by the theoretical analysis.

\begin{figure}[t]
  \centering
  \begin{tikzpicture}
    \pgfplotsset{every axis legend/.append style={
        cells={anchor=west}, at={(0.02,0.98)}, anchor=north west}}
    \pgfplotsset{every axis/.append style={line width=0.5pt}}
    \pgfplotsset{every axis/.append style={mark options=solid, mark
        size=2.5pt}}
    \begin{axis}[xlabel={$\SNR_{dl}$ [dB]}, ylabel={ergodic sum rate
        [bits/s/Hz]},
      grid=major, xmin=0, xmax=35,
      xtick={0,5,...,35}, ymin=0, ymax=120,
      ytick={0,20,...,120}]

      \addplot[red, solid, mark=none] plot coordinates {
        (0.000,11.158) (5.000,22.834) (10.000,38.435) (15.000,55.869)
        (20.000,73.953) (25.000,92.245) (30.000,110.566)
        (35.000,129.019) };

      \addplot[blue, only marks, mark=o] plot coordinates {
        (0.000,10.666) (5.000,21.784) (10.000,35.851) (15.000,50.441)
        (20.000,63.539) (25.000,73.763) (30.000,80.275)
        (35.000,83.668) };

      \addplot[blue, dashed, mark=none] plot coordinates {
        (0.000,10.670) (5.000,21.739) (10.000,35.837) (15.000,50.406)
        (20.000,63.590) (25.000,73.792) (30.000,80.321)
        (35.000,83.632) };

      \addplot[black, dotted, mark=none] plot coordinates {
        (0.000,9.531) (5.000,19.511) (10.000,32.729) (15.000,47.076)
        (20.000,60.897) (25.000,72.354) (30.000,79.828)
        (35.000,83.525) };
      
      \addplot[green, solid, mark=triangle] plot coordinates {
        (0.000,10.674) (5.000,21.634) (10.000,35.147) (15.000,47.813)
        (20.000,56.795) (25.000,61.356) (30.000,63.253)
        (35.000,63.779) };

     \legend{
        {perfect CSIT}\\
        {$T_t\eqq T_{t,\rm zf}^{\star}$}\\
        {$T_t\eqq T_{t,\rm zf}^{\star\circ}$}, \eqref{eq:tdd:6}\\
        {$T_t\eqq T_{t,\rm zf}^{\star\circ}$, \eqref{eq:15}}\\
        {$T_t\eqq K$}\\};
    \end{axis}
  \end{tikzpicture}
  \caption{ZF, ergodic sum rate vs. downlink SNR with $M\eqq 32$, $K\eqq 16$, $\SNR_{ul}\eqq 5$ dB, $\Tt_k=\I_M~\forall k$ and $T\eqq 1\,000$.}
  \label{fig:rzf}
\end{figure}

Figure \ref{fig:rzf} shows the ergodic sum rate under ZF precoding with fixed UL SNR $\SNR_{ul}\eqq 5$ dB for various training intervals. We observe (i) no significant difference in the performance of the schemes employing either optimal training $T_{t,\rm zf}^{\star}$, computed via exhaustive search, or $T_{t,\rm zf}^{\star\circ}$ obtained from a convex optimization of the large system approximation \eqref{eq:tdd:5}, (ii) a small performance loss at low and medium SNR of the (high-SNR) approximation of $T_{t,\rm zf}^{\star\circ}$ in \eqref{eq:15} and (iii) a significant performance loss if the minimum training interval $T_{t,\rm zf}=K$ is used for all SNR. We conclude that our approximation in \eqref{eq:15} achieves very good performance and can therefore be utilized to compute $T_{t,\rm zf}$ very efficiently.

\section{Conclusion}
\label{sec:conclusion}

In this paper, we presented a consistent framework for the study of ZF and RZF precoding schemes based on the theory of large dimensional random matrices. The tools from RMT allowed us to consider a very realistic channel model accounting for per-user channel correlation as well as individual channel gains for each link. The system performance under this general type of channel is extremely difficult to study for finite dimensions but becomes feasible by assuming large system dimensions. Simulation results indicated that these approximations are very accurate even for small system dimensions and reveal the deterministic dependence of the system performance on several important system parameters, such as the transmit correlation, signal powers, SNR, and CSIT quality. Applied to practical optimization problems, the deterministic approximations lead to important insights into the system behavior, which are consistent with previous results, but go further and extend them to more realistic channel models and other linear precoding techniques. Furthermore, the proposed channel-independent performance approximations can be used to simulate the system behavior without having to carry out extensive Monte Carlo simulations. 



\appendices

\section{Proof of Theorem \ref{th:1}}
\label{sec:proof-theorem:1}

The proof is structured as follows: In Appendix \ref{sec:proof-convergence}, we prove that $ m_{\B_N,\Q_N}(z) - \frac1N\tr\D^{-1}\ton 0$ almost surely, where $\D$ is an auxiliary random variable involving the terms $m_{\B_N,\Tt_i}(z)$. Appendix \ref{sec:proof-exist-uniq} shows that the sequence $\{e_{N,i}^{(k)}(z)\}$ defined by \eqref{eq:33} converges to $e_{N,i}$ \eqref{eq:eKz} as $k\to\infty$, if properly initialized. Finally, in Appendix \ref{sec:proof-conv-determ} we demonstrate that $e_{N,i}$ satisfies $|m_{\B_N,\Tt_i}-e_{N,i}|\ton 0$, almost surely.

\subsection{Convergence to an Auxiliary Variable}
\label{sec:proof-convergence}

The objective is to approximate the random variable $m_{\B_N,\Q_N}(z)$ by an appropriate functional $\frac1N\tr\D^{-1}$ such that
\begin{equation}
  \label{eq:24}
  \frac1N\tr\Q_N\left(\B_N - z\I_N\right)^{-1} - \frac1N\tr\D^{-1}
  \ton 0,
\end{equation}
almost surely. Take $z\inn \Cbb^+$. From \eqref{eq:24} we proceed by applying Lemma \ref{resolvent-identity} and obtain
\begin{align}
  &\Q_N\left(\B_N - z\I_N\right)^{-1} - \D^{-1} = \nonumber \\
  \label{eq:2a}
  &\D^{-1}\left[ \D - ( \X_N^\herm\X_N + \S_N - z\I_N)\Q_N^{-1}
  \right] \Q_N\left(\B_N - z\I_N\right)^{-1}.
\end{align}
We choose $\D$ as
\begin{equation}
  \label{eq:26}
  \D\eqq\left(\R + \S_N - z\I_N
  \right)\Q_N^{-1},   
\end{equation}
where $\R$ is to be determined later, and obtain
\begin{align*}
  &\Q_N\left(\B_N - z\I_N\right)^{-1} - \D^{-1}\\ 
  &= \D^{-1}\R\left(\B_N - z\I_N\right)^{-1} - \D^{-1}\X_N^\herm\X_N\left(\B_N - z\I_N\right)^{-1}.
\end{align*}
Consider the term $\D^{-1}\X_N^\herm\X_N\left(\B_N - z\I_N\right)^{-1}$. Taking the trace, together with
$\X_N^\herm\X_N\eqq\sum_{i=1}^n\Pb_i\y_i\y_i^\herm\Pb_i^\herm$, we have
\begin{align*}
  &\frac1N\tr\D^{-1}\X_N^\herm\X_N\left(\B_N - z\I_N\right)^{-1}
   \\
  &=\frac1N\tr
  \D^{-1}\sum_{i=1}^n\Pb_i\y_i\y_i^\herm\Pb_i^\herm\left(\B_N -
    z\I_N\right)^{-1} \\
  &= \frac1N\sum_{i=1}^n\y_i^\herm\Pb_i^\herm\left(\B_N -
    z\I_N\right)^{-1}\D^{-1}\Pb_i\y_i.
\end{align*}
Denoting $\B_{[i]}\eqq \B_N - \Pb_i\y_i\y_i^\herm\Pb_i^\herm$ and applying Lemma \ref{mil}, we obtain
\begin{align*}
  &\frac1N\tr\D^{-1}\X_N^\herm\X_N\left(\B_N -
    z\I_N\right)^{-1}\\ 
  &=
  \frac1N\sum_{i=1}^n\frac{\y_i^\herm\Pb_i^\herm\left(\B_{[i]} -
      z\I_N\right)^{-1}\D^{-1}\Pb_i\y_i}{1 +
    \y_i^\herm\Pb_i^\herm\left(\B_{[i]} -
      z\I_N\right)^{-1}\Pb_i\y_i}. 
\end{align*}
Therefore, the left-hand side of \eqref{eq:24} takes the form
\begin{align}
  &\frac1N\tr\Q_N\left(\B_N - z\I_N\right)^{-1} - \frac1N\tr\D^{-1}
  \nonumber\\
  &=\frac1N\tr\D^{-1}\R\left(\B_N - z\I_N\right)^{-1} \nonumber \\
  \label{eq:6a}
  &-
  \frac1N\sum_{i=1}^n\frac{\y_i^\herm\Pb_i^\herm\left(\B_{[i]} -
      z\I_N\right)^{-1}\D^{-1}\Pb_i\y_i}{1 +
    \y_i^\herm\Pb_i^\herm\left(\B_{[i]} -
      z\I_N\right)^{-1}\Pb_i\y_i}.
\end{align}
The choice of an appropriate value for $\R$, such that \eqref{eq:24} is satisfied, requires some intuition. From Lemma \ref{sec:coll-import-lemm-1} we know that $\y_i^\herm\Pb_i^\herm\left(\B_{[i]}-z\I_N\right)^{-1}\Pb_i\y_i - \frac1N\tr\Tt_i\left(\B_{[i]}-z\I_N\right)^{-1}\ton 0$, almost surely. Then, from Lemma \ref{lemma:rank1-bai}, we surely have $$\frac1N\tr\Tt_i\left(\B_{[i]}-z\I_N\right)^{-1} -
\frac1N\tr\Tt_i\left(\B_N-z\I_N\right)^{-1}\ton 0.$$ From the previous arguments, $\R$ will be chosen as
\begin{equation}
  \label{eq:5a}
  \R = \frac1N\sum_{i=1}^n\frac{\Tt_i}{1+\frac1N\tr\Tt_i\left(\B_N-z\I_N\right)^{-1}}.
\end{equation}
Note that $\R$ is random since it depends on $\B_N$. The remainder of this subsection proves \eqref{eq:24} for the specific choice of $\R$ in \eqref{eq:5a}. Substituting \eqref{eq:5a} into \eqref{eq:6a} we obtain
\begin{align}
  \label{eq:5f}
  w_N &\eqdef w_{\Q_N} \eqdef \frac1N\tr\Q_N\left(\B_N - z\I_N\right)^{-1} - \frac1N\tr\D^{-1}
    \\
 &=\frac1N\sum_{i=1}^n\frac{\frac1N\tr\Tt_i\left(\B_N -
       z\I_N\right)^{-1}\D^{-1}}{1+\frac1N\tr\Tt_i\left(\B_N-z\I_N\right)^{-1}}\nonumber\\
   \label{eq:7a}
   &- \frac1N\sum_{i=1}^n\frac{\y_i^\herm\Pb_i^\herm\left(\B_{[i]} -
       z\I_N\right)^{-1}\D^{-1}\Pb_i\y_i}{1 +
     \y_i^\herm\Pb_i^\herm\left(\B_{[i]} -
       z\I_N\right)^{-1}\Pb_i\y_i}.
\end{align}
In order to prove that $w_N\!\ton\!0$, almost surely, we divide the left-hand side of \eqref{eq:7a} into $4n$ terms, i.e.,
\begin{equation}
  \label{eq:17}
  w_N = \frac1N\sum_{i=1}^n\left[d_i^{(1)} + d_i^{(2)} + d_i^{(3)} + d_i^{(4)} \right].
\end{equation}
It is then easier to show that each $d_i^{(l)}$, ($l\eqq 1,2,3,4$), converges to zero, sufficiently fast, as $N\!\to\!\infty$, which will imply $w_N\!\ton\!0$, almost surely. The $d_i^{(l)}$ are chosen as
\begin{align*}
  d_i^{(1)} &= \frac{\y_i^\herm\Pb_i^\herm\left(\B_{[i]} -
      z\I_N\right)^{-1}\D_{[i]}^{-1}\Pb_i\y_i} {1 +
    \y_i^\herm\Pb_i^\herm\left(\B_{[i]} -
      z\I_N\right)^{-1}\Pb_i\y_i}  \\&-
  \frac{\y_i^\herm\Pb_i^\herm\left(\B_{[i]} -
      z\I_N\right)^{-1}\D^{-1}\Pb_i\y_i} {1 +
    \y_i^\herm\Pb_i^\herm\left(\B_{[i]} -
      z\I_N\right)^{-1}\Pb_i\y_i}\\
  d_i^{(2)} &= \frac{\frac1N\tr\Tt_i\left(\B_{[i]} -
      z\I_N\right)^{-1}\D_{[i]}^{-1}}{1 +
    \y_i^\herm\Pb_i^\herm\left(\B_{[i]} -
      z\I_N\right)^{-1}\Pb_i\y_i} \\
  &-\frac{\y_i^\herm\Pb_i^\herm\left(\B_{[i]} -
      z\I_N\right)^{-1}\D_{[i]}^{-1}\Pb_i\y_i} {1 +
    \y_i^\herm\Pb_i^\herm\left(\B_{[i]} -
      z\I_N\right)^{-1}\Pb_i\y_i}\\
  d_i^{(3)} &= \frac{\frac1N\tr\Tt_i\left(\B_N -
      z\I_N\right)^{-1}\D^{-1}}{1 +
    \y_i^\herm\Pb_i^\herm\left(\B_{[i]} -
      z\I_N\right)^{-1}\Pb_i\y_i}  \\ &-
  \frac{\frac1N\tr\Tt_i\left(\B_{[i]} - z\I_N\right)^{-1}\D_{[i]}^{-1}}{1
    + \y_i^\herm\Pb_i^\herm\left(\B_{[i]} -
      z\I_N\right)^{-1}\Pb_i\y_i}\\
  d_i^{(4)} &= \frac{\frac1N\tr\Tt_i\left(\B_N -
      z\I_N\right)^{-1}\D^{-1}}{1 +
    \frac1N\tr\Tt_i\left(\B_N-z\I_N\right)^{-1} }  \\ &-
  \frac{\frac1N\tr\Tt_i\left(\B_N - z\I_N\right)^{-1}\D^{-1}}{1 +
    \y_i^\herm\Pb_i^\herm\left(\B_{[i]} -
      z\I_N\right)^{-1}\Pb_i\y_i},
\end{align*}
where we defined 
\begin{equation*}
  \label{eq:18b}
  \D_{[i]}^{-1} = \Q_N\left(\frac1N\sum_{i=1}^n\frac{\Tt_i}{1+m_{\B_{[i]},\Tt_i}(z)}
    - z\I_N + \S_N\right)^{-1},
\end{equation*}
where $m_{\B_{[i]},\Tt_i}(z) \eqq \frac1N\tr\Tt_i\left(\B_{[i]}-z\I_N\right)^{-1}$.

In the course of the development of the proof, we require the existence of moments of order $p$ of $w_N$ in \eqref{eq:17}, i.e., $\EE\left[|w_N|^p\right]\neq 0$, for some integer $p$. First we bound \eqref{eq:17} as $\EE[|w_N|^p]\!\leq\!\EE[(\sum_{i=1}^{4n}\tilde{d}_i)^p]$. The application of H\"older's inequality yields
\begin{equation*}
  \label{eq:65}
  \EE\left[|w_N|^p\right] \leq
  \left(\frac{4}{\beta}\right)^{p-1}\frac1N\sum_{i=1}^n\sum_{l=1}^4\EE\left[|d_i^{(l)}|^p\right].
\end{equation*}
Furthermore, for some $T,Q\!<\!\infty$, we can uniformly bound $\Tt_i$ and $\Q_N$ as
\begin{align}
  \label{eq:67}
  &\limsup_{N\to\infty}\sup_{1\leq i\leq n}\|\Tt_i\| \leq T\\
  &\limsup_{N\to\infty}\|\Q_N\| \leq Q.
\end{align}

\begin{proposition}
  \label{prop:Edp}
  Let the following upper bounds be well defined and let the entries of $\y_i$ have eighth order moment of order $O\left(\frac{1}{N^4}\right)$. Then the $p$th order moments $\EE\left[|d_i^{(l)}|^p\right]$, ($l\eqq 1,2,3,4$) can be bounded as
  \begin{align}
    \label{eq:36}
    \EE\left[ |d_i^{(1)}|^p \right] &\leq 2^{p-1}\left(\frac{\beta T^3 Q
      |z|^3}{(\Im z)^7}\right)^p\frac1{N^p}\left(\frac{C_p^{(1)}}{N^{p/2}} + 1\right)\\
    \EE\left[ |d_i^{(2)}|^p \right] &\leq \frac{|z|^4}{(\Im z)^4}
    \frac{C_p^{(2)}}{N^{p/2}},\nonumber\\
    \EE\left[ |d_i^{(3)}|^p \right] &\leq \left(\frac{|z|TQ}{N(\Im
    z)^3} \right)^p\left[1 + \frac{\beta T^2|z|^2}{(\Im
      z)^4}\right]^p,\nonumber \\  
    \EE\left[ |d_i^{(4)}|^p \right] &\leq 2^{p-1}
    \left(\frac{TQ|z|^2}{(\Im z)^4}
    \right)^p\left[\frac{C_p^{(4)}}{N^{p/2}} + \frac{T^p}{N^p(\Im z)^p}\right],\nonumber
  \end{align}  
  where the $C_p^{(i)}$, $i\inn\{1,2,4\}$ are constants depending only on $p$.
\end{proposition}
\begin{IEEEproof}
  The proof is based on various common inequalities. Applying Lemma \ref{lem:2}, $|d_i^{(1)}|$ can be upper-bounded as
  \begin{equation*}
    \label{eq:19a}
    |d_i^{(1)}| \leq \frac{|z|}{\Im z}\left|\y_i^\herm\Pb_i^\herm\left(\B_{[i]} -
      z\I_N\right)^{-1}\left[\D_{[i]}^{-1}-\D^{-1}\right]\Pb_i\y_i\right|.\nonumber
  \end{equation*}
  We further bound $|d_i^{(1)}|$ by applying Lemmas \ref{sec:important-lemmas-2} and \ref{sec:important-lemmas-1} with the fact that $\|(\B_{[i]} - z\I_N)^{-1}\|\!\leq \!\frac1{\Im z}$. Together with \eqref{eq:67} we have
  \begin{equation*}
    \label{eq:30}
    |d_i^{(1)}| \leq \frac{|z|T}{(\Im z)^2} \|\y_i\|_2^2\|\D_{[i]}^{-1}-\D^{-1}\|.
  \end{equation*}
  Similarly, with Lemma \ref{resolvent-identity}, it can be shown that $\|\D_{[i]}^{-1}-\D^{-1}\|\!\leq\!\frac{\beta T^2Q|z|^2}{N(\Im z)^5}$ and thus
  \begin{equation*}
    \label{eq:20}
    |d_i^{(1)}| \leq \frac{\beta T^3Q|z|^3}{N(\Im z)^7}\|\y_i\|_2^2.
  \end{equation*}
  The $p$th order moment of $|d_i^{(1)}|$ thus satisfies
  \begin{align*}
    \EE\left[ |d_i^{(1)}|^p \right] \leq \left[\frac{\beta T^3 Q
      |z|^3}{(\Im z)^7}\right]^p\frac1{N^p}\EE\left[\left\vert
        \y_i^\herm\y_i \right\vert^p\right].
  \end{align*}
  Applying the inequality $|x+y|^p\leq2^{p-1}(|x|^p + |y|^p)$ yields
  \begin{equation*}
    \label{eq:16}
    \EE\left[ |d_i^{(1)}|^p \right] \leq 2^{p-1}\left(\frac{\beta T^3 Q
        |z|^3}{N(\Im z)^7}\right)^p\left(\EE\left[\left\vert\y_i^\herm\y_i
          -1\right\vert^p\right] + 1\right).
  \end{equation*}
  If the moments $\EE[|d_i^{(1)}|^4]$ and $\EE[|d_i^{(1)}|^{2p}]$ exist and are bounded, we can apply Lemma \ref{lemma-bai-silverstein} and obtain \eqref{eq:36}. For the sake of brevity, we omit the derivations of the remaining moments $\EE[ |d_i^{(l)}|^p ]$, $l\eqq\{2,3,4\}$, since the techniques are similar to the previous procedure.
\end{IEEEproof}
From Proposition \ref{prop:Edp}, we conclude that all $\EE[|d_i^{(l)}|^p ]$ are summable if $p\eqq 2+\varepsilon$, $\varepsilon>0$. Therefore, $\EE\left[|w_N|^p\right]$ is summable for $p\eqq 2+\varepsilon$ and hence the Borel-Cantelli Lemma \cite{billingsley2008probability} implies that $w_N\ton 0$, almost surely. Note that with the same approach, the convergence region can be extended to $z\inn\Cbb\setminus \Rbb^+$.

We now prove the existence and uniqueness of a solution to \eqref{eq:eKz}.

\subsection{Proof of Convergence of the Fixed Point Equation}
\label{sec:proof-exist-uniq}

In this section we consider the fixed point equation \eqref{eq:eKz}. We first prove that, properly initialized, the sequence $\{e_{N,i}^{(k)}\}$, $(k\eqq 1,2,\dots$), converges to a limit $e_{N,i}$ as $k\!\to\!\infty$. Subsequently, we show that this limit $e_{N,i}$ satisfies $|m_{\B_N,\Tt_i}-e_{N,i}|\ton 0$, almost surely.

\begin{proposition}
  \label{sec:proof-exist-uniq-1}
  Let $z\inn\Cbb^+$ and $\{e_{N,i}^{(k)}(z)\}$ ($k\!\geq\!0$) be the sequence defined by \eqref{eq:33}. If $\{e_{N,i}^{(0)}(z)\}$ is a Stieltjes transform, then all $\{e_{N,i}^{(k)}(z)\}$ ($k\!>\!0$) are Stieltjes transforms as well.
\end{proposition}
\begin{IEEEproof}
  Suppose \eqref{eq:33} is initialized by $e_{N,i}^{(0)}(z)\eqq -1/z$, which is the Stieltjes transform of a function with a single mass in zero. We demonstrate that at all subsequent iterations $k\!>\!0$ the corresponding $e_{N,i}^{(k)}(z)$ are Stieltjes transforms for all $N$. For ease of notation we omit the dependence on $z$, the $e_{N,i}^{(k+1)}$ are given by
  \begin{align}
    e_{N,i}^{(k+1)} &=
    \frac1N\tr\Tt_i\left(\frac1N\sum_{j=1}^nc_{N,j}^{(k)}\Tt_j + \S_N
      -z\I_N\right)^{-1}\nonumber\\ 
    \label{eq:21}
    &\eqdef \frac1N\tr\Tt_i\A_k,
  \end{align}
  where $c_{N,j}^{(k)}\eqq1/(1+e_{N,j}^{(k)})$. In \eqref{eq:21}, multiplying $\A_k$ from the right by $(\A_k^\herm)^{-1}\A_k^\herm$, we obtain
  \begin{equation}
    \label{eq:22}
    e_{N,i}^{(k+1)} =
    \frac1N\tr\A_k^\herm\Tt_i\A_k\left[\frac1N\sum_{j=1}^nc_{N,j}^{\ast,(k)}\Tt_j\right]
    + v_i^{(k)},
  \end{equation}
  where $v_i^{(k)}\eqq \frac1N\tr\A_k^\herm\Tt_i\A_k\left[\S_N-z^\ast\I_N\right]$. Denoting $\r_i^{(k)}\eqdef \frac1N  [\frac1N\tr\A_k^\herm\Tt_i\A_k\Tt_1,\dots,\frac1N\tr\A_k^\herm\Tt_i\A_k\Tt_n]^\trans$ and $\c_N^{(k)}\eqdef [c_{N,1}^{(k)},\dots,c_{N,n}^{(k)}]^\trans$, \eqref{eq:22} takes the form
  \begin{equation}
    \label{eq:23}
    e_{N,i}^{(k+1)} = \r_i^{\trans,(k)}\c_N^{\herm,(k)} + v_i^{(k)}.
  \end{equation}
  Since the $\Tt_i$ are uniformly bounded w.r.t. $N$, we have $\r_i^{(k)}, v_i^{(k)}\!>\!0$. To show that $e_{N,i}^{(k+1)}$ are Stieltjes transforms of a nonnegative finite measure, the following three conditions must be verified \cite[Proposition 2.2]{hachem2007deterministic}: For $z\inn\Cbb^+$ (i) $e_{N,i}^{(k+1)}(z)\inn\Cbb^+$, (ii) $ze_{N,i}^{(k+1)}(z)\inn\Cbb^+$ and (iii) $\lim_{y\to +\infty}-\i y e_{N,i}^{(k+1)}(\i y)\!<\! \infty$. From \eqref{eq:23} it is easy to verify that all three conditions are met, which completes the proof.
\end{IEEEproof}
We are now in a position to show that any sequence $\{e_{N,i}^{(k)}(z)\}$, $(k\!>\!0)$ converges to a limit $e_{N,i}(z)$ as $k\!\to\!\infty$.

\begin{proposition}
  Any sequence $\{e_{N,i}^{(k)}(z)\}$, $(k\!>\!0)$ defined by \eqref{eq:33} converges to a Stieltjes transform, denoted $e_{N,i}(z)$ as $k\!\to\!\infty$ if $e_{N,i}^{(0)}(z)$ is a Stieltjes transform.
\end{proposition}
\begin{IEEEproof}
  Let $e_{N,i}^{(k)}(z)\eqq \frac1N\tr\Tt_i\A^{(k-1)}$ and $e_{N,i}^{(k+1)}(z)\eqq \frac1N\tr\Tt_i\A^{(k)}$, where
  \begin{align*}
      \A^{(k-1)}&=\left(\frac1N\sum_{j=1}^n\frac{\Tt_j}{1+e_{N,j}^{(k-1)}(z)}+\S_N-z\I_N\right)^{-1},\\
      \A^{(k)}&=\left(\frac1N\sum_{j=1}^n\frac{\Tt_j}{1+e_{N,j}^{(k)}(z)}+\S_N-z\I_N\right)^{-1}.
  \end{align*}
  Applying Lemma \ref{resolvent-identity}, the difference $|e_{N,i}^{(k)}(z)-e_{N,i}^{(k+1)}(z)|$ is
  \begin{align}
    \label{eq:29}
    &|e_{N,i}^{(k)}-e_{N,i}^{(k+1)}|  =\nonumber \\
    &\left\vert\frac1N\tr\A^{(k+1)}\Tt_i\A^{(k)}\left[\frac1N\sum_{j=1}^n\Tt_j\frac{e_{N,j}^{(k)}-e_{N,j}^{(k-1)}}{\left[1+e_{N,j}^{(k)}\right]\left[1+e_{N,j}^{(k-1)}\right]}
      \right] \right\vert
  \end{align}
With Lemmas \ref{lem:2}, \ref{lem-4} and \ref{sec:important-lemmas-1}, \eqref{eq:29} can be bounded as
\begin{equation}
  \label{eq:31}
  |e_{N,i}^{(k)}-e_{N,i}^{(k+1)}| \leq C\sup_{1\leq i\leq n}|e_{N,i}^{(k)}-e_{N,i}^{(k-1)}|,
\end{equation}
where $C\eqq\frac{\beta T^2|z|^2}{(\Im z)^4}$. Clearly, the sequence $\{e_{N,i}^{(k)}\}$ converges to a limit $e_{N,i}$ for $z$ restricted to the set $\{z\inn\Cbb^+:C\!<\!1\}$. Proposition \ref{sec:proof-exist-uniq-1} shows that all $\{e_{N,i}^{(k)}\}$ are uniformly bounded Stieltjes transforms and therefore their limit is analytic. Since $\{e_{N,i}^{(k)}(z)\}$ for $\{z\inn\Cbb^+:C\!<\!1\}$ is at least countable and has a cluster point, Vitali's convergence theorem \cite[Theorem 3.11]{couillet2011rmm} ensures that the sequence $\{e_{N,i}^{(k)}\}$ must converge for all $z\inn\Cbb\!\setminus\!\Rbb^+$ and their limit is $e_{N,i}(z)$.

It is straightforward to verify, that the previous holds also true for $z\inn\Cbb^-$. 
\end{IEEEproof}

\begin{remark}
For $z\!<\!0$, the existence of a unique solution to \eqref{eq:eKz} as well as the convergence of \eqref{eq:33} from any real initial point can be proved within the framework of \textit{standard interference functions} \cite{yates1995fup}. The strategy is as follows. Let $\bar{\e}_N\!\eqdef\!\bar{\e}_{N}(z)\eqq [\bar{e}_{N,1}(z),\bar{e}_{N,2}(z),\dots,\bar{e}_{N,n}(z)]^\trans\inn\Rbb^n$ and $\f(\bar{\e}_N)\eqq [f_1(\bar{\e}_N),f_2(\bar{\e}_N),\dots,f_n(\bar{\e}_N)]^\trans\inn\Rbb^n$, where
\begin{equation*}
  \label{eq:56}
  f_i(\bar{\e}_N) = \frac1N\tr\Tt_i\left(\frac1N\sum_{j=1}^n\frac{\Tt_j}{1 +
      \bar{e}_{N,j}(z)} + \S_N -z\I_N\right)^{-1}.
\end{equation*}
Theorems 1 and 2 in \cite{yates1995fup} prove that, if $\f(\bar{\e}_N)$ is a feasible standard interference function, then \eqref{eq:33} converges to a unique solution $\e_N$ with all nonnegative entries for any initial point $e_{N,i}^{(0)},\dots,e_{N,n}^{(0)}$. The proof that $\f(\bar{\e}_N)$ is feasible as well as a standard interference function is straightforward and details are omitted in this correspondence.
\end{remark}

The uniqueness of $\e_N$, whose entries are Stieltjes transforms of nonnegative finite measures, ensures the functional uniqueness of $e_{N,i}(z),\dots,e_{N,n}(z)$ as a Stieltjes transform solution to \eqref{eq:eKz} for $z\inn\Cbb\setminus\Rbb^+$. This completes the proof of uniqueness.

Denote $m_{\B_N,\Tt_i}(z) \eqdef\frac1N\tr\Tt_i\left(\B_N-z\I_N\right)^{-1}$. In the following section, we prove that $e_{N,i}(z)\eqq\lim_{k\to\infty}e_{N,i}^{(k)}(z)$ satisfies $|m_{\B_N,\Tt_i}(z) - e_{N,i}(z)|\ton 0$, almost surely.

\subsection{Proof of Convergence of the Deterministic Equivalent}
\label{sec:proof-conv-determ}

In Section \ref{sec:proof-convergence} we showed that $w_N\eqq\frac1N\tr\Q_N\left(\B_N - z\I_N\right)^{-1} - \frac1N\tr\Q_N\left(\R + \S_N -z\I_N\right) \ton 0$, almost surely. Furthermore, in Section \ref{sec:proof-exist-uniq} we proved that the sequence defined by \eqref{eq:eKz} converges to a limit $e_{N,i}$. It remains to prove that
\begin{align}
  \label{eq:52}
  &m_{\B_N,\Tt_i}-e_{N,i} = \frac1N\tr\Tt_i\left(\B_N - z\I_N\right)^{-1}
  \nonumber \\ 
  &- \frac1N\tr\Tt_i\left(\frac1N\sum_{j=1}^n\frac{\Tt_j}{1 +
      e_{N,j}(z)} + \S_N - z\I_N\right)^{-1} \ton 0,
\end{align}
almost surely. Denote $w_{N,i}\!\eqdef\!w_{\Tt_i}$ with $w_{\Tt_i}$ defined in \eqref{eq:5f}. Applying Lemma \ref{resolvent-identity}, \eqref{eq:52} can be written as
\begin{align*}
  &m_{\B_N,\Tt_i}-e_{N,i}  \\
  &= w_{N,i} +
  \frac1N\tr\Tt_i\left(\A+\S_N-z\I_N\right)^{-1} -e_{N,i}(z)\\
  &= w_{N,i}-\frac1N\tr\Tt_i\bar{\A}^{-1}\left[\A - \B\right]\bar{\B}^{-1},
\end{align*}
where $\bar{\A}\eqdef\A+\S_N-z\I_N$, $\A\!\eqdef\!\frac1N\sum_{l=1}^n\frac{\Tt_l}{1+\frac1N\tr\Tt_l\left(\B_N-z\I_N\right)^{-1}}$ and $\bar{\B}\eqdef\B+\S_N-z\I_N$, $\B\!\eqdef\!\frac1N\sum_{j=1}^n\frac{\Tt_j}{1+e_{N,j}}$. Applying Lemmas \ref{lem:2} and \ref{lem-4}, $|m_{\B_N,\Tt_i}-e_{N,i}|$ can be bounded as
\begin{align}
  |m_{\B_N,\Tt_i}-e_{N,i}|&\leq |w_{N,i}|+\|\Tt_i\|\|\bar{\A}^{-1}\|\|\bar{\B}^{-1}\|\nonumber \\
  \label{eq:68}
  &\times \left\|\frac1N\sum_{j=1}^n \Tt_j\frac{|m_{\B_N,\Tt_j}-e_{N,j}|}{(1+m_{\B_N,\Tt_j})(1+e_{N,j})}\right\|.
\end{align}
Similar to \eqref{eq:31}, with Lemma \ref{sec:important-lemmas-1}, \eqref{eq:68} can be further bounded as
\begin{equation*}
   |m_{\B_N,\Tt_i}-e_{N,i}|\leq |w_{N,i}|+ C\sup_{1\leq i\leq n}|m_{\B_N,\Tt_i}-e_{N,i}|,
\end{equation*}
where $C\eqq\frac{\beta T^2|z|^2}{(\Im z)^4}$. Taking the supremum over all $i\eqq 1,\dots,n$, we obtain
\begin{equation}
  \label{eq:70}
  \sup_{1\leq i\leq n}|m_{\B_N,\Tt_i}-e_{N,i}|\left[1 - C\right] \leq \sup_{1\leq i\leq n}|w_{N,i}|.
\end{equation}
From \eqref{eq:70}, on the set $\{z\inn\Cbb^+:0\!<\!C\!<\!1\}\neq\emptyset$, it suffices to show that $\sup_{1\leq i\leq n}|w_{N,i}|$ goes to zero sufficiently fast. For any $\varepsilon\!>\!0$ we have
\begin{align}
  \label{eq:71}
  P\left(\sup_{1\leq i\leq n}|w_{N,i}|>\varepsilon\right) &\leq \sum_{i=1}^n P\left(|w_{N,i}|>\varepsilon\right)\nonumber \\
  &= \sum_{i=1}^n P\left(|w_{N,i}|^p>\varepsilon^p\right).
\end{align}
Applying Markov's inequality, \eqref{eq:71} can be further bounded as
\begin{equation*}
  P\left(\sup_{1\leq i\leq n}|w_{N,i}|\geq\varepsilon\right) \leq
  \frac1{\varepsilon^p}\sum_{i=1}^n\EE\left[|w_{N,i}|^p \right].
\end{equation*}
For all $n$ and $p\eqq 4+\varepsilon$ with $\varepsilon\!>\!0$, the term $\sum_{i=1}^n\EE\left[|w_{N,i}|^p\right]$ is summable and we can apply the Borel-Cantelli Lemma which implies $\sup_{1\leq i\leq n}w_{N,i}\ton 0$, almost surely.

On $\{z\inn\Cbb^+:0\!<\!C\!<\!1\}$, the $e_{N,i}(z)$ are summable and have a cluster point. Furthermore, Proposition \ref{sec:proof-exist-uniq-1} assures that the $e_{N,i}(z)$ are Stieltjes transforms and hence uniformly bounded on every closed set in $\Cbb\setminus\Rbb^+$. Therefore, Vitali's convergence theorem \cite[Theorem 3.11]{couillet2011rmm} applies, and extends the convergence region of \eqref{eq:52} to $z\inn\Cbb\setminus\Rbb^+$.

Since \eqref{eq:52} holds true, the following convergence holds almost surely
\begin{align}
  &\frac1N\tr\D^{-1} - \nonumber\\
  \label{eq:5ea}
  &\frac1N\tr\Q_N\left( \frac1N\sum_{i=1}^n\frac{\Tt_i}{1+e_{N,i}}  + \S_N - z\I_N\right)^{-1} \ton 0.
\end{align}
The convergence in \eqref{eq:5ea} implies the convergence in \eqref{eq:mN-mNo}, which completes the proof.

\section{Proof of Theorem \ref{th:2}}
\label{sec:proof-theorem:2}

The strategy is as follows: The SINR $\SINRkrzf$ in \eqref{eq:sinr} consists of three terms, (i) the scaled signal power $|\h_k^{\herm}\Wt\hht_k|^2$: (ii) the scaled interference power $\h_k^{\herm}\Wt\Ht_{[k]}^{\herm}\P_{[k]}\Ht_{[k]}\Wt\h_k$ (both scaled by $\xi^{-2}$) and (iii) the term $\Psi$ of the power normalization. For each of these three terms we will subsequently derive a deterministic equivalent which together constitute the final expression for $\SINRkrzf^\circ$.

\subsection{Deterministic equivalent for $\Psi$}
\label{sec:de-xi2}

The term $\Psi\eqq\tr\P\Ht(\Ht^\herm\Ht + M\alpha\I_M)^{-2}\Ht^{\herm}$ can be written as
\begin{align}
  \label{eq:5}
  \Psi &= \sum_{k=1}^Kp_k\hht_k^\herm\left(\Ht^\herm\Ht + M\alpha\I_M
  \right)^{-2}\hht_k \\
  \label{eq:5-1}
  &\overset{(a)}{=}
  \frac1M\sum_{k=1}^Kp_k\frac{\zh_k^\herm\Tt_k^{1/2}\C_{[k]}^{-2}\Tt_k^{1/2}\zh_k}{\left(1
      +\zh_k^\herm\Tt_k^{1/2}\C_{[k]}^{-1}\Tt_k^{1/2}\zh_k \right)^2},
\end{align}
where $\C_{[k]}\!\eqdef\! \Ga_{[k]}+\alpha\I_M$ with $\Ga_{[k]}\!\eqdef\!\frac1M\Ht_{[k]}^\herm\Ht_{[k]}$ and in $(a)$ we applied Lemma \ref{mil} twice together with \eqref{eq:3}. For $M$ large and under Assumptions \ref{as:0}, we apply Lemma \ref{sec:coll-import-lemm-1} and obtain
\begin{align*}
  &\Psi -
  \frac1M\sum_{k=1}^Kp_k\frac{\frac1M\tr\Tt_k\C_{[k]}^{-2}}{\left(1+\frac1M\tr\Tt_k\C_{[k]}^{-1}\right)^2}\tom
  0\\
  \overset{(b)}{\Leftrightarrow}~& \Psi -
  \frac1M\sum_{k=1}^Kp_k\frac{m_{\Ga,\Tt_k}^\prime(-\alpha)}{\left(1+m_{\Ga,\Tt_k}(-\alpha)\right)^2}
  \tom 0,
\end{align*}
almost surely, where in $(b)$ we applied Lemma \ref{sec:coll-import-lemm-2}, the definition \eqref{eq:det-eq} and denoted $m_{\Ga,\Tt_k}^\prime(-\alpha)$ the derivative of $m_{\Ga,\Tt_k}(z)$ along $z$ at $z\eqq -\alpha$.  Applying Theorem \ref{th:1} to $m_{\Ga,\Tt_k}(z)$, we obtain 
\begin{align*}
  &m_{\Ga,\Tt_k}(-\alpha) - \frac1M\tr\Tt_k\T \tom 0,\\
  &m_{\Ga,\Tt_k}^\prime(-\alpha) - \frac1M\tr\Tt_k\T^\prime \tom 0,
\end{align*}
almost surely, where $\T$ is defined in \eqref{eq:tt} and $\T^\prime$ is given by
\begin{equation}
  \label{eq:157}
  \T^\prime = \T\left[\frac1M\sum_{j=1}^K\frac{\Tt_j e_j^\prime}{(1 + e_j)^2} + \I_M \right]\T.
\end{equation}
Define $\e^\prime=[e_1^\prime,\dots,e_K^\prime]^\trans$ with $e_i^\prime = \frac1M\tr\Tt_i\T^\prime$. The system of $K$ equations formed by the $e_i^\prime$ takes the form $\e^\prime=\J\e^\prime + \v$ and the explicit solution $\e^\prime$ is given in \eqref{eq:166a}. Substituting $m_{\Ga,\Tt_k}(-\alpha)$ and $m_{\Ga,\Tt_k}^\prime(-\alpha)$ by their respective deterministic equivalents $e_k$ and $e_k^\prime$, we obtain $\Psi^\circ$ in \eqref{psi_rzf} such that $\Psi - \Psi^\circ\tom 0$, almost surely.

\subsection{Deterministic equivalent for $\h_k^{\herm}\Wt\hht_k$}

Similar to the derivations in \eqref{eq:5} and \eqref{eq:5-1}, we have
\begin{align*}
  &\h_k^{\herm}\Wt\hht_k =
  \frac{\z_k^\herm\Tt_k^{1/2}\C_{[k]}^{-1}\Tt_k^{1/2}\zh_k}{1+\zh_k^\herm\Tt_k^{1/2}\C_{[k]}^{-1}\Tt_k^{1/2}\zh_k}\\
    &=\frac{\sqrt{1-\tau_k^2}\z_k^\herm\Tt_k^{1/2}\C_{[k]}^{-1}\Tt_k^{1/2}\z_k}{1+\zh_k^\herm\Tt_k^{1/2}\C_{[k]}^{-1}\Tt_k^{1/2}\zh_k}
    + \frac{\tau_k\z_k^\herm\Tt_k^{1/2}\C_{[k]}^{-1}\Tt_k^{1/2}\q_k}{1+\zh_k^\herm\Tt_k^{1/2}\C_{[k]}^{-1}\Tt_k^{1/2}\zh_k}.
\end{align*}
Since $\q_k$ and $\z_k$ are independent, we apply Lemma \ref{sec:coll-import-lemm-3} together with Lemma \ref{sec:coll-import-lemm-1} and \ref{sec:coll-import-lemm-2} and obtain
\begin{equation*}
  \h_k^{\herm}\Wt\hht_k -
  \sqrt{1-\tau_k^2}\frac{m_k^\circ}{1 + m_k^\circ}\tom 0,
\end{equation*}
almost surely.

\subsection{Deterministic equivalent of $\h_k^{\herm}\Wt\Ht_{[k]}^{\herm}\P_{[k]}\Ht_{[k]}\Wt\h_k$}

With \eqref{eq:2} and $\C\!\eqdef\! \Ga+\alpha\I_M$, $\Ga\!\eqdef\!\frac1M\Ht^\herm\Ht$, we have
\begin{align}
  &\h_k^{\herm}\Wt\Ht_{[k]}^{\herm}\P_{[k]}\Ht_{[k]}\Wt\h_k \nonumber\\
  &= \frac1M\z_k^\herm\Tt_k^{1/2}\C^{-1}\Ht_{[k]}^{\herm}\P_{[k]}\Ht_{[k]}\C^{-1}\Tt_k^{1/2}\z_k\\
  &= \frac1M\z_k^\herm\Tt_k^{1/2}\C_{[k]}^{-1}\Ht_{[k]}^{\herm}\P_{[k]}\Ht_{[k]}\C^{-1}\Tt_k^{1/2}\z_k+\nonumber\\
  \label{eq:ip-1}
  &\frac1M\z_k^\herm\Tt_k^{1/2}\left[\C^{-1}-\C_{[k]}^{-1}\right]\Ht_{[k]}^{\herm}\P_{[k]}\Ht_{[k]}\C^{-1}\Tt_k^{1/2}\z_k.
\end{align}
Substituting $\C^{-1}-\C_{[k]}^{-1}\eqq-\C^{-1}(\C -\C_{[k]})\C_{[k]}^{-1}$ with $\C-\C_{[k]}\eqq \Tt_k^{1/2}(c_0\z_k\z_k^\herm+c_1\q_k\q_k^\herm+c_2\z_k\q_k^\herm + c_2\q_k\z_k^\herm)\Tt_k^{1/2}$, where $c_0\!\eqdef\!1-\tau_k^2$, $c_1\!\eqdef\!\tau_k^2$ and $c_2\!\eqdef\!\tau_k\sqrt{1-\tau_k^2}$ into \eqref{eq:ip-1}, we obtain a sum of five terms
\begin{align}
  \label{eq:11b}
  &\h_k^{\herm}\Wt\Ht_{[k]}^{\herm}\P_{[k]}\Ht_{[k]}\Wt\h_k = \frac1M\z_k^\herm\B_k\z_k\nonumber\\
  &\; - \frac{c_0}{M}\z_k^\herm\A_k\z_k\z_k^\herm\B_k\z_k - \frac{c_1}{M}\z_k^\herm\A_k\q_k\q_k^\herm\B_k\z_k\nonumber\\
  &\; - \frac{c_2}{M}\z_k^\herm\A_k\z_k\q_k^\herm\B_k\z_k - \frac{c_2}{M}\z_k^\herm\A_k\q_k\z_k^\herm\B_k\z_k,
\end{align}
where we denoted $\A_k \eqdef \Tt_k^{1/2}\C^{-1}\Tt_k^{1/2}$ and $\B_k\eqdef \Tt_k^{1/2}\C_{[k]}^{-1}\Ht_{[k]}^{\herm}\P_{[k]}\Ht_{[k]}\C^{-1}\Tt_k^{1/2}$. Noting that $c_0 +c_1\eqq 1$ and $c_0c_1-c_2^2\eqq 0$, we apply Lemma \ref{aip-lemma} to each of the four quadratic forms in \eqref{eq:11b}. Under Assumption \ref{as:0}, we obtain
\begin{align*}
  \z_k^\herm\A_k\z_k &- \frac{u(1+c_1u)}{1+u} \tom 0,\\
  \z_k^\herm\A_k\q_k &- \frac{-c_2u^2}{1+u} \tom 0,
\end{align*}
almost surely, where $u = \frac1M\tr\Tt_k\C_{[k]}^{-1}$. Moreover, under Assumptions \ref{as:0}, \ref{as:1a} and $\|\P\|<\infty$ uniformly on $M$, we have
\begin{align*}
  \z_k^\herm\B_k\z_k &- \frac{u^\prime(1+c_1u)}{1+u} \tom 0,\\
  \q_k^\herm\B_k\z_k &- \frac{-c_2uu^\prime}{1+u} \tom 0,
\end{align*}
almost surely, where $u^\prime = \frac1M\tr\P_{[k]}\Ht_{[k]}\C_{[k]}^{-1}\Tt_k\C_{[k]}^{-1}\Ht_{[k]}^{\herm}$. Substituting the random terms in \eqref{eq:11b} by their respective deterministic equivalents yields
\begin{align}
  \label{eq:47}
  &\h_k^{\herm}\Wt\Ht_{[k]}^{\herm}\P_{[k]}\Ht_{[k]}\Wt\h_k - \Bigg[
  \frac1M \frac{u^\prime(1+c_1u)}{1+u} \nonumber  \\
  \;&- \frac1M\frac{c_0(1+c_1u)^2 -c_1c_2^2u^2 - 2c_2^2u}{(1+u)^2}uu^\prime\Bigg]\tom 0,
\end{align}
almost surely. The second term in brackets of \eqref{eq:47} reduces to $\frac1M\frac{1-\tau_k^2}{(1+u)^2}uu^\prime$ and we obtain
\begin{align}
  \label{eq:48}
  &\h_k^{\herm}\Wt\Ht_{[k]}^{\herm}\P_{[k]}\Ht_{[k]}\Wt\h_k- \nonumber \\
    \;&\frac1M\frac{1 -
    \tau_k^2\left[1-(1+u)^2\right]}{(1+u)^2}u^\prime\tom 0,
\end{align}
almost surely. From Lemma \ref{sec:coll-import-lemm-2} we have
\begin{align*}
  u &- m_{\Ga,\Tt_k}(-\alpha) \tom 0,\\
  \frac1M u^\prime &- \Upsilon_k \tom 0,
\end{align*}
almost surely, where $m_{\Ga,\Tt_k}(-\alpha)\eqq\frac1M\tr\Tt_k\C^{-1}$ and $\Upsilon_k\eqq\frac{1}{M^2}\tr\P_{[k]}\Ht_{[k]}\C^{-1}\Tt_k\C^{-1}\Ht_{[k]}^\herm$. Therefore, \eqref{eq:48} becomes
\begin{align*}
  &\h_k^{\herm}\Wt\Ht_{[k]}^{\herm}\P_{[k]}\Ht_{[k]}\Wt\h_k - \\
  &\frac{\Upsilon_k\left[1-\tau_k^2\left(1-(1+m_{\Ga,\Tt_k}(-\alpha))^2\right)
    \right]}{(1+m_{\Ga,\Tt_k}(-\alpha))^2} \tom 0,
\end{align*}
almost surely. We rewrite $\Upsilon_k$ as
\begin{equation*}
  \Upsilon_k 
  = \frac{1}{M}\sum_{j=1,j\neq k}^Kp_j\zh_j^\herm\Tt_j^{1/2}\C^{-1}\Tt_k\C^{-1}\Tt_j^{1/2}\zh_j.
\end{equation*}
Applying Lemmas \ref{mil}, \ref{sec:coll-import-lemm-1} and \ref{sec:coll-import-lemm-2}, we obtain almost surely
\begin{align*}
  \Upsilon_k -
  \frac1M\sum_{j=1,j\neq k}^Kp_j\frac{\frac1M\tr\Tt_j\C^{-1}\Tt_k\C^{-1}}{\left[1
      +  \frac1M\tr\Tt_j\left(\Ga + \alpha\I_M\right)^{-1}\right]^2} \tom 0.
\end{align*}
A deterministic equivalent $e_i$ of $m_{\Ga,\Tt_i}(-\alpha) = \frac1M\tr\Tt_i\left(\Ga + \alpha\I_M\right)^{-1}$ such that $m_{\Ga,\Tt_i}(-\alpha) - e_i\tom 0$, almost surely is given in \eqref{eq:110}. To derive a deterministic equivalent for $\frac1M\tr\Tt_j\C^{-1}\Tt_k\C^{-1}$, we can assume the $\Tt_k$ invertible because the result is also a deterministic equivalent for non-invertible matrices $\Tt_k$, which is proved in \cite[Theorem 4]{hoydis2011mmm}. Define $\bar{\C}\!\eqdef\!\Tt_k^{-1/2}\Ga\Tt_k^{-1/2} + \alpha\Tt_k^{-1}$, we have
\begin{align*}
  \frac1M\tr\Tt_j\C^{-1}\Tt_k\C^{-1} &= \frac1M\tr\Tt_k^{-1/2}\Tt_j\Tt_k^{-1/2}\bar{\C}^{-2}\\
  &= \frac{d}{d z}\frac1M\tr\Tt_j(\Ga+\alpha\I_M-z\Tt_k)^{-1}.
\end{align*}
Denote $m_{\Ga-z\Tt_k,\Tt_j}(-\alpha) = \frac1M\tr\Tt_j(\Ga+\alpha\I_M-z\Tt_k)^{-1}$. Applying Theorem \ref{th:1}, we obtain $m_{\Ga-z\Tt_k,\Tt_j}(-\alpha) - \frac1M\tr\Tt_j\T_k(z)\tom 0$, almost surely, where $\T_k(z)$ is given by
\begin{equation}
  \label{eq:123}
  \T_k(z) = \left(\frac1M\sum_{j=1}^K\frac{\Tt_j}{1 + e_{j,k}(z)} +  \alpha\I_M-z\Tt_k \right)^{-1},
\end{equation}
where $e_{i,k}(z) = \frac1M\tr\Tt_i\T_k(z)$. By differentiating along $z$, we have
\begin{equation}
  \label{eq:155}
   m^\prime_{\Ga-z\Tt_k,\Tt_j}(-\alpha)- \frac1M\tr\Tt_j\T_k^\prime(z) \tom 0,
\end{equation}
almost surely, where $\T_k^\prime(z)=\frac{d}{d z}\T_k(z)$ is given by
\begin{equation*}
  \T_k^\prime(z) = \T_k(z)\left[\frac1M\sum_{j=1}^K\frac{\Tt_j e_{j,k}^\prime(z)}{(1 + e_{j,k}(z))^2} + \Tt_k \right]\T_k(z).
\end{equation*}
Setting $z=0$, we have $e_i = e_{i,k}(0)= \frac1M\tr\Tt_i\T$ with $\T=\T_k(0)$ defined in \eqref{eq:tt} and the $e_{1,k}^\prime,\dots,e_{K,k}^\prime$ are the unique positive solutions of $e_{i,k}^\prime=\frac1M\Tt_i\T_k^\prime(0)$. Define $\e_k^\prime=[e_{1,k}^\prime,\dots,e_{K,k}^\prime]^\trans$ and $\J$ and $\v_k$ as
\begin{align}
  \label{eq:160}
  [\J]_{ij} &= \frac{\frac1M\tr\Tt_i\T\Tt_j\T}{M(1+e_j)^2}, \\
  \v_k &= \left[\frac1M\tr\Tt_1\T\Tt_k\T,\dots,\frac1M\tr\Tt_K\T\Tt_k\T \right]^\trans.
\end{align}
Therefore, $\e_k^\prime$ is given explicitly as 
\begin{equation}
  \label{eq:159}
  \e_k^\prime = \left(\I_K - \J \right)^{-1}\v_k.
\end{equation}
Note that $\I_K - \J$ is always invertible since $\e_k^\prime$ is a unique positive solution. Finally, substituting $m_{\Ga,\Tt_j}(-\alpha)$ and $\frac1M\tr\Tt_j\C^{-1}\Tt_k\C^{-1}$ by their respective deterministic equivalents $e_j$ and $e_{j,k}^\prime$, we obtain $\Upsilon_k^\circ$ in \eqref{ups_rzf} such that $\Upsilon_k - \Upsilon_k^\circ \tom 0$, almost surely.

If all available transmit power is allocated to a single user (i.e., $p_k=P$), both $\Psi^\circ$ and $\Upsilon_k^\circ$ are of order $O(1/M)$ and hence $\SINRkrzf^\circ$ grows unbounded with $M$. Therefore,  we require Assumption \ref{as:0ap} to ensure that the convergence in \eqref{eq:theorem-rzf-sinr-1} holds true, which completes the proof.

\section{Proof of Theorem \ref{th:3}}
\label{sec:proof-theorem:3}

We bound $|\SINRkzf - \SINRkzf^\circ|$ by adding and subtracting $\SINRkrzf(\alpha)$ and $\dSINRkrzf(\alpha)$ and applying the triangle inequality. We obtain
  \begin{align}
    \label{eq:4}
    |\SINRkzf - \SINRkzf^\circ| \leq &|\SINRkzf - \SINRkrzf(\alpha)| + |\SINRkrzf(\alpha) - \dSINRkrzf(\alpha)| \nonumber \\
    &  + |\dSINRkrzf(\alpha) - \dSINRkzf|.
  \end{align}
  To show that $|\SINRkzf - \dSINRkzf|\to 0$ almost surely as $M,K\to\infty$, take $\varepsilon > 0$ arbitrarily small. For $\alpha>0$ small enough, we will demonstrate that $|\SINRkzf - \SINRkrzf(\alpha)|<\frac{\varepsilon}{3}$ almost surely and $|\dSINRkrzf(\alpha) - \dSINRkzf|<\frac{\varepsilon}{3}$ independently of $M$ and $K$. Furthermore, we show that for $M,K$ large enough, $|\SINRkrzf(\alpha) - \dSINRkrzf(\alpha)|<\frac{\varepsilon}{3}$ almost surely, from which we conclude that \eqref{eq:4} can be made as small as desired.

  In order to prove that $|\SINRkzf - \SINRkrzf(\alpha)|<\frac{\varepsilon}{3}$ for $\alpha$ small enough, it suffices to study the matrices $\Wt=(\Ht^\herm\Ht + M\alpha\I_M)^{-1}$ and $\underline{\hat{\W}}=\Ht^\herm(\Ht\Ht^\herm)^{-2}\Ht$ in the SINR of RZF precoding \eqref{eq:sinr} and ZF precoding \eqref{eq:sinr-zf}. Applying the matrix inversion lemma, $\Wt$ takes the form
  \begin{align*}
    \label{eq:28}
    \Wt = \Ht^\herm(\Ht\Ht^\herm + M\alpha\I_K)^{-2}\Ht + M\alpha(\Ht^\herm\Ht + M\alpha\I_M)^{-2}.
  \end{align*}
  Under Assumption \ref{as:1}, $\lambda_{\min}(\Ht\Ht^\herm)>\varepsilon>0$ and, since $\lambda_{\max}(\Ht\Ht^\herm)$ is almost surely bounded for all large $M,K$, for any continuous functional $f(\Wt)$ we have $|f(\Wt) - f(\underline{\hat{\W}})|\toa 0$ with probability one. Therefore, $|\SINRkzf - \SINRkrzf(\alpha)|\toa 0$ uniformly on $M,K$ almost surely. 

From Theorem \ref{th:2}, we have immediately that for any $\alpha>0 $, $|\SINRkrzf(\alpha) - \dSINRkrzf(\alpha)| \tom 0$ almost surely. 


In order to prove $|\dSINRkrzf(\alpha) - \dSINRkzf| < \frac{\varepsilon}{3}$ for $\alpha$ small enough, uniformly on $M$, rewrite $\dSINRkrzf(\alpha)$ as 
  \begin{equation}
    \label{eq:49}
    \dSINRkrzf(\alpha)\! =\! \frac{p_k(1-\tau_k^2)\left(\alpha e_k\right)^2}
    {\Upsilon_k^\circ(\alpha^2 - \tau_k^2[ \alpha^2 - (\alpha + \alpha e_k)^2])+\frac{\Psi^\circ}{\SNR}(\alpha+\alpha e_k)^2}.
  \end{equation}
  To show that $\dSINRkzf = \lim_{\alpha\to 0}\dSINRkrzf(\alpha)$, we need to verify that the limit $\alpha\to 0$ of both numerator and denominator in \eqref{eq:49} exists and that the denominator is uniformly bounded away from zero. Define $\underline{e}_{i} = \lim_{\alpha\to 0}\alpha e_i(\alpha)$. Under Assumption \ref{as:1b}, all $\underline{e}_{i}$ exist and are strictly positive. Since $\alpha e_i(\alpha)$ is holomorphic for $\alpha> 0$, and is bounded away from zero in a neighborhood of zero, by continuity extension in $\alpha=0$, we obtain the limit $\alpha\to 0$ as
  \begin{align}
    \underline{e}_{i} &= \lim_{\alpha\to 0}\left\{\frac1M\tr\Tt_i\left(\frac1M\sum_{j=1}^K\frac{\Tt_j}{\alpha + \alpha e_j(\alpha)} + \I_M \right)^{-1} \right\}\nonumber\\
    \label{eq:40}
      &= \frac1M\tr\Tt_i\underline{\T},
  \end{align}
  where $\underline{\T}$ is given in \eqref{eq:6tzf}. It is easy to verify that $\underline{e} \eqdef \sup_i\underline{e}_{i}$ is uniformly bounded on $M$. We have
  \begin{equation}
    \label{eq:39}
    |\underline{e}| \leq \sup_i\|\Tt_i\|.
  \end{equation}
  Define $\underline{\e}\eqdef [\underline{e}_1,\dots,\underline{e}_K]^\trans$, $f_i: \underline{\e} \mapsto \frac1M\tr\Tt_i\underline{\T}(\underline{\e})$ and $\f(\underline{\e})=[f_1(\underline{\e}),\dots,f_K(\underline{\e})]^\trans$. Under Assumption \ref{as:1b}, there exists a fixed point $\f(\underline{\e}^*)=\underline{\e}^*$, where $\underline{\e}^*\eqdef [\underline{e}^*_1,\dots,\underline{e}^*_K]^\trans$ with $\underline{e}^*_i>0~\forall i$. In this case, we can extend the results in \cite{yates1995fup}\footnote{Since $\f(\underline{\e})$ can be extended by continuity in zero, where it satisfies $\f(0)=0$, the positivity property of $\f(\underline{\e})$, defined in \cite{yates1995fup}, does not hold. We precisely need to show that $\underline{\e}^{(n+1)} = \f(\underline{\e}^{(n)})$ can not converge to the fixed point 0, which unfolds from Assumption  \ref{as:1b} with similar arguments as in \cite{yates1995fup}.} and show that the iterative fixed point algorithm defined by $\underline{\e}^{(n+1)} = \f(\underline{\e}^{(n)})$, ($n\geq 0$), converges to the unique positive solution $\underline{\e}^*$ for any initial point $\underline{\e}^{(0)}$, $\underline{e}_i^{(0)}>0~\forall i$.


Furthermore, we need to show that both $\underline{\Upsilon}_k^\circ = \lim_{\alpha\to 0}\Upsilon_k^\circ$ and $\underline{\Psi}^\circ = \lim_{\alpha\to 0}\Psi^\circ$ exist and are uniformly bounded on $M$. Observe that
\begin{equation}
  \label{eq:25}
  \lim_{\alpha\to 0}\alpha^2e_i^\prime = \underline{e}_{i}
\end{equation}
and we obtain 
\begin{equation}
  \label{eq:27}
  \underline{\Psi}^\circ = \lim_{\alpha\to 0}\frac1M\sum_{j=1}^Kp_j\frac{\alpha^2e_j^\prime}{(\alpha + \alpha e_j)^2} = \frac1M\sum_{j=1}^K\frac{p_j}{\underline{e}_j}.
\end{equation}
Therefore, $0<\underline{\Psi}^\circ<\infty$ for all $\underline{e}_i>0$. Similarly, define $\underline{e}_{j,k}^\prime = \lim_{\alpha\to 0}\alpha^2e_{j,k}^\prime$ given in \eqref{eq:28df} and thus
\begin{equation}
  \label{eq:6}
  \underline{\Upsilon}_k^\circ = \lim_{\alpha\to 0}\frac1M\sum_{j=1,j\neq k}^Kp_j\frac{\alpha^2 e_{j,k}^\prime}{(\alpha + \alpha e_j)^2} = \frac1M\sum_{j=1,j\neq k}^Kp_j\frac{\underline{e}_{j,k}^\prime}{\underline{e}_j^2},\nonumber
\end{equation}
satisfying $0<\underline{\Upsilon}_k^\circ<\infty$ for all $\underline{e}_{i} > 0$. To fulfill the constraints $\underline{e}_i>0$, we have to evoke Assumption \ref{as:1}. The limit $\dSINRkzf = \lim_{\alpha\to 0}\dSINRkrzf(\alpha)$ is given by \eqref{eq:de-zf-puc}, which completes the proof.

\section{Proof of Proposition \ref{pr:1}}
\label{sec:proof-proposition}

The proof is inspired by \cite{muharar2009dbt} with adaptations to account for imperfect CSIT. From Corollary \ref{sec:regul-zero-forc-1} with $p_k=P/K~\forall k$ and $\tau_k=\tau~\forall k$, for large $M,K$, the SINR $\dSINRrzf$ takes the form
\begin{equation*}
  \SINRrzf^\circ = \SNR\beta m^\circ (1-\tau^2)\Gamma,
\end{equation*}
where
\begin{equation*}
  \Gamma = \frac{\frac1\beta e_{22} + \alpha (1+m^\circ)^2e_{12}}{\SNR
    e_{22}(1-\tau^2)+\tau^2\SNR(1+m^\circ)^2e_{22} + (1+m^\circ)^2e_{12}}
\end{equation*}
with $m^\circ$ and $e_{ij}$ defined in \eqref{eq:32} and \eqref{eq:69}, respectively. Taking the derivative along $\alpha$, we obtain
\begin{align}
\label{eq:61}
  \frac{\partial \SINRrzf^\circ}{\partial \alpha} = \SNR\beta m^\circ
  (1-\tau^2)\Gamma\left[\frac{m^{\prime\circ}}{m^\circ} +
    \frac{\Gamma^\prime}{\Gamma} \right],
\end{align}
where 
\begin{equation}
  \label{eq:72}
  m^{\prime\circ} = -\frac{(1+m^\circ)^2e_{12}}{1-\frac{e_{22}}{\beta}}.
\end{equation}
and thus, together with \eqref{eq:59}, we have
\begin{equation*}
  \frac{m^{\prime\circ}}{m^\circ} = -\frac{(1+m^\circ)^2e_{12}}{\frac1\beta e_{22} + \alpha(1+m^\circ)^2e_{12}}.
\end{equation*}
Therefore, \eqref{eq:61} becomes
\begin{align}
  \frac{\partial \SINRrzf^\circ}{\partial \alpha} &= \SNR\beta m^\circ
  (1-\tau^2)\Gamma\nonumber\\
  &\times\Bigg[\frac{2\alpha (1+m^\circ)m^{\prime\circ}
    e_{12}+\alpha(1+m^\circ)^2e_{12}^\prime+ \frac1\beta e_{22}^\prime} {\frac1\beta e_{22} +
    \alpha(1+m^\circ)^2e_{12}}\nonumber \\
  &- \frac{[1-\tau^2+\tau^2(1+m^\circ)^2]\SNR e_{22}^\prime +
    2\tau^2\SNR(1+m^\circ)m^{\prime\circ} e_{22}}
  {[1-\tau^2+\tau^2(1+m^\circ)^2]\SNR
    e_{22} + (1+m^\circ)^2e_{12}}\nonumber\\
  \label{eq:95}
  &- \frac{2(1+m^\circ)m^{\prime\circ} e_{12}+(1+m^\circ)^2e_{12}^\prime}
  {[1-\tau^2+\tau^2(1+m^\circ)^2]\SNR e_{22} + (1+m^\circ)^2e_{12}}\Bigg].
\end{align}
Denoting $\chi\eqdef (1+m^\circ)^2e_{12}$, $\psi\eqdef 2(1+m^\circ)m^{\prime\circ} e_{12}+(1+m^\circ)^2e_{12}^\prime$ and $\phi\eqdef 1-\tau^2+\tau^2(1+m^\circ)^2$, \eqref{eq:95} takes the form
\begin{align}
  \frac{\partial \SINRrzf^\circ}{\partial \alpha} &= \SNR\beta m^\circ
  (1-\tau^2)\Gamma\nonumber\\
  &\times\left[\frac{\frac1\beta
      e_{22}^\prime+\alpha\psi}{\frac1\beta e_{22} + \alpha\chi} -
    \frac{\SNR\phi e_{22}^\prime + \psi + 2\tau^2\SNR(1+m^\circ)m^{\prime\circ}
      e_{22}}{\SNR\phi e_{22} + \chi} \right]\nonumber\\
  &= \frac{\phi\SNR^2\beta m^\circ
    (1-\tau^2)\Gamma}{Z}\Bigg[\left(\alpha-\frac{1}{\beta\SNR\phi}
  \right)(e_{22}\psi-e_{22}^\prime\chi)\nonumber\\
  &\qquad-\frac{2\tau^2(1+m^\circ)m^{\prime\circ}e_{22}[\frac{e_{22}}{\beta}+\alpha\chi]}{\phi} \Bigg]\nonumber,
\end{align}
where $Z = (\frac1\beta e_{22} + \alpha\chi)(\SNR\phi e_{22} + \chi)$. Denoting 
\begin{align}
\Omega&\eqdef\frac{2\phi\SNR^2\beta m^\circ (1-\tau^2)(1+m^\circ)m^{\prime\circ} e_{12}e_{22}\Gamma}{Z} \nonumber \\
  \label{eq:73}
\nu &\eqdef \frac{(1+m^\circ)^2[e_{12}^\prime e_{22} - e_{12}e_{22}^\prime]}{2(1+m^\circ)m^{\prime\circ}e_{12}e_{22}},
\end{align}
we obtain  
\begin{align}
  \label{eq:98}
   \frac{\partial \SINRrzf^\circ}{\partial \alpha} = \Omega \left[\left(\alpha-\frac{1}{\beta\SNR\phi}\right)(1+\nu)-\frac{\tau^2[\frac{e_{22}}{\beta}+\alpha\chi]}{\phi e_{12}}\right].
\end{align}
Rewriting the term in brackets in \eqref{eq:98}, we have
\begin{equation*}
  \frac{\partial \SINRrzf^\circ}{\partial \alpha} = \Omega\left[\alpha - \frac{[1+\nu+\tau^2\SNR \frac{e_{22}}{e_{12}}]\frac{1}{\beta\SNR}}{(1-\tau^2)(1+\nu) + \tau^2\nu(1+m^\circ)^2}\right]=0.
\end{equation*}
Since $\Omega\neq 0$ for $\SNR>0$ and $\tau^2<1$, the optimal regularization parameter $\alpha^{\star\circ}$ is given by \eqref{eq:alpha-optimal}. Substituting \eqref{eq:72} into \eqref{eq:73}, the term $\nu$ takes the form
\begin{equation}
  \label{eq:74}
  \nu=\frac{1-\frac{e_{22}}{\beta}}{2(1+m^\circ)e_{12}}\frac{e^\prime_{12}}{e_{22}}\left[\frac{e_{22}^\prime}{e_{12}^\prime} - \frac{e_{22}}{e_{12}} \right].
\end{equation}
With \eqref{eq:59} and \eqref{eq:72}, we obtain $e_{12}^\prime=\frac{-2e_{13}}{1-e_{22}/\beta}$ and $e_{22}^\prime=\frac{-2e_{23}}{1-e_{22}/\beta}$. Substituting these terms into \eqref{eq:74} yields \eqref{eq:12a}, which completes the proof.

\section{Proof of Proposition \ref{sec:high-snr-regime}}
\label{sec:proof-proposition-1}

The sum rate $\Rsumh$ can be written as a function of the per-user rate under perfect CSIT $\bar{R}^\circ$ and the per-user rate gap $\Delta R^\circ$ as
\begin{equation*}
  \Rsumh = K\left(1 - \frac{T_t}{T}\right)\left[\bar{R}^\circ - \Delta R^\circ \right],
\end{equation*}
where for ZF and RZF-CDA we have $\bar{R}^\circ_{\rm zf}\eqq\log(1+\SNR_{dl}(\beta-1))$ and
$\bar{R}^\circ_{\rm rzf}\eqq\log(\frac12 +\frac12 \SNR_{dl}(\beta-1) + \frac{\chi(1)}{2})$, respectively, and
\begin{align*}
  \Delta R^\circ_{\rm zf} &=
  \log\left(\frac{(\beta-1)(\SNR_{dl}+1)}{1+\frac1{\SNR_{dl}}+T_{t,{\rm
          zf}}[\frac1c+\SNR_{ul}(\beta-1)]} \right),\\
  \Delta R^\circ_{\rm rzf} &=\log\left(\frac{1+\SNR_{dl}(\beta-1)+\chi(1)}{1+\omega\SNR_{dl}(\beta-1)+\chi(\omega)} \right),
\end{align*}
where $\chi(\omega)$ is defined in \eqref{eq:tdd:5-d}. Denoting $\psi\!\eqdef\!1+\frac1{\SNR_{dl}}+T_{t,{\rm zf}}[\frac1c+\SNR_{ul}(\beta-1)]$, the derivatives take the form
\begin{align}
  \label{eq:10}
  \frac{\partial\Rsumh^{\rm zf}}{\partial T_{t,{\rm zf}}} =
  &-\frac{K}{T}(\bar{R}^\circ_{\rm zf}-\Delta R^\circ_{\rm zf})+K\left(1-\frac{T_{t,{\rm zf}}}{T}\right)\nonumber\\ 
  &\times \frac{(\beta-1)(\SNR_{dl}+1)[\frac1c+\SNR_{ul}(\beta-1)]}{\psi^2
  + (\beta-1)(\SNR_{dl}+1)\psi}, \\
\label{eq:11}
 \frac{\partial\Rsumh^{\rm rzf}}{\partial T_{t,{\rm rzf}}} =
 &-\frac{K}{T}(\bar{R}^\circ_{\rm rzf}-\Delta R^\circ_{\rm rzf})\nonumber\\ &+ 
K\left(1-\frac{T_{t,{\rm rzf}}}{T}\right)\frac{\omega^\prime\SNR_{dl}(\beta-1)+\chi^\prime}{1+\omega\SNR_{dl}(\beta-1)+\chi},
\end{align}
where $\omega^\prime\eqq\partial \omega/\partial T_{t,{\rm rzf}}\eqq(1/\SNR_{ul}+c)/(T_{t,{\rm rzf}}+1/\SNR_{ul}+c)^2$ and $\chi^\prime\eqq\partial \chi/\partial T_{t,{\rm  rzf}}\eqq[(\beta-1)^2\omega\omega^\prime\SNR_{dl}^2+\omega^\prime\SNR_{dl}(1+\beta)+1]/\chi$.
In \eqref{eq:10} and \eqref{eq:11} the per-user rate-gap $\Delta R^\circ_{\rm zf}$ and $\Delta R^\circ_{\rm rzf}$ can be neglected, since at high SNR $\Delta R^\circ_{\rm zf}\!\ll\! \bar{R}^\circ_{\rm zf}$ and $\Delta R^\circ_{\rm rzf}\!\ll\! \bar{R}^\circ_{\rm rzf}$, respectively. Treating $\bar{R}^\circ_{\rm zf},\bar{R}^\circ_{\rm rzf}$ as constant, for $\SNR_{dl},\SNR_{ul}\to\infty$ and $c\eqq\SNR_{dl}/\SNR_{ul}$ finite, solving \eqref{eq:10} and \eqref{eq:11} for $T_{t,{\rm zf}}$ and $T_{t,{\rm rzf}}$, respectively, yields \eqref{eq:12} and \eqref{eq:13}, respectively, which completes the proof.

\section{Important Lemmas}
\label{sec:coll-import-lemm}

\begin{lemma}[Matrix Inversion Lemma]
  \label{mil}\cite[Lemma 2.2]{silverstein1995empirical} Let $\U$ be an $N\times N$ invertible matrix and $\x\inn\Cbb^N$, $c\inn\Cbb$ for which $\U+c\x\x^\herm$ is invertible. Then
  \begin{equation*}
    \x^\herm\left(\U+c\x\x^\herm \right)^{-1} = \frac{\x^\herm\U^{-1}}{1+c\x^\herm\U^{-1}\x}.
  \end{equation*}
\end{lemma}

\begin{lemma}[Resolvent Identity]
  \label{resolvent-identity}
  Let $\U$ and $\V$ be two invertible complex matrices of size $N\! \times \!N$. Then
  \begin{equation*}
    \label{eq:resolvent-identity}
    \U^{-1} - \V^{-1} = -\U^{-1}(\U-\V)\V^{-1}.
  \end{equation*}
\end{lemma}


\begin{lemma}
  \label{lemma-bai-silverstein}\cite[Lemma B.26]{bai2010sal} Let $\A\in\Cbb^{N\times N}$ be a deterministic matrix and $\x\inn\Cbb^N$ have i.i.d. complex entries of zero mean, variance $1/N$ and bounded $l$th order moment $\EE|x_i|^l\leq \nu_l$. Then for any $p\geq 1$ 
  \begin{equation}
    \EE\left\vert \x^{\herm}\A\x-\frac{1}{N} \tr\A \right\vert^p
    \leq
    \frac{C_p}{N^{p/2}}\left(\frac{1}{N}\tr\A\A^\herm\right)^{p/2}\left[\nu_4^{p/2}
      + \nu_{2p}\right],
  \end{equation}
  where $C_p$ is a constant solely depending on $p$.
\end{lemma}



\begin{lemma}
  \label{sec:coll-import-lemm-1}\cite[Lemma 14.2]{couillet2011rmm}
  Let $\A_1, \A_2,\dots$, with $\A_N\inn\Cbb^{N\times N}$, be a series of random matrices generated by the probability space $(\Omega,\mathcal{F},P)$ such that, for $\omega\inn A\subset\Omega$, with $P(A)=1$, $\|\A_N(\omega)\|<K(\omega)<\infty$, uniformly on $N$. Let $\x_1, \x_2,\dots$, with $\x_N\inn\Cbb^N$, be random vectors of i.i.d.\ entries with zero mean, variance $1/N$ and eighth order moment of order $O(1/N^4)$, independent of $\A_N$. Then
  \begin{equation*}
    \x_N^{\herm}\A_N\x_N - \frac{1}{N} \tr\A_N \ton 0,
  \end{equation*}
  almost surely.
\end{lemma}
\begin{IEEEproof}
  The proof unfolds from a direct application of the Tonelli theorem, \cite[Theorem 18.3]{billingsley2008probability}. Denoting $(X, \mathcal{X}, P_X)$ the probability space that generates the series $\x_1, \x_2,\ldots,$ we have that for every $\omega\in A$ (i.e., for every realization $\A_1(\omega), \A_2(\omega),\ldots)$, the trace lemma, \cite[Theorem 3.4]{couillet2011rmm}, holds true. From \cite[Theorem 18.3]{billingsley2008probability}, the space $B$ of couples $(x,\omega)\in Y\eqdef X\times \Omega$ for which the trace lemma holds, satisfies
\begin{equation*}
  \int_{Y}1_B(x,\omega)dP_{Y}(x,\omega)\!=\!\int_\Omega\int_X1_B(x,\omega)dP_X(x)dP_\Omega(\omega).
\end{equation*}
If $\omega\in A$, then $1_B(x,\omega)=1$ on a subset of $X$ of probability one. Therefore, the inner integral equals one whenever $\omega\in A$. As for the outer integral, since $P(A)=1$, it also equals one, and the result is proved.
\end{IEEEproof}

\begin{lemma}
  \label{sec:coll-import-lemm-3}
  Let $\A_N$ be as in Lemma \ref{sec:coll-import-lemm-1} and $\x_N,\y_N\inn\Cbb^N$ be random, mutually independent with standard i.i.d.\ entries of zero mean, variance $1/N$ and eighth order moment of order $O(1/N^4)$, independent of $\A_N$.
\begin{equation*}
  \y_N^{\herm}\A_N\x_N \overset{N\to \infty}{\longrightarrow} 0,
\end{equation*}
almost surely.
\end{lemma}
\begin{IEEEproof}
  Remark that $\EE\left[|\y_N^{\herm}\A_N\x_N|^4\right]\!<\!c/N^2$ for some constant $c\!>\!0$ independent of $N$. The result then unfolds from the Markov inequality the Borel-Cantelli Lemma \cite{billingsley2008probability} and the Tonelli Theorem \cite[Theorem 18.3]{billingsley2008probability}.
\end{IEEEproof}

\begin{lemma}
  \label{sec:coll-import-lemm-2}\cite[Lemma 14.3]{couillet2011rmm}
  Let $\A_1,\A_2,\ldots$, with $\A_N\inn\Cbb^{N\times N}$, be deterministic with uniformly bounded spectral norm and $\B_1,\B_2,\ldots$, with $\B_N\inn\Cbb^{N\times N}$, be random Hermitian, with eigenvalues $\lambda_1^{\B_N}\leq \ldots\leq \lambda_N^{\B_N}$ such that, with probability one, there exist $\varepsilon>0$ for which $\lambda^{\B_N}_1>\varepsilon$ for all large $N$. Then for $\v\inn\Cbb^N$
\begin{equation*}
  \frac1N\tr \A_N\B_N^{-1} - \frac1N \tr\A_N(\B_N+\v\v^\herm)^{-1} \ton 0
\end{equation*}
almost surely, where $\B_N^{-1}$ and $(\B_N+\v\v^\herm)^{-1}$ exist with probability one.
\end{lemma}
\begin{IEEEproof}
  The proof unfolds similarly as above, with some particular care to be taken. For $\omega\in B$, the smallest eigenvalue of $\B_N (\omega)$ is uniformly greater than $\varepsilon(\omega)$. Therefore, with $\B_N(\omega)$ and $\B_N(\omega) + \v\v^\herm$ invertible and, taking $z=-\varepsilon(\omega)/2$, we can write
\begin{align*}
  &\frac1N\tr \A_N\B_N^{-1}(\omega) \\ 
  &=\frac1N\tr
  \A_N\left(\left[\B_N(\omega) -\frac{\varepsilon(\omega)}{2}\I_N \right] +         \frac{\varepsilon(\omega)}{2}\I_N\right)^{-1}
\end{align*}
and 
\begin{align*}
  &\frac1N\tr\A_N\left(\B_N(\omega)+\v\v^\herm\right)^{-1} \\
  &= \frac1N\tr
  \A_N\left(\left[\B_N(\omega) +\v\v^\herm -\frac{\varepsilon(\omega)}{2}\I_N \right] +     \frac{\varepsilon(\omega)}{2}\I_N\right)^{-1}.
\end{align*}
Under these notations, $\B_N(\omega)-\frac{\varepsilon(\omega)}{2}\I_N$ and $\B_N(\omega) +\v\v^\herm -\frac{\varepsilon(\omega)}{2}\I_N$ are still nonnegative definite for all $N$. Therefore, the rank-1 perturbation lemma, \cite[Lemma 2.1]{bai2007sir}, can be applied for this $\omega$. But then, from the Tonelli theorem again, in the space that generates the couples $((\x_1, \x_2,\ldots), (\B_1, \B_2,\dots))$, the subspace where the rank-1 perturbation lemma applies has probability one, which completes the proof.
\end{IEEEproof}

\begin{lemma}
  \label{aip-lemma}
  Let $\U,\V,\Tt\in\Cbb^{N\times N}$ be of uniformly bounded spectral norm with respect to $N$ and let $\V$ be invertible. Further, define $\x\eqdef\Tt^{1/2}\z$ and $\y\eqdef\Tt^{1/2}\q$ where $\z,\q\inn\Cbb^N$ have i.i.d.\ complex entries of zero mean, variance $1/N$ and finite $8$th order moment and be mutually independent as well as independent of $\U,\V$. Define $c_0,c_1,c_2\in\Rbb^+$ such that $c_0c_1-c_2^2\geq 0$ and let $u\eqdef\frac{1}{N}\tr\Tt\V^{-1}$ and $u^\prime\eqdef\frac{1}{N}\tr\Tt\U\V^{-1}$. Then we have
  \begin{align*}
    &\x^\herm\U\left(\V + c_0\x\x^\herm + c_1\y\y^\herm + c_2\x\y^\herm + c_2\y\x^\herm\right)^{-1}\x \\
    &-\frac{u^\prime(1+c_1u)}{(c_0c_1-c_2^2)u^2+(c_0+c_1)u+1}
    \overset{N\to \infty}{\longrightarrow} 0,
  \end{align*}
  almost surely. Furthermore,  
  \begin{align*}
    &\x^\herm\U\left(\V + c_0\x\x^\herm + c_1\y\y^\herm + c_2\x\y^\herm + c_2\y\x^\herm\right)^{-1}\y \\     &-\frac{-c_2 u u^\prime}{(c_0c_1-c_2^2)u^2+(c_0+c_1)u+1}
    \overset{N\to \infty}{\longrightarrow} 0,
  \end{align*}
  almost surely.
\end{lemma}
\begin{IEEEproof}
Denote $\V\eqq(\A\! +\! c_0\x\x^\herm\! +\! c_1\y\y^\herm\! +\! c_2\x\y^\herm
 \! + \!c_2\y\x^\herm)^{-1}$. Now $\x^\herm\U\V\x$ can be resolved
using Lemma \ref{resolvent-identity}
\begin{align}
  &\x^\herm\U\V\x - \x^\herm\U\A^{-1}\x = \x^\herm\U\V\left(\V^{-1} - \A\right)\A^{-1}\x\nonumber\\
  \label{eq:proof-aip-resolvent}
  &= -\x^\herm\U\V(c_0\x\x^\herm + c_1\y\y^\herm + c_2\x\y^\herm +
  c_2\y\x^\herm)\A^{-1}\x.
\end{align}
Rewrite \eqref{eq:proof-aip-resolvent} as
\begin{align*}
  \x^\herm\U\V\x \!=\! \frac{\x^\herm\U\A^{-1}\x \!-\!
  \x^\herm\U\V\y(c_1\y^\herm\A^{-1}\x+c_2\x^\herm\A^{-1}\x)}{1+c_0\x^\herm\A^{-1}\x
    + c_2\y^\herm\A^{-1}\x}.
\end{align*}
Similarly to \eqref{eq:proof-aip-resolvent}, we apply Lemma \ref{resolvent-identity} to $\x^\herm\U\V\y$. Thus, we obtain an expression involving the terms $\x^\herm\U\A^{-1}\x$, $\y^\herm\A^{-1}\y$, $\x^\herm\U\A^{-1}\y$ and $\y^\herm\A^{-1}\x$. To complete the proof, we apply Lemma \ref{sec:coll-import-lemm-1} and Lemma \ref{sec:coll-import-lemm-3}, with $u\eqq\frac1{N}\tr\Tt\A^{-1}$ and $u^\prime\eqq\frac1{N}\tr\Tt\U\A^{-1}$ and obtain
\begin{align}
  \label{eq:proof-aip-lim:1}
  \x^\herm\U\V\x -
  \frac{u^\prime(1+c_1u)}{(c_0c_1-c_2^2)u^2 + (c_0+c_1)u + 1}\overset{N\to
    \infty}{\longrightarrow} 0, 
\end{align}
almost surely. Similarly we have
\begin{align}
  \label{eq:proof-aip-lim:2}
  \x^\herm\U\V\y -
  \frac{-c_2 u u^\prime}{(c_0c_1-c_2^2)u^2 + (c_0+c_1)u + 1}\overset{N\to
    \infty}{\longrightarrow} 0,
\end{align}
almost surely. Note that as $c_0,c_1,c_2\inn\Rbb^+$ and $c_0c_1\geq c_2^2$, the convergence in \eqref{eq:proof-aip-lim:1} and \eqref{eq:proof-aip-lim:2} still holds since $(c_0c_1-c_2^2)u^2 + (c_0+c_1)u + 1$ is bounded away from zero, which completes the proof.
\end{IEEEproof}

\begin{lemma}
  \label{lemma:rank1-bai}
  \cite[Lemma 2.1]{bai2007sir} Let $\zeta>0$, $\B,\A \inn \Cbb^{N\times N}$ with $\B$ Hermitian nonnegative definite, $\tau\inn \Rbb$ and $\q\inn\Cbb^N$. Then
  \begin{align*}
    \left\vert \tr\A\left[ (\B+\zeta\I_N)^{-1} -
        (\B+\tau\q\q^\herm+\zeta\I_N)^{-1} \right] \right\vert \leq
    \frac{\|\A\|}{\zeta}.
  \end{align*}
\end{lemma}

\begin{lemma}
  \label{lem:2}\cite[Corollary 2.2]{couillet2011rmm}
  Let $z\inn\Cbb^+$, $t\!>\!0$, $\q\inn\Cbb^N$ and $\B\inn\Cbb^{N\times N}$ Hermitian nonnegative definite. Then
  \begin{equation*}
    \left|\frac{1}{1+t\q^\herm\left(\B + z\I_N \right)^{-1}\q}\right|
    \leq \frac{|z|}{\Im z}.
  \end{equation*}
\end{lemma}

\begin{lemma}
  \label{sec:important-lemmas-2}
  Let $\q\inn\Cbb^N$ and $\A\inn\Cbb^{N\times N}$ Hermitian nonnegative definite, then
  \begin{equation*}
    \q^\herm\A\q \leq \|\A\|\|\q\|_2^2.
  \end{equation*}
\end{lemma}

\begin{lemma}
  \label{lem-4}
  Let $\A\inn\Cbb^{N\times N}$ be Hermitian nonnegative-definite, then
  \begin{equation*}
    \frac1N\tr\A\leq\|\A\|.
  \end{equation*}
\end{lemma}

\begin{lemma}
  \label{sec:important-lemmas-1}
  Let $\A,\B\inn\Cbb^{N\times N}$ Hermitian nonnegative-definite, then
  \begin{equation*}
    \|\A\B\|\leq\|\A\|\|\B\|.
  \end{equation*}  
\end{lemma}

\section*{Acknowledgements}
This research has been partially supported by EU NoE Newcom++ and ANR
project SESAME.


\bibliographystyle{/home/sebastian/texmf/tex/latex/IEEEtran/IEEEtran}
\bibliography{/home/sebastian/texmf/tex/latex/IEEEtran/IEEEabrv,/home/sebastian/texmf/tex/latex/IEEEtran/IEEEconf,/home/sebastian/work/references/references}

\end{document}